\pdfoutput=1

\documentclass[11pt]{article}
\usepackage[dvipsnames,table,xcdraw]{xcolor}
\usepackage[super,sort&compress,comma]{natbib} 
\usepackage[version=3]{mhchem}
\usepackage[left=1.5cm, right=1.5cm, top=1.785cm, bottom=2.0cm]{geometry}
\usepackage{doi}
\usepackage{balance}
\usepackage{amssymb}
\usepackage{mathptmx}
\usepackage{fancyvrb}
\usepackage{inconsolata}[scaled=0.95]
\usepackage{sectsty}
\usepackage{lastpage}
\usepackage[format=plain,justification=justified,singlelinecheck=false,font={stretch=1.125,small,sf},labelfont=bf,labelsep=space]{caption}
\usepackage{subcaption}
\usepackage{float}
\usepackage{fancyhdr}
\usepackage{fnpos}
\usepackage[english]{babel}
\addto{\captionsenglish}{%
  
}
\usepackage{array}
\usepackage{droidsans}
\usepackage{charter}
\usepackage[T1]{fontenc}
\usepackage{orcidlink}
\usepackage{siunitx}
\DeclareSIUnit{\atom}{atom}
\usepackage{setspace}
\usepackage[compact]{titlesec}
\usepackage{hyperref}

\usepackage{epstopdf}
\usepackage{multicol}
\usepackage{tabularray}
\usepackage{booktabs}
\usepackage{booktabs} 

\title{Accelerating the discovery of high-performance nonlinear optical materials using active learning and high-throughput screening}

\author{Victor~Trinquet,\orcidlink{0009-0001-2188-4824}\textit{$^{a}$} \and
Matthew~L.~Evans,\orcidlink{0000-0002-1182-9098}\textit{$^{ab}$} \and
Gian-Marco~Rignanese \orcidlink{0000-0002-1422-1205}$^{acd}$}

\date{}

\begin{document}


\renewcommand{\thefootnote}{\fnsymbol{footnote}}
\setcounter{secnumdepth}{5}

\makeatletter 
\renewcommand\@biblabel[1]{#1}            
\renewcommand\@makefntext[1]%
{\noindent\makebox[0pt][r]{\@thefnmark\,}#1}
\makeatother 
\renewcommand{\figurename}{\small{Fig.}~}
\sectionfont{\sffamily\Large}
\subsectionfont{\normalsize}
\subsubsectionfont{\bf}
\setstretch{1.125} 
\setlength{\skip\footins}{0.8cm}
\setlength{\footnotesep}{0.25cm}
\setlength{\jot}{10pt}
\titlespacing*{\section}{0pt}{4pt}{4pt}
\titlespacing*{\subsection}{0pt}{15pt}{1pt}

\makeatletter 
\newlength{\figrulesep} 
\setlength{\figrulesep}{0.5\textfloatsep}

\makeatother

\maketitle

\begin{abstract}
    \noindent{Due to their abundant use in all-solid-state lasers, nonlinear optical (NLO) crystals are needed for many applications across diverse fields such as medicine and communication. However, because of conflicting requirements, the design of suitable inorganic crystals with strong second-harmonic generation (SHG) has proven to be challenging to both experimentalists and computational scientists. In this work, we leverage a data-driven approach to accelerate the search for high-performance NLO materials. We construct an extensive pool of candidates using databases within the OPTIMADE federation and employ an active learning strategy to gather optimal data while iteratively improving a machine learning model. The result is a publicly accessible dataset of $\sim$2,200 computed SHG tensors using density-functional perturbation theory. We further assess the performance of machine learning models on SHG prediction and introduce a multi-fidelity correction-learning scheme to refine data accuracy. This study represents a significant step towards data-driven materials discovery in the NLO field and demonstrates how new materials can be screened in an automated fashion.}
\end{abstract}


\renewcommand*\rmdefault{bch}\normalfont\upshape
\rmfamily
\section*{}
\vspace{-1cm}


\footnotetext{\textit{$^{a}$~UCLouvain, Institut de la Mati{\`e}re Condens{\'e}e et des Nanosciences (IMCN), Chemin des {\'E}toiles 8, Louvain-la-Neuve 1348, Belgium}}
\footnotetext{\textit{$^{b}$~Matgenix SRL, 185 Rue Armand Bury, 6534 Goz{\'e}e, Belgium}}
\footnotetext{\textit{$^{c}$~WEL Research Institute, Avenue Pasteur 6, 1300 Wavre, Belgium}}
\footnotetext{\textit{$^{d}$~School of Materials Science and Engineering, Northwestern Polytechnical University, No. 127 Youyi West Road, Xi’an, 710072 Shaanxi, China}}


\onecolumn  

\section{Introduction}

Thanks to their frequency conversion properties, nonlinear optical (NLO) materials play a significant role in modern optoelectronics~\cite{boyd2020nonlinear}. Their ability to produce coherent light by up- or down-converting incident electromagnetic waves has found applications in a variety of fields, from laser technologies and optical communication to biomedical imaging and quantum information processing~\cite{Kang2022Jul,Legres2014Mar,Jia2023Jan,Dutt2024May}. As is often the case with functional materials, a good NLO compound needs to meet several requirements such that it turns out to be a multi-objective optimisation. This ends up limiting the number of efficient materials, especially in the deep ultraviolet (DUV), the mid-, and the far-infrared (IR) ranges~\cite{Rondinelli2015,Aslam2023Dec}. It is thus of interest to accelerate the discovery of novel NLO materials, both for academic and industrial purposes.

As things stand, experimental studies lack the speed and cost-efficiency to freely consider the whole compositional and structural space. For this reason, computational methods are increasingly being used to navigate the almost endless possibilities~\cite{Wang2025Feb}. In practice, the search for NLO materials is translated into a search for appropriate compounds displaying strong second-harmonic generation (SHG), which enables a doubling of the incident frequency. With the use of density-functional theory (DFT), the SHG tensor can be calculated and investigated with respect to the chemistry and structure of a given compound. Many studies have thus focused on the efficient design of novel NLO crystals~\cite{Jiang2014Oct,Lin2020Mar,Dong2025Jan}. Another approach relies on high-throughput screenings of existing databases to identify promising materials~\cite{zhang2020first,Chu2023Apr,Wang2024Aug,Alkabakibi2025Feb}. The latter can then be used to suggest other candidates and investigate unexplored families of compounds. However, large open-access databases do not readily provide the SHG tensors~\cite{Jain2013,Schmidt2021}. Other basic properties are usually used to restrict the DFT computations of the SHG tensors to stable non-centrosymmetric (NCS) crystals with an electronic band gap in the range of interest. Although this procedure has led to the emergence of a few datasets with SHG information, the domain is definitely lacking significant NLO datasets that could be used for efficient screening or materials informatics~\cite{yu2020finding,zhang2020first,wang2020data}. To address this issue, \citet{Xie2023Nov} computed 1,500 SHG tensors of stable NCS semiconductors from the Materials Project (MP)~\cite{Jain2013} in 2023. Combined with 900 materials generated via an evolutionary algorithm, this dataset is a first step towards big data in the NLO field. In 2024, \citet{Wang2024Aug} also performed a screening of the MP involving the computation of $\sim$2,400 SHG tensors.

Recent years have seen a significant increase in the amount of available data related to materials properties. Existing experimental and computational databases are continuously growing while new actors and initiatives appear~\cite{Bergerhoff1983,Zagorac2019,Jain2013,Kirklin2015,Schmidt2021}. This trend has been accelerating with the emergence of data-driven and machine learning approaches that are able to generate hypothetical compounds, of which many are predicted to be stable, to some definition~\cite{Merchant2023,Zeni2024,Cheetham2024}.
Although this growth in data presents new opportunities for materials discovery, it also provides new challenges that require rational screening methods to help efficiently allocate experimental resources within this growing design space.
This is where data standardisation and federation can play an important role. 
The OPTIMADE consortium~\cite{Andersen2021, Evans2024} consists of several leading crystal structure database providers and datasets that have agreed upon a common data format and query language, enabling seamless access to over 60 million structures across 30 decentralised databases.
Several of these databases are targeted towards assessing materials stability, typically using DFT, providing a fruitful and growing pool for screening compounds with potentially exemplary properties in order to prioritise costly synthesis attempts.

In this work, we aim at propelling the NLO field into the era of big(ger) data while addressing the above challenge when generating and navigating the candidates design space. The end goal is the discovery of NLO bulk inorganic crystals with strong second-harmonic conversion. In practice, the search is translated into a multi-objective optimisation involving conflicting physical quantities, namely the SHG tensor and the band gap. For a given strength of SHG, maximising the band gap ensures a broad transparency window while promoting higher laser damage thresholds, an important practical consideration. By leveraging the common application programming interface (API) designed by the OPTIMADE federation, we easily build a large pool of candidate materials that will continue to grow as more structures and databases come online.  
This design space is then searched for good NLO materials with a cheap machine-learned model trained on an existing dataset of SHG tensors. Since DFT computations of SHG tensors are resource intensive and the size of the initial dataset is limited, we adopt an active learning (AL) procedure for the “training–predicting–selecting–computing” steps in the data acquisition process. 
This allows us to efficiently target promising materials in this large search pool, whether they are interesting for the combined SHG-band gap optimisation, or for improving the accuracy of the machine learning (ML) model.

This paper first describes the computational workflow for computing the static SHG tensors, the details of the active learning procedure and the candidate pool generation. The end result is a dataset of $\sim$2,200 static SHG tensors, which is made publicly available on the Materials Cloud Archive, itself accessible via an OPTIMADE API~\cite{Talirz2020,SHG-25-MaterialsCloud}. Thanks to this new dataset, we explore the performance of various ML algorithms on the present SHG task and we investigate a multi-fidelity correction-learning scheme to alleviate the inherent limitation of our data. Finally, we list the most promising materials uncovered in our dataset and look onward to the continued screening of large databases of hypothetical materials.

\section{Methods}

\subsection{First-principles calculations}\label{sec:FP}
The quantity of interest in this work is the third-rank tensor responsible for second-harmonic generation~\cite{boyd2020nonlinear, zernike2006applied, new2011introduction}.
This nonlinear optical phenomenon naturally appears in the framework of perturbation theory when the macroscopic polarisation, $\mathbf{P}$, is expressed as a power series of the incident electric field, $\mathbf{E}$, such that
\begin{equation}\label{eq:perturbation}
    P_i = \epsilon_0 \sum_j \chi_{ij}^{(1)} E_j + \epsilon_0 \sum_{jk} \chi_{ijk}^{(2)} E_j E_k + \text{higher order terms},
\end{equation}
with $\epsilon_0$, the vacuum permittivity, and $\mathbf{\chi}^{(1)}$, the linear susceptibility. The nonlinear susceptibility, $\mathbf{\chi}^{(2)}$, is responsible for SHG in the case of two incident fields at the same frequency. By convention, this tensor is halved and is commonly referred to as the SHG tensor, $\mathbf{d}$. By symmetry, the Voigt form can be adopted, thereby reducing it to a 3$\times$6 second-rank tensor. It is important to note that only NCS compounds can display non-zero components of the SHG tensor. To facilitate visualisation and comparison across different materials, an effective scalar coefficient, $d_\text{KP}$, can be derived following the Kurtz-Perry (KP) powder method~\cite{kurtz1968powder}.

In the present work, the open-source first-principles software ABINIT is used to compute the static limit of the SHG tensor in the framework of density-functional perturbation theory (DFPT)~\cite{gonze2016recent,gonze2020abinit,romero2020abinit,Gonze1995Aug,veithen2005nonlinear}. The exchange-correlation energy is modelled in the local-density approximation (LDA) by using the norm-conserving pseudopotentials from the PseudoDojo (scalar relativistic v0.4.1), which also provides the cutoff values ("standard" accuracy with hint "normal")~\cite{van2018pseudodojo,majewski1992self}. The Brillouin zone is sampled with a reciprocal density of 3,000 points per reciprocal atom. This sampling respects the symmetry of the system. 

These high-throughput calculations are performed with the \texttt{ShgFlowMaker} class implemented in the \emph{atomate2} Python package~\cite{Ganose2025} as \emph{jobflow} workflows~\cite{Rosen2024}. Since it defaults to the aforementioned $k$-point grid, only the type of pseudopotentials must be explicitly set to reproduce our results. This workflow is similar to the one presented in \citet{Trinquet2024Jul} apart from the pseudopotentials version and a revised algorithm to generate the $k$-points. Combined with the \texttt{FireWorks} workflows manager and the MongoDB database engine, this tool handles calculation submission and retrieval of the results~\cite{jain2015fireworks}. Sometimes, the SHG tensor requires a rotation in order for its components to match the conventional form set by the IEEE~\cite{ieee1987,roberts1992}. Both raw and post-processed tensors are made available.

The materials exhibiting a good balance between the KP coefficient and the band gap are selected for additional calculations to further refine their SHG tensors.
Higher accuracy can indeed be achieved by including a rigid shift of the conduction bands. Up to this point, the band gaps were directly taken from the source databases at the Perdew-Burke-Ernzerhof generalised-gradient approximation (GGA-PBE) level~\cite{Perdew1996Oct, Perdew1997Feb}. In order to obtain the values of the scissor shifts, the band gaps of the crystals of interest are computed at two different levels. 
First, the LDA band gaps are computed thanks to the ABINIT \texttt{BandStructureMaker} class of \emph{atomate2} using the same set of pseudopotentials as the SHG calculations. The number of divisions to sample the smallest segment of the high-symmetry path is set to 10. The band gap is then taken as the lowest gap value across both the self-consistent field (SCF) and the non-SCF calculations.
Second, the higher accuracy band gap is computed with the Heyd-Scuseria-Ernzerhof (HSE) hybrid functional~\cite{Heyd2003,Heyd2006} as implemented in VASP with Projector Augmented Wave pseudopotentials~\cite{Blochl1994Dec, Kresse1994Oct, Kresse1999Jan} (PBE\_64). A first SCF step is performed with PBE beforehand to help the convergence. This process is automated by linking the \texttt{HSEBSMaker} class to the VASP \texttt{StaticMaker} class of \emph{atomate2}. The electronic self-consistent loops are considered converged when a difference in energy lower than \SI{1e-6}{\electronvolt} is reached. The self-interaction energy is corrected with element-specific Hubbard $U$ values~\cite{Dudarev1998Jan} recommended by the MP~\cite{Jain2011Jun}.
In both the initial PBE and the LDA SCF steps, the Brillouin Zone is sampled using a uniform grid with a density of 1,500 points per reciprocal atom.
The difference between the HSE and LDA gaps provides a scissor shift to refine the SHG tensors. These band gap corrections can be given to the \texttt{ShgFlowMaker} class to correct the DFPT computation. 
It was decided not to perform any structural optimisation of the crystals, since the source databases already performed such relaxations at the GGA-PBE level using compatible settings.

\subsection{Active learning}

Similarly to our previous work, an active learning loop is adopted to optimally guide the acquisition of new data~\cite{Trinquet2025Jan}. In practice, cheap machine learning predictions of the KP coefficient are used to select materials whose SHG tensor will be computed with the more expensive DFPT method, thus extending the available SHG dataset for training. While the end-goal is the discovery of new materials boasting high SHG coefficients for a given band gap, it is still of interest to spend computing resources on suboptimal compounds, provided their addition in the training set significantly improves the performance of the surrogate model. Since the methodology and the choice of the ML model are similar to \citet{Trinquet2025Jan}, only the differences are described hereafter. 
In contrast to the case of the refractive index, we are not aware of any effective quantity, whose maximisation could replace the optimisation of the ($E_g, d_\text{KP}$) Pareto front; instead, here we sample explicitly from the Pareto front of our candidates.

At each iteration, a MODNet ensemble is trained on $T_i$, the training set at the $i_{th}$ iteration of the AL process. This ML model yields a prediction of the KP coefficient, $p_i(d_\text{KP}|x)\sim \mathcal{N}(\mu_{d_\text{KP},i},\sigma_{d_\text{KP},i})$ for a material $x$ with mean ensemble model prediction $\mu_{d_\text{KP},i}(x)$ and uncertainty $\sigma_{d_\text{KP},i}(x)$~\cite{DeBreuck2021,DeBreuck2021a,DeBreuck2022}. This allows us to define an upper bound to the target for each material as follows:

\begin{equation}
  d_{U,i}(x) = \mu_{d_\text{KP},i}(x) +\lambda \cdot \sigma_{d_\text{KP},i}(x),
\end{equation}
where the balance between exploration and exploitation is determined by the dimensionless parameter, $\lambda$. In order to diversify the selected compounds, the following regimes can be adopted:
\begin{alignat*}{2}
  &0 &&\to \text{highest mean (uncertainty-agnostic exploitation)} \\
  -&1 &&\to \text{highest mean with lowest uncertainty} \\
  &1 &&\to \text{highest mean with highest uncertainty} \\
  &\lambda_\text{cal.} &&\to \text{highest mean with high calibrated uncertainty} \\
  &\infty &&\to \text{highest uncertainty (exploration)}
\end{alignat*}
The calibrating factor, $\lambda_\text{cal.}$, is obtained by minimising the miscalibration area on a hold-out set and is then averaged over a 5-fold splits. It was found to consistently lie between $1.2$ and $1.5$ across all AL cycles.

The compounds are selected based on the following acquisition function:
\begin{equation}
    \alpha_i(x) = \left\{ \begin{array}{ll} 1 \quad \text{if} \; x \in \mathcal{F}_{U,i} \\ 0 \quad \text{else} \end{array} \right.
\end{equation}
where $\mathcal{F}_{U,i}$ is the Pareto front of the ($E_g, d_{U,i}$) distribution built from the entire candidate pool of materials, $\mathcal{P}$. This front is determined purely geometrically working from high to low band gap, after removing candidates with greater than 50 atoms in the primitive unit cell. Since $d_{U,i}$ can be defined according to several regimes, the Pareto front for each $\lambda$ regime, $\mathcal{F}_{U,i}(\lambda)$, is found and they are all merged to form the selected subset, $\mathcal{F}_{\forall, i}$. If the latter is not large enough, it is removed from the distribution and extended by the front of this new distribution. This acquisition function effectively classifies $\mathcal{P}$ at each AL cycle. After running the DFPT calculations on $\mathcal{F}_{\forall,i}$, a compound is flagged as an outlier if its $d_\text{KP}$ is greater than \SI{170}{\pico\meter\per\volt} or if its static refractive index is greater than \SI{20}{}. The cleaned DFPT results (without outliers) are then added to the training set for the next iteration:
\begin{align}
  T_{i+1} &= T_i\,\cup\, \mathcal{F}_{\text{DFPT},i} \quad \text{with} \quad \mathcal{F}_{\text{DFPT},i} \in \mathcal{F}_{\forall,i}.
\end{align}
A new MODNet model is then trained on $T_{i+1}$, thus initiating a new iteration. This process can be stopped based on arbitrary criteria involving the model accuracy, the size of the training set, the coverage of the materials, their performance, or the available computing resources.

\subsection{Training and candidate data}\label{sec:AL}
The dataset from \citet{Trinquet2024Jul} serves as the initial training set, $T_0$.
It comprises 579 SHG tensors of inorganic semiconductors computed with ABINIT using the DFPT procedure outlined in \autoref{sec:FP}. It should be noted that these calculations used an older set of pseudopotentials than presented here.

The MODNet model, feature selection algorithm, hyperparameters optimisation and training procedure follow those described in \citet{Trinquet2025Jan}.
A first set of ~200 descriptors was generated using the \emph{matminer} Python package via the \texttt{Matminer2024FastFeaturizer} preset implemented in MODNet v0.4.3~\cite{Ward2018}. 
A second set of ~1000 features, referred to as pGNN, is derived from the latent space of graph neural networks (GNN) models trained on different target properties~\cite{Gouvea2025}, as implemented in the \href{https://github.com/rogeriog/pGNN}{rogeriog/pGNN} GitHub repository.
Moreover, the final MODNet model of \citet{Trinquet2025Jan} predicts refractive indices and their uncertainties, which are included in the set of features, along with the band gap found in the source databases (typically computed with PBE).
These additional descriptors were considered due to the known relationship they share with the SHG strength~\cite{Miller1964Jul}. 
It is important to note that the actual set of features used in the AL loop was not fixed from the start as these iterations were refined over several months.
Additional descriptors were tested during this process and added to the feature set, if they were deemed useful, as illustrated in \autoref{fig: raw_bmk_al_mae}.
Since the resulting dataset of \citet{Trinquet2025Jan} was built to target the refractive index - band gap Pareto front, the SHG coefficient of each of its constituent NCS materials was computed, independently of the AL selection scheme.

Two source databases are considered to form the initial search pool, $\mathcal{P}$.
The first one is the Materials Project (MP) with $\sim$160k materials (v2023.11.1)~\cite{Jain2013}, resulting from DFT relaxations of primarily experimentally determined crystal structures from the Inorganic Crystal Structure Database (ICSD)~\cite{Zagorac2019}.
Using a combination of the MP API and their corresponding OPTIMADE API, the MP was filtered for NCS inorganic crystal structures possessing a PBE-computed band gap greater than \SI{0.05}{\electronvolt} and a distance from the MP convex hull (by the latest mixed GGA+U/mGGA workflow) less than \SI{50}{\milli\electronvolt\per\text{atom}}.
The resulting set of compounds is further reduced by excluding any lanthanide- or actinide-containing compounds, effectively reducing the MP to a subset of $\sim$13.5k relevant crystal structures relaxed with GGA-PBE.
The second database is Alexandria~\cite{Schmidt2021, Schmidt2023} with its $\sim$4.5M PBE-relaxed structures (v2023.12.29).
Thanks to the OPTIMADE API, this vast amount of entries is filtered for the same criteria as the MP. This query added $\sim$30.6k relevant structures to the pool of trial materials available to the AL process.

Duplicates across these two databases were removed by combining entries that share the same composition and space group, in which case the MP entry was preferred.
The final candidate pool, $\mathcal{P}$, spans $\sim$33.5k NCS stable semiconductors.

It should be noted at this stage that both MP and Alexandria now contain additional entries matching our criteria that were not present at the initiation of our AL procedure.
Additionally, new databases have been made available through OPTIMADE, such as the GNoME dataset~\cite{Merchant2023}, which contains several hundred thousand hypothetically stable compounds.
This study could thus be viewed as an intermediate step of a broader screening which will continue as new hypothetical compounds are suggested, and can act to prioritise experimental resources towards verifying the computed structures.

\subsection{Benchmarking ML models for SHG}\label{sec:method-bmk}

The second part of this work investigates the performance of various ML models on the prediction of the computed $d_\text{KP}$ coefficient. 
Both $T_0$ and the newly acquired SHG tensors are considered. 
The dataset is cleaned by removing any outliers or duplicates found by the default \texttt{StructureMatcher} of the \emph{pymatgen} Python package~\cite{ong2013python}. 
After removing the materials that fall abnormally far away from the data distribution (indicating, e.g., a convergence issue), $\sim$2,600 instances remain. 

The different models were benchmarked on four different holdout test sets. 
Two of them are randomly drawn with a size of 125 (\emph{random\_125}) and 250 (\emph{random\_250}) entries. 
The other two contain the same number of entries, but are sampled such that the distribution of the target values mimics the one of the full dataset, forming the \emph{distribution\_125} and \emph{distribution\_250} sets. Given the large range of target values and the clear bias of the dataset towards low values, this procedure allows for a more robust comparison than a single test set while being less computationally intensive than full cross-validation.
These datasets comprise four independent tasks, with independent models and training loops.
In the presentation of these benchmarks, we will focus on \emph{distribution\_250}, as shown in \autoref{fig:top-10pc-distribution_250}.

When needed for hyperparameter optimisation, a validation set was sampled from the training set with the same algorithm that was used to generate the test set. The resulting set of hyperparameters was then adopted for training the model on the whole training set before assessing it on the holdout sets. 
For descriptor-based models, both the \texttt{Matminer2024FastFeaturizer} preset and the pGNN features~\cite{Gouvea2025} were considered.
Three sets of features are derived: \emph{mmf} with only the former, \emph{pgnn} with only the latter, and \emph{mmf\_pgnn} merging both of them.

Several classes of ML models were investigated, from simple feed-forward neural networks like MODNet~\cite{DeBreuck2021,DeBreuck2021a}, to tree-based methods (Extra Trees and LGBM)~\cite{scikit-learn,Geurts2006Apr,Ke2017}, graph neural networks (co(N)GN~\cite{Ruff2024Mar}, MEGNet~\cite{Chen2019}, TensorNet~\cite{Simeon2023} for scalar predictions and Matten~\cite{Wen2024May} for full tensor predictions) as well as several commercial (GPT-4o, Claude Sonnet 3.7) and open (DARWIN 1.5~\cite{Xie2025}) large language models (LLMs).
A description of each model and any specifics of the training procedure or hyperparameter optimisation for each model are provided in \autoref{sec:ml-appendix}.

Model performance was assessed using standard metrics: MAE, RMSE, $R^2$, and most relevant for screening studies, Spearman's rank correlation coefficient.
In addition to these simple metrics, enrichment factors and discovery curves were computed for each model and holdout set.
An enrichment factor ($\text{EF}$) defined at a given percentage, say $\text{EF}(10\%)$, corresponds to the reduction in the number of oracle evaluations (in this case DFT calculations) required to find the top 10\% of materials.
For example, for a set of 100 candidate materials, if following the model's predicted ranking would allow the top 10\% to be found after 20 evaluations, the EF(10\%) would be 5, out of at theoretical maximum of 10, or to compare across different thresholds, this can be normalised to 0.5.
This metric is particularly important given the skewed nature of our dataset; a model could achieve reasonable performance in the low-SHG regime without being an effective discriminator of exemplary materials and vice versa.
Discovery curves provide a generalisation of the enrichment factor, by spanning the entire range of percentiles; they are conceptually similar to receiver-operating characteristic (ROC) curves, extended to a global ranking rather than binary classification at different probability thresholds. 

\subsection{Multi-fidelity correction-learning}\label{sec:method-multifidelity}

The ($E_g$,$d_\text{KP}$) Pareto front of SHG-25 was determined and selected to investigate the efficacy of the scissor correction. These entries are then removed from the dataset and a new Pareto front is determined and merged with the already selected subset. This process is repeated until $\sim$1,000 compounds are gathered. As described in \autoref{sec:FP}, both their ABINIT LDA and their VASP HSE gaps are computed to derive scissor shifts, which are used in subsequent DFPT SHG computations. Compounds with an HSE gap lower than \SI{1}{\electronvolt} are discarded. In addition, the selected entries from $T_0$ are also computed at the LDA level to ensure that their KP coefficient is consistent with the ones of SHG-25.

By combining the final dataset from the AL and its subset of SHG tensors computed with a scissor shift of the conduction bands, a multi-fidelity correction-learning task is investigated. Using MODNet, this supervised learning scheme targets:
\begin{align}\label{eq:d_corr}
    d_{\text{corr}} = d_\text{LDA} - d_\text{HSE},
\end{align}
where $d_\text{LDA}$ is the usual LDA KP coefficient from the main dataset and $d_\text{HSE}$ is the scissor-corrected KP coefficient as introduced above. The features generation and selection are similar to the AL section. The performance of the model is determined via a nested 5-folds cross-validation scheme, in which the inner loop corresponds to the hyperparameter optimisation with the native genetic algorithm implemented in the MODNet package~\cite{DeBreuck2022}.

\section{Results}

\subsection{Conclusion of the active learning procedure}
Following the methodology of \autoref{sec:AL}, almost 20 AL iterations were carried out, two of which consisted of adding materials from \citet{Trinquet2025Jan}.
The maximal and minimal number of oracle evaluations per iteration were $\sim$280 and $\sim$50, respectively.
The performance of the ML model is monitored at each cycle and plots showing the raw metrics are provided in the \autoref{sec:al-appendix}. While Figures~\ref{fig: raw_bmk_al_mae}, \ref{fig: raw_bmk_al_rmse}, and \ref{fig: raw_bmk_al_spr} correspond to an estimation of the performance from a nested 5-folds cross-validation scheme, the parity plots in Figures~\ref{fig: raw_bmk_al_mae_selection}, \ref{fig: raw_bmk_al_rmse_selection}, and \ref{fig: raw_bmk_al_spr_selection} are a better reflection of the reality as they correspond to the selected set of materials at each cycle. Although the curves are not monotonically decreasing, both illustrations show the improvement of the model with the increasing dataset size for all considered metrics. 

\begin{figure}\centering
    \includegraphics[width=1.0\linewidth]{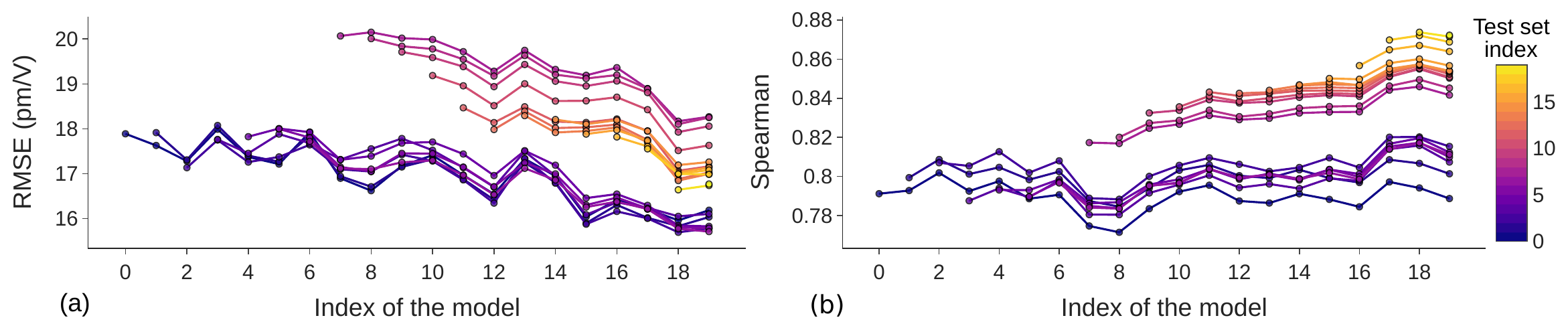}
    \caption{Evolution of the average RMSE (pm/V) (a) and Spearman's rank coefficient (b) over the AL process. The index of the model refers to its training set as the AL goes on. Each curve with index $i$ corresponds to the test sets of a 5-folds splitting of the dataset at the $i^\text{th}$ iteration of the AL procedure such that the same test sets are kept for the whole curve.}
    \label{fig:bmk_al}
\end{figure}

However, one shortcoming of these raw performance checks is the modification of the test sets throughout the AL scheme. To alleviate this issue, a post-processing approach was used to rationalise model performance. The starting training set, $T_0$, is first divided into 5 folds, $t_{0,j}$. Each of them is then extended by a part of the new DFPT data of each AL iteration:
\begin{align}
  t_{i+1,j} &= t_{i,j}\,\cup\, f_{i,j},.
\end{align}
where $f_{i,j}$ results from a 5-fold splitting of $\mathcal{F}_{\text{DFPT},i}$. Finally, a nested cross-validation scheme is applied on all $T_i$ using the $t_{i,j}$ splitting, which yielded fitted MODNet models, $m_{i,j}$. Each of these sets of models can then be used to perform a cross-validation (without training) of the other $T_i$. The set of features was restricted to the \emph{mmf} descriptors. After compiling the results, Figures~\ref{fig:bmk_al}, \ref{fig:bmk_al_mae}, and \ref{fig:bmk_al_r2} are obtained. The horizontal axis refers to the index of the models over the AL iterations and the vertical axis indicates the metric. As indicated by the colour, each curve corresponds to a training set $T_i$. This testing procedure ensures fixed test sets across the AL iterations while avoiding any data leakage. The figures show that all metrics experience an improvement when increasing the training data seen by the models (x-axis). Except for the coefficient of determination, the other metrics present a significant jump when going from the $T_6$ to the $T_7$ curves. Both the MAE and RMSE worsen while the Spearman correlation coefficient improves. In the 7th iteration, 239 materials from \citet{Trinquet2025Jan} were added in the training set, which amounted to \SI{27}{\percent} of $T_6$. Moreover, it contained a relatively greater number of high $d_\text{KP}$ values than the previous additions, which explains the noticeable worsening of performance evidenced by the MAE and RMSE, despite the positive effect on the Spearman coefficient.

\begin{figure}\centering
\includegraphics[width=0.9\linewidth]{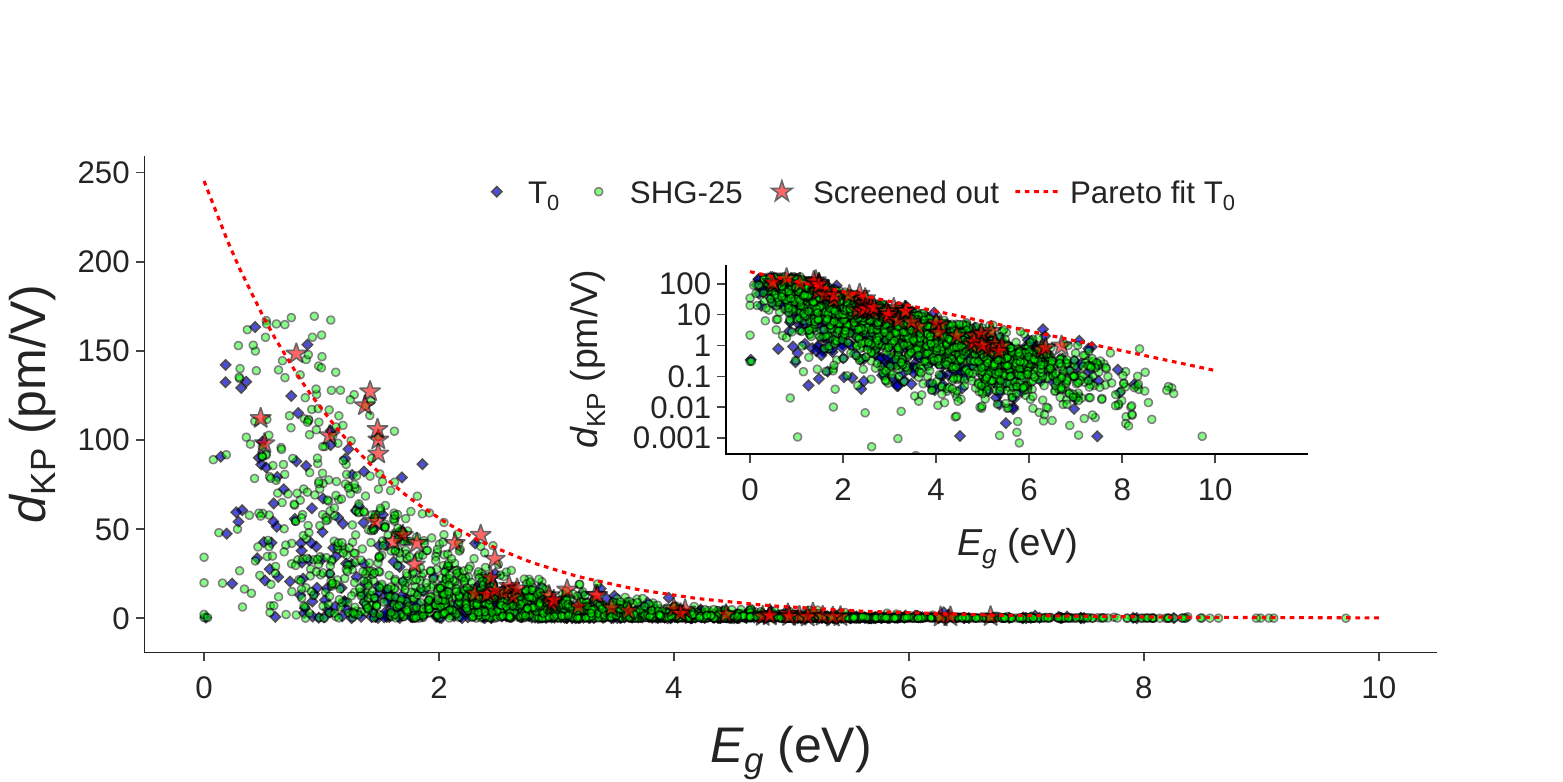}
\caption{Representation of the new dataset, SHG-25, and the starting one, $T_0$, in the ($E_g$,$d_\text{KP}$) space. The red dashed line illustrates the fit of the Pareto front of $T_0$ while the red stars highlights materials, which are suggested as promising based on several criteria (see text). The band gaps are taken from the source databases (GGA-PBE).}
\label{fig:global_shg25_T0_screened_out}
\end{figure}

The main contribution of the present work to the quest for new NLO inorganic crystals is a new dataset of $\sim$2,200 static SHG tensors computed within DFPT, which will be named SHG-25 hereafter. \autoref{fig:global_shg25_T0_screened_out} represents it in the ($E_g$,$d_\text{KP}$) space along with the starting dataset, $T_0$. From this plot, it is difficult to assess if SHG-25 properly targets the "Pareto materials" in this space, as intended by the AL procedure. 

\begin{figure}\centering
\includegraphics[width=0.7\linewidth]{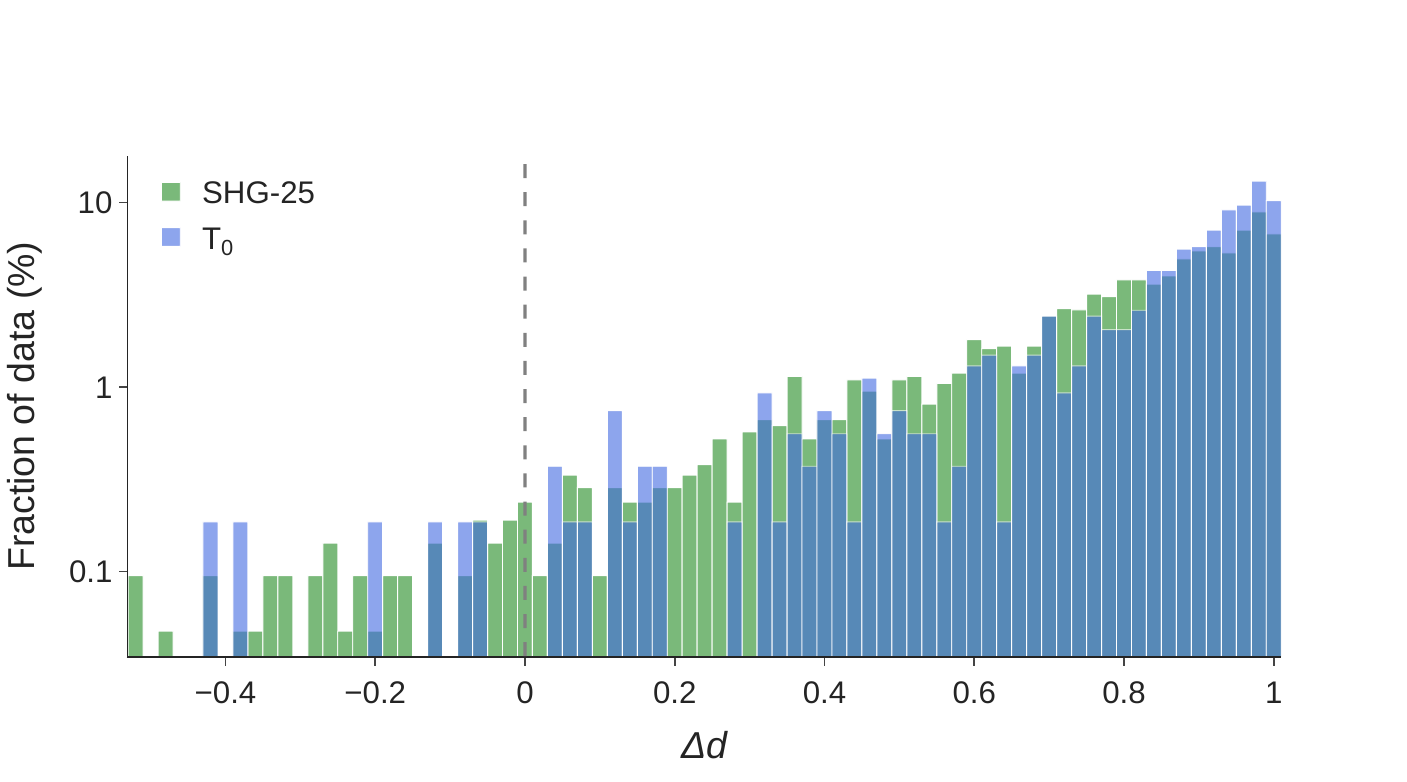}
\caption{Percentage of the data binned over the normalised distance from the fitted Pareto front of $T_0$ as defined in the text.}
\label{fig:histogram_distance}
\end{figure}

To appraise it, the Pareto front of $T_0$ is found and used to fit a function of the form:
\begin{align}
    f_\text{KP}\left(E_g\right) = a \cdot \exp{\left(b \cdot E_g\right)}.
\end{align}
For each entry $x$ of both $T_0$ and SHG-25, a normalised distance to this fitted front is then derived as:
\begin{align}
    \Delta d \left( x \right) = \dfrac{ f_\text{KP}\left(E_g\left( x \right)\right) - d_\text{KP}\left( x \right)}{f_\text{KP}\left(E_g\left( x \right)\right)}.
\end{align}
If it is close to 1, this distance implies that the material is far below the Pareto front, while if it is close to or below 0, the material is in the targeted range of screening. \autoref{fig:histogram_distance} illustrates the distribution of this proxy target for the two datasets. It shows that SHG-25 contains relatively more compounds close to or above $f_\text{KP}$ than $T_0$. The difference is, however, not as striking as it was in \citet{Trinquet2025Jan}, which can be explained by the low accuracy of the SHG ML model, especially at the start of the data acquisition process. The sampling pool, $\mathcal{P}$, is also more restricted and might not be large enough to effectively push or sample the Pareto front. 
In addition to this histogram, it is possible to consider the individual data contribution of each AL iteration separately from $T_0$. To do so, we introduce $\kappa$, the fraction of instances with $\Delta d$ greater than an arbitrary threshold. The latter is set to 0.5 in order to focus on the data closer to the $T_0$ Pareto fit than to a zero SHG response. In the case of $T_0$ and SHG-25, $\kappa$ is equal to \SI{9}{\percent} and \SI{14}{\percent}, respectively. When averaged over the first five data contributions of the AL, $\kappa$ is also \SI{9}{\percent}, confirming that the first few iterations are almost equivalent to a random selection, as in $T_0$. However, the last five iterations yield an average $\kappa$ of \SI{19}{\percent}, despite materials with high uncertainties being also selected. This demonstrates the performance of our ML model and validates the need to iteratively improve the ML model as the amount of the available training data increases. It is interesting to note that the additions of materials from \citet{Trinquet2025Jan} display a $\kappa$ of around \SI{16}{\percent}, thus confirming the usefulness of targeting compounds with a high refractive index when possible.

The new dataset, SHG-25, is made publicly available on the Materials Cloud Archive~\cite{Talirz2020,SHG-25-MaterialsCloud} and on the MPContribs~\cite{huck2015community} when possible, in the hope that it fosters high-throughput screenings as well as the development of reliable ML models. Combined with $T_0$, this new dataset amounts to 2,700 entries and is enough to achieve qualitative predictions of the KP coefficient as shown from the above analyses. While high-throughput screenings might already benefit from such accuracy, it is desirable to further improve the performance of cheap ML predictions. Increasing the amount of SHG data is thus of paramount importance. 

\subsection{Machine-learning the SHG coefficient}

In addition to the amount and diversity of training data, the choice of the ML model is another critical factor in the reliability of the $d_\text{KP}$ predictions.
This section presents the results of the ML benchmarks following the methodology introduced in \autoref{sec:method-bmk} in 4 different holdout test sets derived from SHG-25 and $T_0$.
\autoref{table:ml-benchmark-results} presents the top-level metrics on the largest and most diverse holdout set, \emph{distribution\_250}, sorted by decreasing Spearman's rank correlation coefficient.
This performance metric is emphasized as we consider the relative ranking of the predictions to be the most important criterion for screening purposes.

Based on Spearman's rank correlation alone, we find MODNet to be the most performant model ($r_s=0.87$), also possessing the lowest MAE of $\text{5.76 pm/V}$ and highest $R^2$ of 0.70. However, we also find that several models perform competitively with MODNet at this data set size, both those with increased complexity, namely the co(N)GN series of GNNs, and simpler tree-based methods that use the same descriptors as MODNet, namely Extra Trees and Light Gradient Boosting Machines, in agreement with \citet{An2025Mar}.
Given both the skewed distribution of $d_\text{KP}$ values in SHG-25, and the multi-objective nature of our materials design problem (i.e., finding materials on the ($E_g, d_\text{KP}$) Pareto front), we also compute enrichment factors (EF) and discovery curves for each model.
Using the procedure outlined in \autoref{sec:AL}, a figure of merit (FOM) for discovery was computed as the distance of a given candidate material from the fitted $T_0$ Pareto front, $\Delta d$.

\begin{figure}\centering
\includegraphics[width=0.6\linewidth]{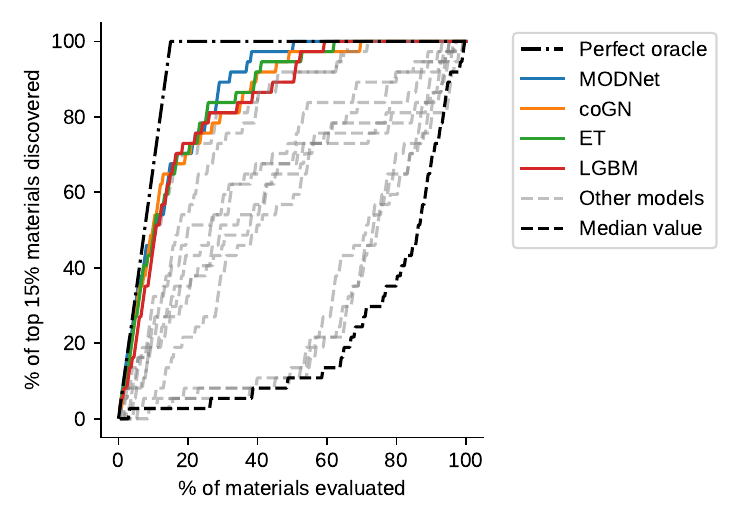}
\caption{Discovery curves for the benchmarked models on the top 15\% of compounds in the \emph{distribution\_250} holdout set.
\label{fig:discovery-curves}}
\end{figure}

\autoref{fig:discovery-curves} shows the discovery curves for the top 15\% of materials according to the computed FOM, highlighting the performance of the best models.
Once again, MODNet, the tree-based methods and the co(N)GN series outcompete all the rest on this metric, achieving normalised EF(15\%) values between 0.61 and 0.67, i.e., after evaluating 15\% of the dataset following these model's rankings, between 61\% and 67\% of the top 15\% of materials can be recovered.
This metric is better suited for capturing model performance specifically when used as a discriminator for potential SHG materials.
The clustering around this value perhaps indicates a reasonable maximum enrichment for this holdout set, given the small sample size involved (250 candidates in the holdout set, 37 in the top 15\% and thus ~12 ``missing'' from the predictions).
MODNet is marginally more efficient at spanning the entire top 15\%, requiring around 40\% of all materials to be evaluated.
Interestingly, even models that perform reasonably well when looking at simple metrics like MAE and $r_s$ appear much less effective at this task, with significantly reduced enrichment factors at this threshold.

The threshold of benchmarking against the top 15\% of materials is somewhat arbitrary and dataset-dependent.
\autoref{fig:top-10pc-distribution_250} shows the materials that were selected as the top 15\%  of this holdout set using the computed FOM.
Given the small holdout set size, the choice of threshold is affected by aliasing, however the best-performing models came out on top for all tested thresholds, providing a \emph{post hoc} rationalisation of our choice to use MODNet during the AL procedure.

\begin{table}
\centering
\begin{tabular}{lccccc}
\toprule
 & MAE (pm/V) & RMSE (pm/V) & $r_s$ & $R^2$ & EF(15\%) \\
 \midrule
MODNet & {\cellcolor[HTML]{FDE725}} 5.80 & {\cellcolor[HTML]{E5E419}} 15.30 & {\cellcolor[HTML]{FDE725}}0.87 & {\cellcolor[HTML]{FDE725}}0.70 & {\cellcolor[HTML]{FDE725}}0.67 \\
coNGN & {\cellcolor[HTML]{EFE51C}} 6.00 & {\cellcolor[HTML]{DDE318}} 15.50 & {\cellcolor[HTML]{F8E621}}0.86 & {\cellcolor[HTML]{C5E021}}0.62 & {\cellcolor[HTML]{C0DF25}}0.61 \\
coGN & {\cellcolor[HTML]{EFE51C}} 6.00 & {\cellcolor[HTML]{EFE51C}} 15.10 & {\cellcolor[HTML]{F1E51D}}0.85 & {\cellcolor[HTML]{D5E21A}}0.64 & {\cellcolor[HTML]{DFE318}}0.64 \\
ET & {\cellcolor[HTML]{B0DD2F}} 6.70 & {\cellcolor[HTML]{D0E11C}} 15.80 & {\cellcolor[HTML]{F1E51D}}0.85 & {\cellcolor[HTML]{BDDF26}}0.61 & {\cellcolor[HTML]{DFE318}}0.64 \\
LGBM & {\cellcolor[HTML]{CAE11F}} 6.40 & {\cellcolor[HTML]{FDE725}} 14.70 & {\cellcolor[HTML]{E5E419}}0.83 & {\cellcolor[HTML]{E2E418}}0.66 & {\cellcolor[HTML]{DFE318}}0.64 \\
TensorNet & {\cellcolor[HTML]{4EC36B}} 7.90 & {\cellcolor[HTML]{A2DA37}} 16.80 & {\cellcolor[HTML]{C8E020}}0.79 & {\cellcolor[HTML]{B8DE29}}0.60 & {\cellcolor[HTML]{20A386}} \textcolor{white}{0.41} \\
Matten & {\cellcolor[HTML]{3BBB75}} \textcolor{white}{8.20} & {\cellcolor[HTML]{25AC82}} \textcolor{white}{20.60} & {\cellcolor[HTML]{C8E020}}0.79 & {\cellcolor[HTML]{24AA83}} \textcolor{white}{0.34} & {\cellcolor[HTML]{1F988B}} \textcolor{white}{0.38} \\
MEGNet & {\cellcolor[HTML]{1F988B}} \textcolor{white}{9.30} & {\cellcolor[HTML]{54C568}} 18.80 & {\cellcolor[HTML]{6ECE58}}0.66 & {\cellcolor[HTML]{4CC26C}}0.44 & {\cellcolor[HTML]{46337F}} \textcolor{white}{0.14} \\
Claude Sonnet 3.5 & {\cellcolor[HTML]{3E4C8A}} \textcolor{white}{11.60} & {\cellcolor[HTML]{3C4F8A}} \textcolor{white}{26.30} & {\cellcolor[HTML]{4CC26C}}0.60 & {\cellcolor[HTML]{472E7C}} \textcolor{white}{-0.10} & {\cellcolor[HTML]{2E6D8E}} \textcolor{white}{0.27} \\
GPT-4o & {\cellcolor[HTML]{443B84}} \textcolor{white}{12.00} & {\cellcolor[HTML]{443A83}} \textcolor{white}{27.40} & {\cellcolor[HTML]{29AF7F}} \textcolor{white}{0.52} & {\cellcolor[HTML]{481467}} \textcolor{white}{-0.17} & {\cellcolor[HTML]{29798E}} \textcolor{white}{0.30} \\
DARWIN 1.5 & {\cellcolor[HTML]{440154}} \textcolor{white}{13.30} & {\cellcolor[HTML]{440154}} \textcolor{white}{30.00} & {\cellcolor[HTML]{440154}} \textcolor{white}{-0.08} & {\cellcolor[HTML]{440154}} \textcolor{white}{-0.22} & {\cellcolor[HTML]{440154}} \textcolor{white}{0.05} \\
\bottomrule
\end{tabular}
\caption{Performance metrics for the benchmarked models on the SHG-25 dataset for the \emph{distribution\_250} holdout set, sorted by Spearman's rank correlation coefficient, $r_s$.
In cases where multiple hyperparameter sets or architectures were benchmarked for the same model type, the table presents the model with the best performance.
The normalised enrichment factor for the top 15\% of materials, $\text{EF}(15\%)$ is a relevant metric for the application of these models for materials discovery (e.g., continuing the active learning procedure in this study).
Standard metrics, mean absolute errors (MAE), root-mean-square errors (RMSE), coefficient of determination ($R^2$) are also provided for completeness.
\label{table:ml-benchmark-results}}
\end{table}

\subsection{Correcting the band gap}

Few computational SHG datasets exist in the literature, and even fewer are publicly available~\cite{yu2020finding, zhang2020first, wang2020data, Xie2023Nov}. However, a common trait that most of them share is the adoption of scissor shifts to match high accuracy band gaps obtained with hybrid functionals. By artificially opening the gap, one can alleviate the overestimation of the SHG components caused by the usual underestimation of the band gap by DFT. In our work, contrary to those datasets, neither $T_0$ nor SHG-25 include scissor shifts. One could argue that this choice limits the impact of our databases, which would be true if the relative ranking of the materials were very different when considering a scissor shift, as any high-throughput screening involving our data would then be meaningless. In this section, we show that this is not the case, i.e., that the uncorrected SHG coefficients are sufficient for screening.

\begin{figure}\centering
\includegraphics[width=0.75\linewidth]{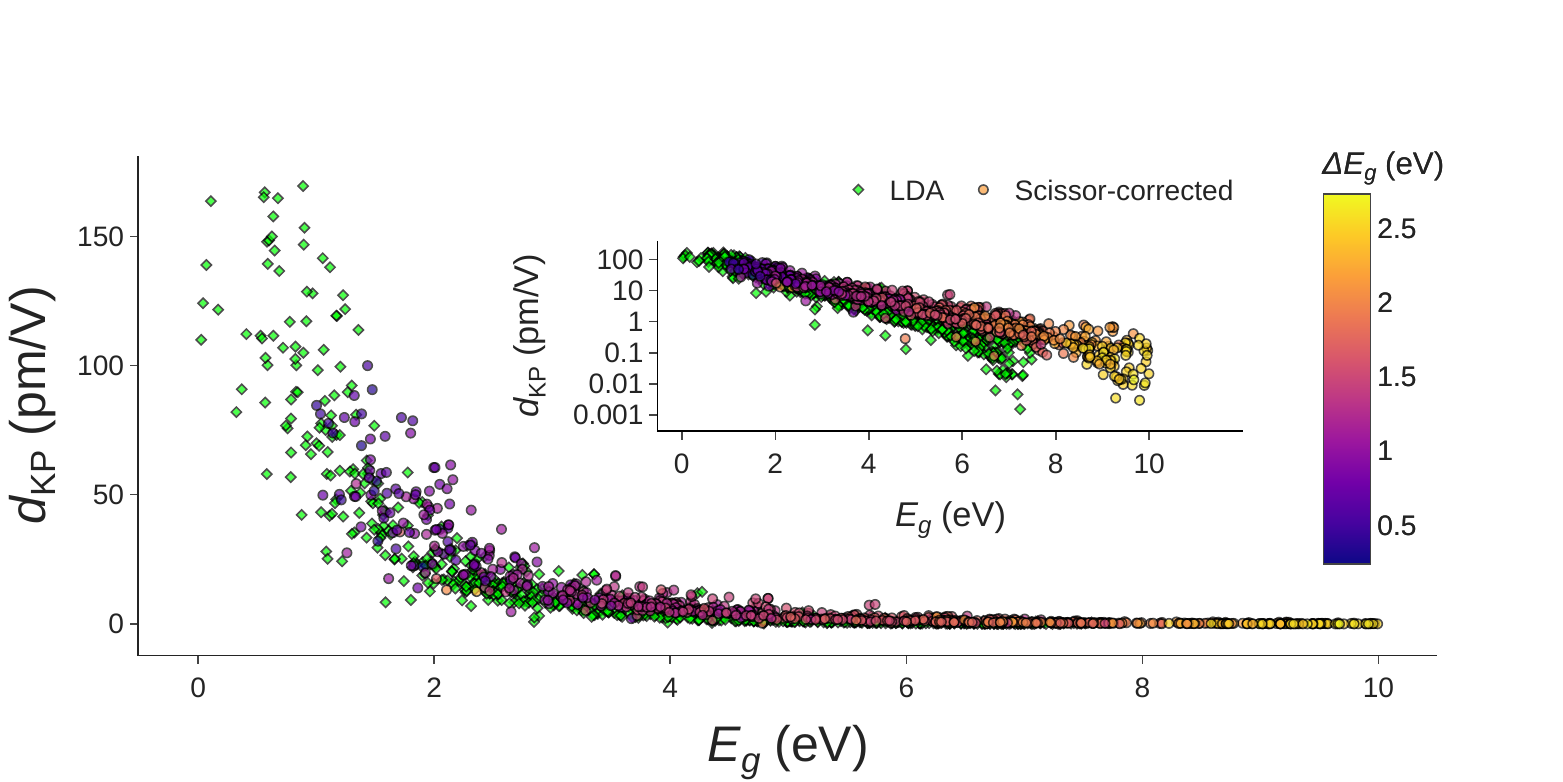}
\caption{The $\sim$700 materials subset selected for scissor correction in the ($E_g$,$d_\text{KP}$) space. The inset shows a log scale for a clearer visualisation. Both the LDA and the scissor-corrected values for the KP coefficients and band gaps are displayed for comparison. The colour bar indicates the scissor of each compound to go from the LDA to the HSE gap.}
\label{fig:effect of scissor}
\end{figure}

\begin{figure}\centering
\includegraphics[width=0.5\linewidth]{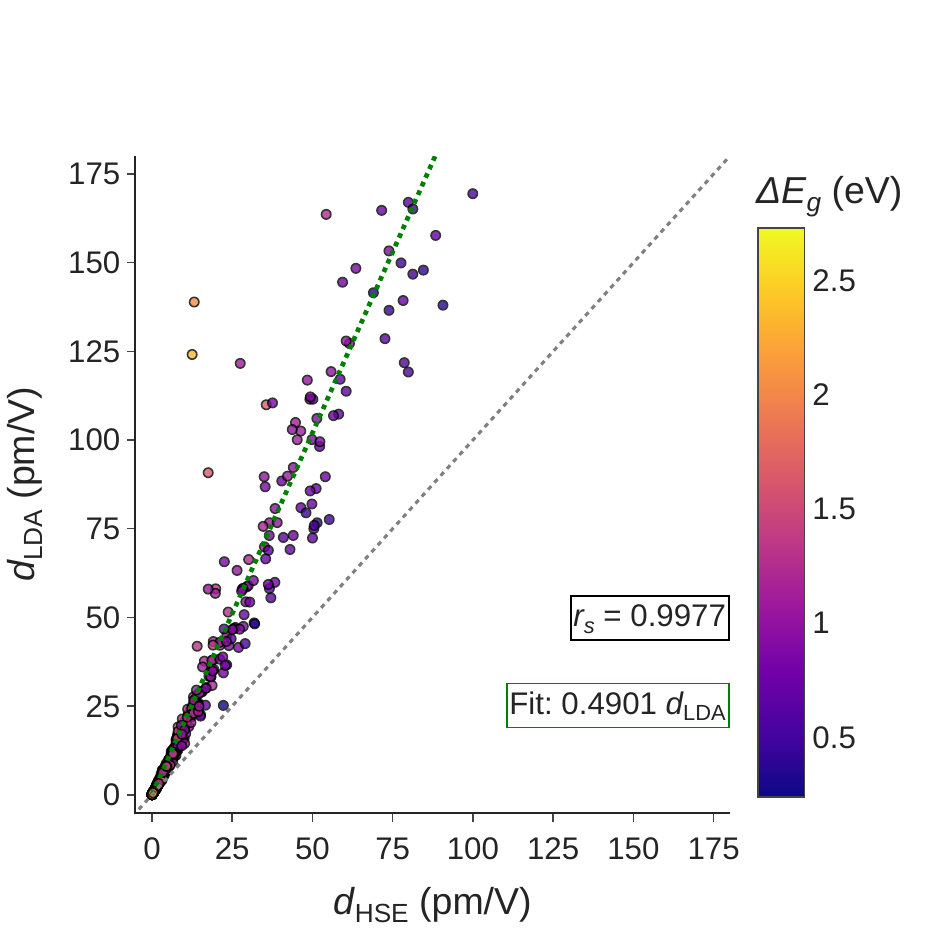}
\caption{Parity plot showing the effect of the scissor shift on the KP coefficients. The colour bar indicates the scissor of each compound to go from the LDA to the HSE gap. A linear regression is fitted on those points as shown by the green dotted line and the formula in the box. The Spearman's rank correlation coefficient, $r_s$, is included as well.}
\label{fig: parity_plot_dKP_scissor}
\end{figure}

Following the selection and computations described in \autoref{sec:method-multifidelity}, $\sim$700 pairs of LDA and HSE band gaps as well as the corresponding scissor-corrected SHG tensors are obtained and made publicly available along with SHG-25. This new dataset is represented in \autoref{fig:effect of scissor}. As expected, the band gap correction induces a blue-shift of the band gap and decreases the KP coefficient. It can already be seen from this plot that the distribution of $d_\text{HSE}$ is similar to the non-corrected one. These observations are confirmed by the parity plots in Figures~\ref{fig: parity_plot_dKP_scissor} and \ref{fig: parity_plot_gaps}. As indicated by the high Spearman's rank correlation coefficients, both figures show that the relative rankings stay the same for both the low and high-fidelity coefficients. This implies that any screening performed at the LDA level is equivalent to screening HSE results. Moreover, the HSE band gaps and the scissor-corrected $d_\text{KP}$ can both be modelled with a linear regression of their LDA counterpart as a first approximation. Figures~\ref{fig: parity_plot_dKP_scissor} and \ref{fig: parity_plot_gaps} illustrate this simple fit by the green dotted line, whose parameters are given in the green box. Given a material with its LDA band gap and its KP coefficient, it is thus possible to approximate its HSE gap and its corresponding KP coefficient at a very low cost.

Unfortunately, these linear regressions present an obvious limitation. For example, any two different materials with two different gap corrections would have the same corrected KP coefficient if their $d_\text{LDA}$ are equal. It is thus necessary to go one step further. Since SHG tensors at both the LDA and the "HSE" level are now available, machine learning algorithms can be used to leverage this kind of multi-fidelity data~\cite{Kim2025Jan}. In the present work, a correction learning (CL) scheme is investigated. This method consists in learning the difference between the low- (LDA) and the high-fidelity (HSE) data. Although conceptually simple, this multi-fidelity technique was shown to outperform others when modelling the band gap with MODNet~\cite{DeBreuck2022}. Since the band gap task has already been addressed in the literature, MODNet is chosen to explore the SHG correction by targeting $d_\text{corr}$ as defined in \autoref{eq:d_corr}. In addition to the \emph{mmf\_pgnn} set of features, the inclusion of the following quantities as descriptors is considered: the LDA gap ($E_g^\text{LDA}$), the HSE gap ($E_g^\text{HSE}$), the scissor shift ($\Delta E_g$) and $d_\text{LDA}$.

\begin{table}
\resizebox{\textwidth}{!}{
\centering
\begin{tabular}{lcccccc}
\toprule
 & MAE (pm/V) & RMSE (pm/V) & Spearman & $R^2$ & $\eta$ (\%) & $\zeta$ (\%) \\
\midrule
Linear regression & {\cellcolor[HTML]{54C568}} 1.6742 & {\cellcolor[HTML]{3FBC73}} \textcolor{white}{4.6544} & {\cellcolor[HTML]{FDE725}}0.9972 & {\cellcolor[HTML]{7FD34E}}0.9173 & {\cellcolor[HTML]{FDE725}} 0.0000 & {\cellcolor[HTML]{FDE725}} 0.0000 \\
\emph{mmf\_pgnn} & {\cellcolor[HTML]{440154}} \textcolor{white}{3.5342} & {\cellcolor[HTML]{460B5E}} \textcolor{white}{9.3651} & {\cellcolor[HTML]{440154}} \textcolor{white}{0.8895} & {\cellcolor[HTML]{48186A}} \textcolor{white}{0.6861} & {\cellcolor[HTML]{AADC32}} 0.1481 & {\cellcolor[HTML]{440154}} \textcolor{white}{10.3355} \\
\emph{mmf\_pgnn} $\cup$ $d_\text{LDA}$ & {\cellcolor[HTML]{81D34D}} 1.4700 & {\cellcolor[HTML]{56C667}} 4.3213 & {\cellcolor[HTML]{AADC32}}0.9837 & {\cellcolor[HTML]{98D83E}}0.9285 & {\cellcolor[HTML]{2C718E}} \textcolor{white}{0.7397} & {\cellcolor[HTML]{26828E}} \textcolor{white}{5.7614} \\
\emph{mmf\_pgnn} $\cup$ $d_\text{LDA}$ $\cup$ $\Delta E_g$ & {\cellcolor[HTML]{E5E419}} 1.0984 & {\cellcolor[HTML]{D5E21A}} 2.9003 & {\cellcolor[HTML]{C0DF25}}0.9870 & {\cellcolor[HTML]{EAE51A}}0.9659 & {\cellcolor[HTML]{21908D}} \textcolor{white}{0.5893} & {\cellcolor[HTML]{22A884}} \textcolor{white}{4.1307} \\
\emph{mmf\_pgnn} $\cup$ $\Delta E_g$ $\cup$ $d_\text{LDA}$ $\cup$ $E_g^\text{LDA}$ $\cup$ $E_g^\text{HSE}$ & {\cellcolor[HTML]{EAE51A}} 1.0766 & {\cellcolor[HTML]{E7E419}} 2.6977 & {\cellcolor[HTML]{C8E020}}0.9882 & {\cellcolor[HTML]{F4E61E}}0.9711 & {\cellcolor[HTML]{440154}} \textcolor{white}{1.1776} & {\cellcolor[HTML]{3DBC74}} \textcolor{white}{3.2484} \\
\bottomrule
\end{tabular}
}
\caption{Performance of MODNet on the $d_\text{corr}$ task when using different set of features under a nested 5-folds cross-validation. The quantities $\eta$ (\%) and $\zeta$ (\%) correspond to the fraction of $d_\text{corr}$ with a wrong sign and of negative ($d_\text{LDA}-d_\text{corr}$), respectively.}
\label{tab: correction learning}
\end{table}

Following a nested 5-fold cross-validation, the results of the different features sets are tabulated in \autoref{tab: correction learning} (and fully in \autoref{tab: correction learning full}). The linear regression introduced above is indicated as a baseline. The scores are derived from $d_\text{LDA}-d_\text{corr}$ instead of just the correction. This simplifies the interpretation and allows to consider the fraction of predicted corrections with a wrong sign ($\eta$) or inducing a negative "HSE" KP coefficient ($\zeta$). These two quantities act as a safeguard against non-physical predictions.

As expected from \autoref{fig: parity_plot_dKP_scissor}, the LDA KP coefficient is a necessary feature for the ML model to perform as well as the linear regression, while the band gaps and the scissor are not sufficient. This is not an issue in itself, since $d_\text{LDA}$ is a prerequisite for using the correction. To further improve MODNet, the band gaps and scissor are separately added as features. As intuition would suggest, the scissor results in a significant improvement to model performance. Further combining all of our custom features only slightly reduces the errors. Unfortunately, the predictions of MODNet are not constrained, as reflected in its $\eta$ and $\zeta$ of \SI{1}{\percent} and \SI{3}{\percent}-\SI{6}{\percent}, respectively. In contrast, the linear regression reaches \SI{0}{\percent} by definition. For this reason, the low values of $d_\text{HSE}$ are better represented by the linear regression than by MODNet while the higher values benefit from the flexibility of the ML model as illustrated in Figures~\ref{fig: parity_plot_dKP_corr_fit}, \ref{fig: parity_plot_dKP_corr_modnet}, \ref{fig: parity_plot_dKP_corr_fit_log}, and \ref{fig: parity_plot_dKP_corr_modnet_log}, although this can be remedied by a simple output rescaling. The significant reduction of the RMSE also supports this interpretation. Thanks to the close relationship between $d_\text{LDA}$, $d_\text{HSE}$ and the custom features, only less than 700 data entries are enough to reasonably correct the LDA KP coefficient. This limited size of dataset suggests that increasing the data will significantly improve the correction.

As many studies have shown before, the HSE band gaps and the scissor-corrected SHG tensors can successfully be used to screen promising NLO materials with balanced properties~\cite{Chu2023Apr,Wang2024Aug}. Here, this approach is illustrated on our high-fidelity subset of SHG tensors, which contains optimal materials in the ($E_g, d_\text{KP}$) space with a gap greater than \SI{1}{\electronvolt}. The screening is based on the following criteria:
\begin{itemize}
    \item good theoretical stability ($E_\text{hull}\leq$ \SI{10}{\milli\electronvolt\per\atom} with respect to the DFT-predicted convex hull of known materials),
    \item a scissor-corrected KP coefficient ($d_\text{HSE}$) greater than \SI{0.33}{\pico\meter\per\volt},
    \item a birefringence ($\Delta n_\text{HSE}$) larger than \SI{0.03}{\pico\meter\per\volt},
    \item non-toxic and sustainable elements.
\end{itemize}

The KP coefficient threshold corresponds to the effective coefficient of the experimental SHG tensor component for the widely used material \ce{KH2PO4} (KDP), which sets a lower bound for DUV crystals~\cite{Dmitriev1999}. The birefringence is also restricted by the minimal value for practical application. This condition is challenging because our DFPT calculations only compute the static limit of the electronic contribution to the dielectric tensor. One could argue that the dispersion of the refractive indices is weak below the band gap, thus limiting the difference between the birefringence in the static limit and at a finite frequency. Although \citet{Wang2024Aug} showed that static birefringence underestimates its counterpart at finite frequencies, the relationship between the two quantities warrants further investigation. To further reduce the selection, compositions with toxic elements (\ce{Pb}, \ce{As}, \ce{Be}, \ce{Hg}, \ce{Cd}) or hydrogen were discarded. Moreover, only sustainable elements were retained as characterised by Herfindahl–Hirschman Indices (HHIs) lower than 6,000 for both production and reserves~\cite{Herfindahl1950, Gaultois2013, Kim2023}\footnote{Namely Li, B, C, N, O, F, Na, Mg, Al, Si, P, S, Cl, Ca, Zn, Ga, Ge, As, Se, Sr, Cd, In, Sn, Te, I, Ba, Hg, Pb}. In the end, 59 materials remain, as listed in \autoref{tab: best materials}. Many of the entries originating from the Materials Project have already been experimentally observed and/or highlighted as potential NLO materials at the HSE level by \citet{Chu2023Apr} and \citet{Wang2024Aug}, as indicated. This highlights the importance of diversifying the original sources of the compounds as well the ability to periodically reassess the screening with machine-actionable queries of updated databases (via OPTIMADE or otherwise).
As initially desired, the materials selected by these criteria span a broad band gap range, from 1.3 to \SI{9.2}{\electronvolt}, allowing the potentially exemplary materials to be suggested in the relevant portion of the spectrum for a given application.

\section{Conclusions and outlook}

\begin{figure}\centering
\includegraphics[width=0.8\linewidth]{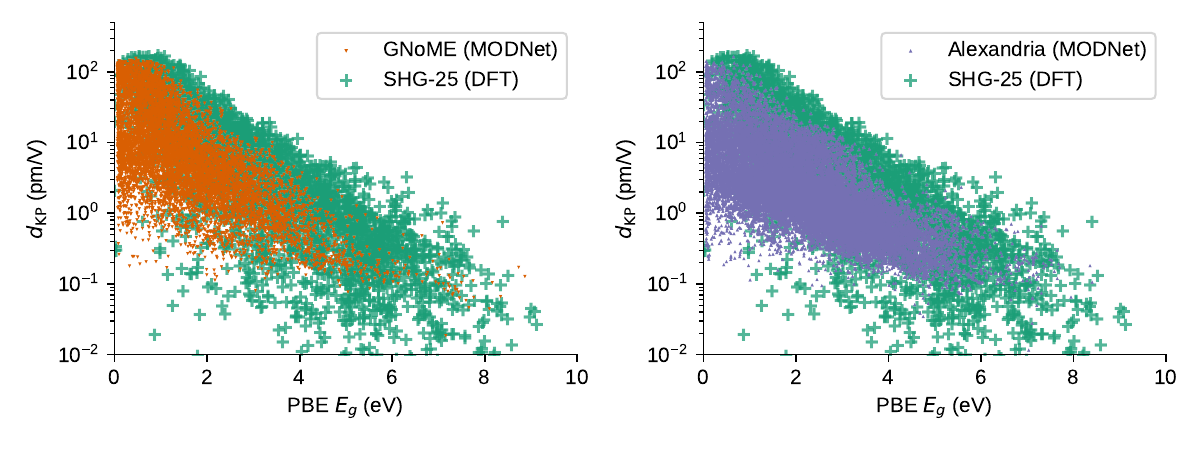}
\caption{MODNet-predicted $d_\text{KP}$ values for hypothetical structures added to GNoME (left) and Alexandria (right) since the conclusion of the active learning study, plotted against the database-reported band gaps computed at the PBE level, alongside the DFT-computed SHG values in SHG-25 (purple). The structures considered were limited to those that: i) are near the predicted convex hull reported by the database ($\leq 0.05\text{eV/atom}$), ii) have PBE band gaps greater than \SI{0.05}{eV}, iii) are non-centrosymmetric, iv) do not contain lanthanides or actinides, and v) have compositions that are not present in the computed SHG set. This left 9,657 structures from GNoME and 22,438 from Alexandria.}
\label{fig:latest-modnet-optimade}
\end{figure}

Despite a large and active community, the field of nonlinear optical materials is still looking for appropriate compounds in specific electromagnetic ranges, such as the deep UV and the mid- and far-infrared, that could be used in industrial applications. Today, this search can be driven by computations in order to accelerate the discovery of promising compounds~\cite{Wang2025Feb}. In order to navigate the rapidly growing design space offered by curated databases of hypothetical compounds, it is imperative to use fast screening methods to avoid wasting computational resources on suboptimal materials. A solution is to train cheap machine learning models on the target property to efficiently guide the allocation of DFT resources. However, this approach necessitates a large enough pre-existing dataset of the target property to attain a reasonable predictive power. Since the field of NLO materials is lacking in datasets, the present work adopted an active learning framework to acquire new static SHG tensors. By leveraging a relatively small existing dataset, this procedure resulted in $\sim$2,200 newly computed SHG tensors, which is made openly available on the Materials Could Archive~\cite{SHG-25-MaterialsCloud} and is itself accessible via an OPTIMADE API~\cite{Andersen2021}. The ML proxy allowed us to bias the data acquisition towards compounds exhibiting high SHG coefficients given their bandgap. Thanks to this new dataset, we were able to test a variety of ML models on this SHG task and its relationship with higher-fidelity data was also investigated. The tools used throughout this work enable periodic reassessment of the decentralised design space with minimal modifications to the code. This has already begun, as shown in \autoref{fig:latest-modnet-optimade}, where the GNoME dataset~\cite{Merchant2023} (~10,000 relevant entries) and new entries to Alexandria (~20,000 relevant entries) have been screened using our latest model.

Although the effective KP coefficient can be qualitatively predicted with the present ML model, it is of interest to the community to improve its performance. We believe that the first step to achieving this is to increase the number of training data. Thanks to the OPTIMADE API, we plan on continually extending our SHG dataset by querying unexplored databases providing either experimentally verified compounds or hypothetical compounds with the proper thermodynamic information. If the source does not provide band gaps, then one of the many ML models in the literature can be used to approximate it. In parallel to the screening of existing data, we could try to generate our own pool of hypothetical compounds. On the one hand, more targeted searches for hypothetical stable materials can make use of an evolutionary algorithm before being filtered on predicted SHG coefficients~\cite{Glass2006Dec,An2025Mar}. On the other hand, inverse design via constrained generation might quickly offer suggestions of promising compositions and/or structures~\cite{Noh2020May, Ren2022Jan}. As explored in other works, it would be of interest to include the lattice thermal conductivity as a quantity to maximise in our search~\cite{Chu2023Apr,Wu2025Jan,Liu2025Jan}, as it should help promote higher laser damage thresholds. Another improvement will be to consider the birefringence, which is of primary importance to achieve angular phase-matching~\cite{Zhang2017Apr,Lou2025Jan}. 

Finally, the hope is to derive physical insights from this dataset to better understand the characteristics behind a good NLO material. 
Whilst a close investigation of the promising materials is not within the scope for this work, we invite the community to consider these compounds, pending more detailed calculations of their suitability in future work.

\clearpage

\section*{Author Contributions}

\textbf{V.T.}: Conceptualisation, Methodology, Software, Validation, Investigation, Data curation, Writing - Original Draft, Writing - Review and Editing, Visualisation.
\textbf{M.L.E.}: Conceptualisation, Methodology, Software, Validation, Investigation, Data curation, Writing - Original Draft, Writing - Review and Editing, Visualisation.
\textbf{G-M.R.}: Conceptualisation, Resources, Writing - Review and Editing, Supervision, Funding acquisition.

\section*{Conflicts of interest}
G.-M.R. is a shareholder and Chief Innovation Officer of Matgenix SRL.

\section*{Data availability}

The crystal structures, DFT band gaps, and DFPT SHG tensors of both SHG-25 and its scissor-corrected subset have been deposited to the Materials Cloud Archive (\url{https://doi.org/10.24435/materialscloud:wk-qm}), with a corresponding OPTIMADE API available at \url{https://optimade.materialscloud.org/archive/materialscloud:wk-qm}. The code repository \url{https://github.com/modl-uclouvain/shg-ml-benchmarks} (archived at \url{https://doi.org/XXXX}) contains the ML benchmarking code, results, and the scripts used for screening the latest OPTIMADE structures. It also contains scripts used for the AL procedure. The code is released under the permissive MIT license. 

\section*{Acknowledgements}
Computational resources have been provided by the supercomputing facilities of the Université catholique de Louvain (CISM/UCL) and the Consortium des Équipements de Calcul Intensif en Fédération Wallonie Bruxelles (CÉCI) funded by the Fond de la Recherche Scientifique de Belgique (F.R.S.-FNRS) under convention 2.5020.11 and by the Walloon Region. The present research benefited from computational resources made available on Lucia, the Tier-1 supercomputer of the Walloon Region, infrastructure funded by the Walloon Region under the grant agreement n°1910247. V.T. acknowledges the support from the FRS-FNRS through a FRIA Grant. M.L.E. thanks the BEWARE scheme of the Wallonia-Brussels Federation for funding under the European Commission's Marie Curie-Skłodowska Action (COFUND 847587).

\balance

\bibliography{main} 

\begin{thebibliography}{100}
\providecommand{\natexlab}[1]{#1}
\providecommand{\url}[1]{\texttt{#1}}
\expandafter\ifx\csname urlstyle\endcsname\relax
  \providecommand{\doi}[1]{doi: #1}\else
  \providecommand{\doi}{doi: \begingroup \urlstyle{rm}\Url}\fi

\bibitem[Boyd(1992)]{boyd2020nonlinear}
Robert~W Boyd.
\newblock \emph{{Nonlinear Optics}}.
\newblock Academic press, San Diego, 1992.

\bibitem[Kang and Lin(2022)]{Kang2022Jul}
Lei Kang and Zheshuai Lin.
\newblock {Deep-ultraviolet nonlinear optical crystals: concept development and materials discovery}.
\newblock \emph{Light Sci. Appl.}, 11\penalty0 (201):\penalty0 1--12, July 2022.
\newblock ISSN 2047-7538.
\newblock \doi{10.1038/s41377-022-00899-1}.

\bibitem[Legres et~al.(2014)Legres, Chamot, Varna, and Janin]{Legres2014Mar}
Luc~G. Legres, Christophe Chamot, Mariana Varna, and Anne Janin.
\newblock {The Laser Technology: New Trends in Biology and Medicine}.
\newblock \emph{Journal of Modern Physics}, 5\penalty0 (5):\penalty0 267--279, March 2014.
\newblock \doi{10.4236/jmp.2014.55037}.

\bibitem[Jia et~al.(2023)Jia, Wu, Zhang, Qu, Jia, and Moss]{Jia2023Jan}
Linnan Jia, Jiayang Wu, Yuning Zhang, Yang Qu, Baohua Jia, and David~J. Moss.
\newblock {Third-Order Optical Nonlinearities of 2D Materials at Telecommunications Wavelengths}.
\newblock \emph{Micromachines}, 14\penalty0 (2):\penalty0 307, January 2023.
\newblock ISSN 2072-666X.
\newblock \doi{10.3390/mi14020307}.

\bibitem[Dutt et~al.(2024)Dutt, Mohanty, Gaeta, and Lipson]{Dutt2024May}
Avik Dutt, Aseema Mohanty, Alexander~L. Gaeta, and Michal Lipson.
\newblock {Nonlinear and quantum photonics using integrated optical materials}.
\newblock \emph{Nat. Rev. Mater.}, 9:\penalty0 321--346, May 2024.
\newblock ISSN 2058-8437.
\newblock \doi{10.1038/s41578-024-00668-z}.

\bibitem[Rondinelli and Kioupakis(2015)]{Rondinelli2015}
James~M. Rondinelli and Emmanouil Kioupakis.
\newblock Predicting and {{Designing Optical Properties}} of {{Inorganic Materials}}.
\newblock \emph{Annual Review of Materials Research}, 45\penalty0 (1):\penalty0 491--518, July 2015.
\newblock ISSN 1531-7331, 1545-4118.
\newblock \doi{10.1146/annurev-matsci-070214-021150}.

\bibitem[Aslam et~al.(2023)Aslam, Doane, Yeung, and Akopov]{Aslam2023Dec}
Hafiz~Zohaib Aslam, Joseph~T. Doane, Michael~T. Yeung, and Georgiy Akopov.
\newblock {Advances in Solid-State Nonlinear Optical Materials: From Fundamentals to Applications}.
\newblock \emph{ACS Appl. Opt. Mater.}, 1\penalty0 (12):\penalty0 1898--1921, December 2023.
\newblock \doi{10.1021/acsaom.3c00352}.

\bibitem[Wang et~al.(2025)Wang, Mutailipu, Yang, Pan, and Li]{Wang2025Feb}
Hongshan Wang, Miriding Mutailipu, Zhihua Yang, Shilie Pan, and Junjie Li.
\newblock {Computer-Aided Development of New Nonlinear Optical Materials}.
\newblock \emph{Angew. Chem. Int. Ed.}, 64\penalty0 (6):\penalty0 e202420526, February 2025.
\newblock ISSN 1433-7851.
\newblock \doi{10.1002/anie.202420526}.

\bibitem[Jiang et~al.(2014)Jiang, Kang, Luo, Gong, Lee, and Lin]{Jiang2014Oct}
Xingxing Jiang, Lei Kang, Siyang Luo, Pifu Gong, Ming-Hsien Lee, and Zheshuai Lin.
\newblock {Development of nonlinear optical materials promoted by density functional theory simulations}.
\newblock \emph{Int. J. Mod. Phys B}, 28\penalty0 (27):\penalty0 1430018, October 2014.
\newblock ISSN 0217-9792.
\newblock \doi{10.1142/S0217979214300187}.

\bibitem[Lin et~al.(2020)Lin, Wei, Chen, Wu, and Zhu]{Lin2020Mar}
Hua Lin, Wen-Bo Wei, Hong Chen, Xin-Tao Wu, and Qi-Long Zhu.
\newblock {Rational design of infrared nonlinear optical chalcogenides by chemical substitution}.
\newblock \emph{Coord. Chem. Rev.}, 406:\penalty0 213150, March 2020.
\newblock ISSN 0010-8545.
\newblock \doi{10.1016/j.ccr.2019.213150}.

\bibitem[Dong et~al.(2025)Dong, Huang, and Zou]{Dong2025Jan}
Xuehua Dong, Ling Huang, and Guohong Zou.
\newblock {Rational Design and Controlled Synthesis of High-Performance Inorganic Short-Wave UV Nonlinear Optical Materials}.
\newblock \emph{Acc. Chem. Res.}, 58\penalty0 (1):\penalty0 150--162, January 2025.
\newblock ISSN 0001-4842.
\newblock \doi{10.1021/acs.accounts.4c00704}.

\bibitem[Zhang et~al.(2020)Zhang, Zhang, Yu, Wang, Wu, and Lee]{zhang2020first}
Bingbing Zhang, Xiaodong Zhang, Jin Yu, Ying Wang, Kui Wu, and Ming-Hsien Lee.
\newblock {First-Principles High-Throughput Screening Pipeline for Nonlinear Optical Materials: Application to Borates}.
\newblock \emph{Chem. Mater.}, 32\penalty0 (15):\penalty0 6772--6779, August 2020.
\newblock ISSN 0897-4756.
\newblock \doi{10.1021/acs.chemmater.0c02583}.

\bibitem[Chu et~al.(2023)Chu, Huang, Xie, Tikhonov, Kruglov, Li, Pan, and Yang]{Chu2023Apr}
Dongdong Chu, Yi~Huang, Congwei Xie, Evgenii Tikhonov, Ivan Kruglov, Guangmao Li, Shilie Pan, and Zhihua Yang.
\newblock {Unbiased Screening of Novel Infrared Nonlinear Optical Materials with High Thermal Conductivity: Long-neglected Nitrides and Popular Chalcogenides}.
\newblock \emph{Angew. Chem. Int. Ed.}, 62\penalty0 (16):\penalty0 e202300581, April 2023.
\newblock ISSN 1433-7851.
\newblock \doi{10.1002/anie.202300581}.

\bibitem[Wang et~al.(2024)Wang, Ye, Guo, Li, Zou, Li, Zhang, Zhao, Xu, Chen, Wu, Bao, Xu, and Duan]{Wang2024Aug}
Jizhang Wang, Meng Ye, Xiaomi Guo, Yang Li, Nianlong Zou, He~Li, Zetao Zhang, Sibo Zhao, Zhiming Xu, Haowei Chen, Dezhao Wu, Ting Bao, Yong Xu, and Wenhui Duan.
\newblock {Unbiased screening of deep-ultraviolet and mid-infrared nonlinear optical crystals: Long-neglected covalent and mixed-cation motifs}.
\newblock \emph{Phys. Rev. Mater.}, 8\penalty0 (8):\penalty0 085202, August 2024.
\newblock \doi{10.1103/PhysRevMaterials.8.085202}.

\bibitem[Alkabakibi et~al.(2025)Alkabakibi, Barma, Rybkovskiy, Tudi, Xie, and Oganov]{Alkabakibi2025Feb}
Y.~Alkabakibi, D.~D. Barma, D.~V. Rybkovskiy, A.~Tudi, C.~Xie, and A.~R. Oganov.
\newblock {Computational Identification of Four Promising Nonlinear Optical Materials for Near and Middle Ultraviolet Operation}.
\newblock \emph{JETP Lett.}, pages 1--6, February 2025.
\newblock ISSN 1090-6487.
\newblock \doi{10.1134/S0021364024605074}.

\bibitem[Jain et~al.(2013)Jain, Ong, Hautier, Chen, Richards, Dacek, Cholia, Gunter, Skinner, Ceder, and Persson]{Jain2013}
Anubhav Jain, Shyue~Ping Ong, Geoffroy Hautier, Wei Chen, William~Davidson Richards, Stephen Dacek, Shreyas Cholia, Dan Gunter, David Skinner, Gerbrand Ceder, and Kristin~A. Persson.
\newblock Commentary: {{The Materials Project}}: {{A}} materials genome approach to accelerating materials innovation.
\newblock \emph{APL Materials}, 1\penalty0 (1):\penalty0 011002, July 2013.
\newblock \doi{10.1063/1.4812323}.

\bibitem[Schmidt et~al.(2021)Schmidt, Pettersson, Verdozzi, Botti, and Marques]{Schmidt2021}
Jonathan Schmidt, Love Pettersson, Claudio Verdozzi, Silvana Botti, and Miguel A.~L. Marques.
\newblock Crystal graph attention networks for the prediction of stable materials.
\newblock \emph{Science Advances}, 7\penalty0 (49):\penalty0 eabi7948, December 2021.
\newblock \doi{10.1126/sciadv.abi7948}.

\bibitem[Yu et~al.(2020)Yu, Zhang, Zhang, Wang, Wu, and Lee]{yu2020finding}
Jin Yu, Bingbing Zhang, Xiaodong Zhang, Ying Wang, Kui Wu, and Ming-Hsien Lee.
\newblock {Finding Optimal Mid-Infrared Nonlinear Optical Materials in Germanates by First-Principles High-Throughput Screening and Experimental Verification}.
\newblock \emph{ACS Appl. Mater. Interfaces}, 12\penalty0 (40):\penalty0 45023--45035, October 2020.
\newblock ISSN 1944-8244.
\newblock \doi{10.1021/acsami.0c15728}.

\bibitem[Wang et~al.(2020)Wang, Liang, and Lin]{wang2020data}
Rui Wang, Fei Liang, and Zheshuai Lin.
\newblock {Data-driven prediction of diamond-like infrared nonlinear optical crystals with targeting performances}.
\newblock \emph{Sci. Rep.}, 10\penalty0 (3486):\penalty0 1--8, February 2020.
\newblock ISSN 2045-2322.
\newblock \doi{10.1038/s41598-020-60410-x}.

\bibitem[Xie et~al.(2023)Xie, Tikhonov, Chu, Wu, Kruglov, Pan, and Yang]{Xie2023Nov}
Congwei Xie, Evgenii Tikhonov, Dongdong Chu, Mengfan Wu, Ivan Kruglov, Shilie Pan, and Zhihua Yang.
\newblock {A prediction-driven database to enable rapid discovery of nonlinear optical materials}.
\newblock \emph{Sci. China Mater.}, 66\penalty0 (11):\penalty0 4473--4479, November 2023.
\newblock ISSN 2199-4501.
\newblock \doi{10.1007/s40843-023-2592-x}.

\bibitem[Bergerhoff et~al.(1983)Bergerhoff, Hundt, Sievers, and Brown]{Bergerhoff1983}
G.~Bergerhoff, R.~Hundt, R.~Sievers, and I.~D. Brown.
\newblock The inorganic crystal structure data base.
\newblock \emph{Journal of Chemical Information and Modeling}, 23\penalty0 (2):\penalty0 66--69, May 1983.
\newblock \doi{10.1021/ci00038a003}.

\bibitem[Zagorac et~al.(2019)Zagorac, M{\"u}ller, Ruehl, Zagorac, and Rehme]{Zagorac2019}
D.~Zagorac, H.~M{\"u}ller, S.~Ruehl, J.~Zagorac, and S.~Rehme.
\newblock Recent developments in the {{Inorganic Crystal Structure Database}}: theoretical crystal structure data and related features.
\newblock \emph{Journal of Applied Crystallography}, 52\penalty0 (5):\penalty0 918--925, October 2019.
\newblock \doi{10.1107/S160057671900997X}.

\bibitem[Kirklin et~al.(2015)Kirklin, Saal, Meredig, Thompson, Doak, Aykol, R{\"u}hl, and Wolverton]{Kirklin2015}
Scott Kirklin, James~E. Saal, Bryce Meredig, Alex Thompson, Jeff~W. Doak, Muratahan Aykol, Stephan R{\"u}hl, and Chris Wolverton.
\newblock The {{Open Quantum Materials Database}} ({{OQMD}}): assessing the accuracy of {{DFT}} formation energies.
\newblock \emph{npj Computational Materials}, 1\penalty0 (1):\penalty0 1--15, December 2015.
\newblock \doi{10.1038/npjcompumats.2015.10}.

\bibitem[Merchant et~al.(2023)Merchant, Batzner, Schoenholz, Aykol, Cheon, and Cubuk]{Merchant2023}
Amil Merchant, Simon Batzner, Samuel~S. Schoenholz, Muratahan Aykol, Gowoon Cheon, and Ekin~Dogus Cubuk.
\newblock Scaling deep learning for materials discovery.
\newblock \emph{Nature}, 624\penalty0 (7990):\penalty0 80--85, December 2023.
\newblock ISSN 1476-4687.
\newblock \doi{10.1038/s41586-023-06735-9}.

\bibitem[Zeni et~al.(2024)Zeni, Pinsler, Zügner, Fowler, Horton, Fu, Shysheya, Crabbé, Sun, Smith, Nguyen, Schulz, Lewis, Huang, Lu, Zhou, Yang, Hao, Li, Tomioka, and Xie]{Zeni2024}
Claudio Zeni, Robert Pinsler, Daniel Zügner, Andrew Fowler, Matthew Horton, Xiang Fu, Sasha Shysheya, Jonathan Crabbé, Lixin Sun, Jake Smith, Bichlien Nguyen, Hannes Schulz, Sarah Lewis, Chin-Wei Huang, Ziheng Lu, Yichi Zhou, Han Yang, Hongxia Hao, Jielan Li, Ryota Tomioka, and Tian Xie.
\newblock {{MatterGen}}: A generative model for inorganic materials design.
\newblock \penalty0 (arXiv:2312.03687), January 2024.
\newblock \doi{10.48550/arXiv.2312.03687}.

\bibitem[Cheetham and Seshadri(2024)]{Cheetham2024}
Anthony~K. Cheetham and Ram Seshadri.
\newblock Artificial {{Intelligence Driving Materials Discovery}}? {{Perspective}} on the {{Article}}: {{Scaling Deep Learning}} for {{Materials Discovery}}.
\newblock \emph{Chemistry of Materials}, April 2024.
\newblock ISSN 0897-4756.
\newblock \doi{10.1021/acs.chemmater.4c00643}.

\bibitem[Andersen et~al.(2021)Andersen, Armiento, Blokhin, Conduit, Dwaraknath, Evans, Fekete, Gopakumar, Gražulis, Merkys, Mohamed, Oses, Pizzi, Rignanese, Scheidgen, Talirz, Toher, Winston, Aversa, Choudhary, Colinet, Curtarolo, Di~Stefano, Draxl, Er, Esters, Fornari, Giantomassi, Govoni, Hautier, Hegde, Horton, Huck, Huhs, Hummelshøj, Kariryaa, Kozinsky, Kumbhar, Liu, Marzari, Morris, Mostofi, Persson, Petretto, Purcell, Ricci, Rose, Scheffler, Speckhard, Uhrin, Vaitkus, Villars, Waroquiers, Wolverton, Wu, and Yang]{Andersen2021}
Casper~W. Andersen, Rickard Armiento, Evgeny Blokhin, Gareth~J. Conduit, Shyam Dwaraknath, Matthew~L. Evans, Adam Fekete, Abhijith Gopakumar, Saulius Gražulis, Andrius Merkys, Fawzi Mohamed, Corey Oses, Giovanni Pizzi, Gian-Marco Rignanese, Markus Scheidgen, Leopold Talirz, Cormac Toher, Donald Winston, Rossella Aversa, Kamal Choudhary, Pauline Colinet, Stefano Curtarolo, Davide Di~Stefano, Claudia Draxl, Suleyman Er, Marco Esters, Marco Fornari, Matteo Giantomassi, Marco Govoni, Geoffroy Hautier, Vinay Hegde, Matthew~K. Horton, Patrick Huck, Georg Huhs, Jens Hummelshøj, Ankit Kariryaa, Boris Kozinsky, Snehal Kumbhar, Mohan Liu, Nicola Marzari, Andrew~J. Morris, Arash~A. Mostofi, Kristin~A. Persson, Guido Petretto, Thomas Purcell, Francesco Ricci, Frisco Rose, Matthias Scheffler, Daniel Speckhard, Martin Uhrin, Antanas Vaitkus, Pierre Villars, David Waroquiers, Chris Wolverton, Michael Wu, and Xiaoyu Yang.
\newblock {{OPTIMADE}}, an {{API}} for exchanging materials data.
\newblock \emph{Scientific Data}, 8\penalty0 (1):\penalty0 217, August 2021.
\newblock ISSN 2052-4463.
\newblock \doi{10.1038/s41597-021-00974-z}.

\bibitem[Evans et~al.(2024)Evans, Bergsma, Merkys, Andersen, Andersson, Beltrán, Blokhin, Boland, Balderas, Choudhary, Díaz, García, Eckert, Eimre, {Fuentes-Montero}, Krajewski, Mortensen, {Nápoles-Duarte}, Pietryga, Qi, Carrillo, Vaitkus, Yu, Zettel, de~Castro, Carlsson, Cerqueira, Divilov, Hajiyani, Hanke, Jose, Oses, Riebesell, Schmidt, Winston, Xie, Yang, Bonella, Botti, Curtarolo, Draxl, {Fuentes-Cobas}, Hospital, Liu, Miguel~A., Marzari, Morris, Ong, Orozco, Persson, Thygesen, Wolverton, Scheidgen, Toher, Conduit, Pizzi, Grazulis, Rignanese, and Armiento]{Evans2024}
Matthew Evans, Johan Bergsma, Andrius Merkys, Casper Andersen, Oskar~B. Andersson, Daniel Beltrán, Evgeny Blokhin, Tara~M. Boland, Rubén~Castañeda Balderas, Kamal Choudhary, Alberto~Díaz Díaz, Rodrigo~Domínguez García, Hagen Eckert, Kristjan Eimre, Maria~Elena {Fuentes-Montero}, Adam~M. Krajewski, Jens~Jørgen Mortensen, José~Manuel {Nápoles-Duarte}, Jacob Pietryga, Ji~Qi, Felipe de Jesús~Trejo Carrillo, Antanas Vaitkus, Jusong Yu, Adam Zettel, Pedro~Baptista de~Castro, Johan~Martin Carlsson, Tiago F.~T. Cerqueira, Simon Divilov, Hamidreza Hajiyani, Felix Hanke, Kevin Jose, Corey Oses, Janosh Riebesell, Jonathan Schmidt, Donald Winston, Christen Xie, Xiaoyu Yang, Sara Bonella, Silvana Botti, Stefano Curtarolo, Claudia Draxl, Luis E.~E. {Fuentes-Cobas}, Adam Hospital, Zi-Kui Liu, L.~Marques Miguel~A., Nicola Marzari, Andrew~James Morris, Shyue~Ping Ong, Modesto Orozco, Kristin Persson, Kristian~Sommer Thygesen, Chris~M. Wolverton, Markus Scheidgen, Cormac Toher, Gareth Conduit, Giovanni Pizzi,
  Saulius Grazulis, Gian-Marco Rignanese, and Rickard Armiento.
\newblock Developments and applications of the {{OPTIMADE API}} for materials discovery, design, and data exchange.
\newblock \emph{Digital Discovery}, April 2024.
\newblock ISSN 2635-098X.
\newblock \doi{10.1039/D4DD00039K}.

\bibitem[Talirz et~al.(2020)Talirz, Kumbhar, Passaro, Yakutovich, Granata, Gargiulo, Borelli, Uhrin, Huber, Zoupanos, Adorf, Andersen, Schütt, Pignedoli, Passerone, VandeVondele, Schulthess, Smit, Pizzi, and Marzari]{Talirz2020}
Leopold Talirz, Snehal Kumbhar, Elsa Passaro, Aliaksandr~V. Yakutovich, Valeria Granata, Fernando Gargiulo, Marco Borelli, Martin Uhrin, Sebastiaan~P. Huber, Spyros Zoupanos, Carl~S. Adorf, Casper~Welzel Andersen, Ole Schütt, Carlo~A. Pignedoli, Daniele Passerone, Joost VandeVondele, Thomas~C. Schulthess, Berend Smit, Giovanni Pizzi, and Nicola Marzari.
\newblock Materials {{Cloud}}, a platform for open computational science.
\newblock \emph{Scientific Data}, 7\penalty0 (1):\penalty0 299, September 2020.
\newblock ISSN 2052-4463.
\newblock \doi{10.1038/s41597-020-00637-5}.

\bibitem[Trinquet et~al.(2025{\natexlab{a}})Trinquet, Evans, and Rignanese]{SHG-25-MaterialsCloud}
V.~Trinquet, M.~L. Evans, and G-M. Rignanese.
\newblock {Research data supporting ``Accelerating the discovery of high-performance nonlinear optical materials using active learning and high-throughput screening''}, 2025{\natexlab{a}}.
\newblock URL \url{https://doi.org/10.24435/materialscloud:wk-qm}.

\bibitem[Zernike and Midwinter(1973)]{zernike2006applied}
Frits Zernike and John~E Midwinter.
\newblock \emph{{Applied Nonlinear Optics}}.
\newblock Wiley, New York, 1973.

\bibitem[New(2011)]{new2011introduction}
Geoffrey New.
\newblock \emph{{Introduction to Nonlinear Optics}}.
\newblock Cambridge University Press, New York, 2011.

\bibitem[Kurtz and Perry(1968)]{kurtz1968powder}
S.~K. Kurtz and T.~T. Perry.
\newblock {A Powder Technique for the Evaluation of Nonlinear Optical Materials}.
\newblock \emph{J. Appl. Phys.}, 39\penalty0 (8):\penalty0 3798--3813, July 1968.
\newblock ISSN 0021-8979.
\newblock \doi{10.1063/1.1656857}.

\bibitem[Gonze et~al.(2016)Gonze, Jollet, Abreu~Araujo, Adams, Amadon, Applencourt, Audouze, Beuken, Bieder, Bokhanchuk, Bousquet, Bruneval, Caliste, C{\ifmmode\hat{o}\else\^{o}\fi}t{\ifmmode\acute{e}\else\'{e}\fi}, Dahm, Da~Pieve, Delaveau, Di~Gennaro, Dorado, Espejo, Geneste, Genovese, Gerossier, Giantomassi, Gillet, Hamann, He, Jomard, Laflamme~Janssen, Le~Roux, Levitt, Lherbier, Liu, Luka{\ifmmode\check{c}\else\v{c}\fi}evi{\ifmmode\acute{c}\else\'{c}\fi}, Martin, Martins, Oliveira, Ponc{\ifmmode\acute{e}\else\'{e}\fi}, Pouillon, Rangel, Rignanese, Romero, Rousseau, Rubel, Shukri, Stankovski, Torrent, Van~Setten, Van~Troeye, Verstraete, Waroquiers, Wiktor, Xu, Zhou, and Zwanziger]{gonze2016recent}
X.~Gonze, F.~Jollet, F.~Abreu~Araujo, D.~Adams, B.~Amadon, T.~Applencourt, C.~Audouze, J.-M. Beuken, J.~Bieder, A.~Bokhanchuk, E.~Bousquet, F.~Bruneval, D.~Caliste, M.~C{\ifmmode\hat{o}\else\^{o}\fi}t{\ifmmode\acute{e}\else\'{e}\fi}, F.~Dahm, F.~Da~Pieve, M.~Delaveau, M.~Di~Gennaro, B.~Dorado, C.~Espejo, G.~Geneste, L.~Genovese, A.~Gerossier, M.~Giantomassi, Y.~Gillet, D.~R. Hamann, L.~He, G.~Jomard, J.~Laflamme~Janssen, S.~Le~Roux, A.~Levitt, A.~Lherbier, F.~Liu, I.~Luka{\ifmmode\check{c}\else\v{c}\fi}evi{\ifmmode\acute{c}\else\'{c}\fi}, A.~Martin, C.~Martins, M.~J.~T. Oliveira, S.~Ponc{\ifmmode\acute{e}\else\'{e}\fi}, Y.~Pouillon, T.~Rangel, G.-M. Rignanese, A.~H. Romero, B.~Rousseau, O.~Rubel, A.~A. Shukri, M.~Stankovski, M.~Torrent, M.~J. Van~Setten, B.~Van~Troeye, M.~J. Verstraete, D.~Waroquiers, J.~Wiktor, B.~Xu, A.~Zhou, and J.~W. Zwanziger.
\newblock {Recent developments in the ABINIT software package}.
\newblock \emph{Comput. Phys. Commun.}, 205:\penalty0 106--131, August 2016.
\newblock ISSN 0010-4655.
\newblock \doi{10.1016/j.cpc.2016.04.003}.

\bibitem[Gonze et~al.(2020)Gonze, Amadon, Antonius, Arnardi, Baguet, Beuken, Bieder, Bottin, Bouchet, Bousquet, Brouwer, Bruneval, Brunin, Cavignac, Charraud, Chen, C{\ifmmode\hat{o}\else\^{o}\fi}t{\ifmmode\acute{e}\else\'{e}\fi}, Cottenier, Denier, Geneste, Ghosez, Giantomassi, Gillet, Gingras, Hamann, Hautier, He, Helbig, Holzwarth, Jia, Jollet, Lafargue-Dit-Hauret, Lejaeghere, Marques, Martin, Martins, Miranda, Naccarato, Persson, Petretto, Planes, Pouillon, Prokhorenko, Ricci, Rignanese, Romero, Schmitt, Torrent, van Setten, Van~Troeye, Verstraete, Z{\ifmmode\acute{e}\else\'{e}\fi}rah, and Zwanziger]{gonze2020abinit}
Xavier Gonze, Bernard Amadon, Gabriel Antonius, Fr{\ifmmode\acute{e}\else\'{e}\fi}d{\ifmmode\acute{e}\else\'{e}\fi}ric Arnardi, Lucas Baguet, Jean-Michel Beuken, Jordan Bieder, Fran{\ifmmode\mbox{\c{c}}\else\c{c}\fi}ois Bottin, Johann Bouchet, Eric Bousquet, Nils Brouwer, Fabien Bruneval, Guillaume Brunin, Th{\ifmmode\acute{e}\else\'{e}\fi}o Cavignac, Jean-Baptiste Charraud, Wei Chen, Michel C{\ifmmode\hat{o}\else\^{o}\fi}t{\ifmmode\acute{e}\else\'{e}\fi}, Stefaan Cottenier, Jules Denier, Gr{\ifmmode\acute{e}\else\'{e}\fi}gory Geneste, Philippe Ghosez, Matteo Giantomassi, Yannick Gillet, Olivier Gingras, Donald~R. Hamann, Geoffroy Hautier, Xu~He, Nicole Helbig, Natalie Holzwarth, Yongchao Jia, Fran{\ifmmode\mbox{\c{c}}\else\c{c}\fi}ois Jollet, William Lafargue-Dit-Hauret, Kurt Lejaeghere, Miguel A.~L. Marques, Alexandre Martin, Cyril Martins, Henrique P.~C. Miranda, Francesco Naccarato, Kristin Persson, Guido Petretto, Valentin Planes, Yann Pouillon, Sergei Prokhorenko, Fabio Ricci, Gian-Marco Rignanese,
  Aldo~H. Romero, Michael~Marcus Schmitt, Marc Torrent, Michiel~J. van Setten, Benoit Van~Troeye, Matthieu~J. Verstraete, Gilles Z{\ifmmode\acute{e}\else\'{e}\fi}rah, and Josef~W. Zwanziger.
\newblock {The Abinit project: Impact, environment and recent developments}.
\newblock \emph{Comput. Phys. Commun.}, 248:\penalty0 107042, March 2020.
\newblock ISSN 0010-4655.
\newblock \doi{10.1016/j.cpc.2019.107042}.

\bibitem[Romero et~al.(2020)Romero, Allan, Amadon, Antonius, Applencourt, Baguet, Bieder, Bottin, Bouchet, Bousquet, Bruneval, Brunin, Caliste, C{\ifmmode\hat{o}\else\^{o}\fi}t{\ifmmode\acute{e}\else\'{e}\fi}, Denier, Dreyer, Ghosez, Giantomassi, Gillet, Gingras, Hamann, Hautier, Jollet, Jomard, Martin, Miranda, Naccarato, Petretto, Pike, Planes, Prokhorenko, Rangel, Ricci, Rignanese, Royo, Stengel, Torrent, van Setten, Van~Troeye, Verstraete, Wiktor, Zwanziger, and Gonze]{romero2020abinit}
Aldo~H. Romero, Douglas~C. Allan, Bernard Amadon, Gabriel Antonius, Thomas Applencourt, Lucas Baguet, Jordan Bieder, Fran{\ifmmode\mbox{\c{c}}\else\c{c}\fi}ois Bottin, Johann Bouchet, Eric Bousquet, Fabien Bruneval, Guillaume Brunin, Damien Caliste, Michel C{\ifmmode\hat{o}\else\^{o}\fi}t{\ifmmode\acute{e}\else\'{e}\fi}, Jules Denier, Cyrus Dreyer, Philippe Ghosez, Matteo Giantomassi, Yannick Gillet, Olivier Gingras, Donald~R. Hamann, Geoffroy Hautier, Fran{\ifmmode\mbox{\c{c}}\else\c{c}\fi}ois Jollet, G{\ifmmode\acute{e}\else\'{e}\fi}rald Jomard, Alexandre Martin, Henrique P.~C. Miranda, Francesco Naccarato, Guido Petretto, Nicholas~A. Pike, Valentin Planes, Sergei Prokhorenko, Tonatiuh Rangel, Fabio Ricci, Gian-Marco Rignanese, Miquel Royo, Massimiliano Stengel, Marc Torrent, Michiel~J. van Setten, Benoit Van~Troeye, Matthieu~J. Verstraete, Julia Wiktor, Josef~W. Zwanziger, and Xavier Gonze.
\newblock {ABINIT: Overview and focus on selected capabilities}.
\newblock \emph{J. Chem. Phys.}, 152\penalty0 (12), March 2020.
\newblock ISSN 0021-9606.
\newblock \doi{10.1063/1.5144261}.

\bibitem[Gonze(1995)]{Gonze1995Aug}
Xavier Gonze.
\newblock {Perturbation expansion of variational principles at arbitrary order}.
\newblock \emph{Phys. Rev. A}, 52\penalty0 (2):\penalty0 1086--1095, August 1995.
\newblock ISSN 2469-9934.
\newblock \doi{10.1103/PhysRevA.52.1086}.

\bibitem[Veithen et~al.(2005)Veithen, Gonze, and Ghosez]{veithen2005nonlinear}
M.~Veithen, X.~Gonze, and {\relax Ph}.~Ghosez.
\newblock {Nonlinear optical susceptibilities, Raman efficiencies, and electro-optic tensors from first-principles density functional perturbation theory}.
\newblock \emph{Phys. Rev. B}, 71\penalty0 (12):\penalty0 125107, March 2005.
\newblock ISSN 2469-9969.
\newblock \doi{10.1103/PhysRevB.71.125107}.

\bibitem[van Setten et~al.(2018)van Setten, Giantomassi, Bousquet, Verstraete, Hamann, Gonze, and Rignanese]{van2018pseudodojo}
M.~J. van Setten, M.~Giantomassi, E.~Bousquet, M.~J. Verstraete, D.~R. Hamann, X.~Gonze, and G.-M. Rignanese.
\newblock {The PseudoDojo: Training and grading a 85 element optimized norm-conserving pseudopotential table}.
\newblock \emph{Comput. Phys. Commun.}, 226:\penalty0 39--54, May 2018.
\newblock ISSN 0010-4655.
\newblock \doi{10.1016/j.cpc.2018.01.012}.

\bibitem[Majewski and Vogl(1992)]{majewski1992self}
J.~A. Majewski and P.~Vogl.
\newblock {Self-interaction-corrected density-functional formalism. I. Ground-state properties of the Hubbard-Peierls model}.
\newblock \emph{Phys. Rev. B}, 46\penalty0 (19):\penalty0 12219--12234, November 1992.
\newblock ISSN 2469-9969.
\newblock \doi{10.1103/PhysRevB.46.12219}.

\bibitem[Ganose et~al.(2025)Ganose, Sahasrabuddhe, Asta, Beck, Biswas, Bonkowski, Bustamante, Chen, Chiang, Chrzan, Clary, Cohen, Ertural, George, Gallant, George, Gerits, Goodall, Guha, Hautier, Horton, Kaplan, Kingsbury, Kuner, Li, Linn, McDermott, Mohanakrishnan, Naik, Neaton, Persson, Petretto, Purcell, Ricci, Rich, Riebesell, Rignanese, Rosen, Scheffler, Schmidt, Shen, Sobolev, Sundararaman, Tezak, Trinquet, Varley, Vigil-Fowler, Wang, Waroquiers, Wen, Yang, Zheng, Zheng, Zhu, and Jain]{Ganose2025}
Alex Ganose, Hrushikesh Sahasrabuddhe, Mark Asta, Kevin Beck, Tathagata Biswas, Alexander Bonkowski, Joana Bustamante, Xin Chen, Yuan Chiang, Daryl Chrzan, Jacob Clary, Orion Cohen, Christina Ertural, Janine George, Max Gallant, Janine George, Sophie Gerits, Rhys Goodall, Rishabh Guha, Geoffroy Hautier, Matthew Horton, Aaron Kaplan, Ryan Kingsbury, Matthew Kuner, Bryant Li, Xavier Linn, Matthew McDermott, Rohith~Srinivaas Mohanakrishnan, Aakash Naik, Jeffrey Neaton, Kristin Persson, Guido Petretto, Thomas Purcell, Francesco Ricci, Benjamin Rich, Janosh Riebesell, Gian-Marco Rignanese, Andrew Rosen, Matthias Scheffler, Jonathan Schmidt, Jimmy-Xuan Shen, Andrei Sobolev, Ravishankar Sundararaman, Cooper Tezak, Victor Trinquet, Joel Varley, Derek Vigil-Fowler, Duo Wang, David Waroquiers, Mingjian Wen, Han Yang, Hui Zheng, Jiongzhi Zheng, Zhuoying Zhu, and Anubhav Jain.
\newblock {Atomate2: Modular workflows for materials science}.
\newblock \emph{ChemRxiv}, pages 1--66, 2025.
\newblock \doi{10.26434/chemrxiv-2025-tcr5h}.

\bibitem[Rosen et~al.(2024)Rosen, Gallant, George, Riebesell, Sahasrabuddhe, Shen, Wen, Evans, Petretto, Waroquiers, Rignanese, Persson, Jain, and Ganose]{Rosen2024}
Andrew~S. Rosen, Max Gallant, Janine George, Janosh Riebesell, Hrushikesh Sahasrabuddhe, Jimmy-Xuan Shen, Mingjian Wen, Matthew~L. Evans, Guido Petretto, David Waroquiers, Gian-Marco Rignanese, Kristin~A. Persson, Anubhav Jain, and Alex~M. Ganose.
\newblock Jobflow: {{Computational Workflows Made Simple}}.
\newblock \emph{Journal of Open Source Software}, 9\penalty0 (93):\penalty0 5995, January 2024.
\newblock ISSN 2475-9066.
\newblock \doi{10.21105/joss.05995}.

\bibitem[Trinquet et~al.(2024)Trinquet, Naccarato, Brunin, Petretto, Wirtz, Hautier, and Rignanese]{Trinquet2024Jul}
Victor Trinquet, Francesco Naccarato, Guillaume Brunin, Guido Petretto, Ludger Wirtz, Geoffroy Hautier, and Gian-Marco Rignanese.
\newblock {Second-harmonic generation tensors from high-throughput density-functional perturbation theory}.
\newblock \emph{Sci. Data}, 11\penalty0 (757):\penalty0 1--10, July 2024.
\newblock ISSN 2052-4463.
\newblock \doi{10.1038/s41597-024-03590-9}.

\bibitem[Jain et~al.(2015)Jain, Ong, Chen, Medasani, Qu, Kocher, Brafman, Petretto, Rignanese, Hautier, Gunter, and Persson]{jain2015fireworks}
Anubhav Jain, Shyue~Ping Ong, Wei Chen, Bharat Medasani, Xiaohui Qu, Michael Kocher, Miriam Brafman, Guido Petretto, Gian-Marco Rignanese, Geoffroy Hautier, Daniel Gunter, and Kristin~A. Persson.
\newblock {FireWorks: a dynamic workflow system designed for high-throughput applications}.
\newblock \emph{Concurrency Computat.: Pract. Exper.}, 27\penalty0 (17):\penalty0 5037--5059, December 2015.
\newblock ISSN 1532-0626.
\newblock \doi{10.1002/cpe.3505}.

\bibitem[iee(1988)]{ieee1987}
{IEEE Standard on Piezoelectricity}.
\newblock \emph{ANSI/IEEE Std 176-1987}, pages 0\_1--, 1988.
\newblock \doi{10.1109/IEEESTD.1988.79638}.

\bibitem[Roberts(1992)]{roberts1992}
D.A. Roberts.
\newblock Simplified characterization of uniaxial and biaxial nonlinear optical crystals: a plea for standardization of nomenclature and conventions.
\newblock \emph{IEEE Journal of Quantum Electronics}, 28\penalty0 (10):\penalty0 2057--2074, 1992.
\newblock \doi{10.1109/3.159516}.

\bibitem[Perdew et~al.(1996)Perdew, Burke, and Ernzerhof]{Perdew1996Oct}
John~P. Perdew, Kieron Burke, and Matthias Ernzerhof.
\newblock {Generalized Gradient Approximation Made Simple}.
\newblock \emph{Phys. Rev. Lett.}, 77\penalty0 (18):\penalty0 3865--3868, October 1996.
\newblock ISSN 1079-7114.
\newblock \doi{10.1103/PhysRevLett.77.3865}.

\bibitem[Perdew et~al.(1997)Perdew, Burke, and Ernzerhof]{Perdew1997Feb}
John~P. Perdew, Kieron Burke, and Matthias Ernzerhof.
\newblock {Generalized Gradient Approximation Made Simple [Phys. Rev. Lett. 77, 3865 (1996)]}.
\newblock \emph{Phys. Rev. Lett.}, 78\penalty0 (7):\penalty0 1396, February 1997.
\newblock ISSN 1079-7114.
\newblock \doi{10.1103/PhysRevLett.78.1396}.

\bibitem[Heyd et~al.(2003)Heyd, Scuseria, and Ernzerhof]{Heyd2003}
Jochen Heyd, Gustavo~E. Scuseria, and Matthias Ernzerhof.
\newblock Hybrid functionals based on a screened coulomb potential.
\newblock \emph{The Journal of Chemical Physics}, 118\penalty0 (18):\penalty0 8207–8215, May 2003.
\newblock ISSN 1089-7690.
\newblock \doi{10.1063/1.1564060}.
\newblock URL \url{http://dx.doi.org/10.1063/1.1564060}.

\bibitem[Heyd et~al.(2006)Heyd, Scuseria, and Ernzerhof]{Heyd2006}
Jochen Heyd, Gustavo~E. Scuseria, and Matthias Ernzerhof.
\newblock Erratum: “hybrid functionals based on a screened coulomb potential” [j. chem. phys. 118, 8207 (2003)].
\newblock \emph{The Journal of Chemical Physics}, 124\penalty0 (21), June 2006.
\newblock ISSN 1089-7690.
\newblock \doi{10.1063/1.2204597}.
\newblock URL \url{http://dx.doi.org/10.1063/1.2204597}.

\bibitem[Bl{\ifmmode\ddot{o}\else\"{o}\fi}chl(1994)]{Blochl1994Dec}
P.~E. Bl{\ifmmode\ddot{o}\else\"{o}\fi}chl.
\newblock {Projector augmented-wave method}.
\newblock \emph{Phys. Rev. B}, 50\penalty0 (24):\penalty0 17953--17979, December 1994.
\newblock ISSN 2469-9969.
\newblock \doi{10.1103/PhysRevB.50.17953}.

\bibitem[Kresse and Hafner(1994)]{Kresse1994Oct}
G.~Kresse and J.~Hafner.
\newblock {Norm-conserving and ultrasoft pseudopotentials for first-row and transition elements}.
\newblock \emph{J. Phys.: Condens. Matter}, 6\penalty0 (40):\penalty0 8245, October 1994.
\newblock ISSN 0953-8984.
\newblock \doi{10.1088/0953-8984/6/40/015}.

\bibitem[Kresse and Joubert(1999)]{Kresse1999Jan}
G.~Kresse and D.~Joubert.
\newblock {From ultrasoft pseudopotentials to the projector augmented-wave method}.
\newblock \emph{Phys. Rev. B}, 59\penalty0 (3):\penalty0 1758--1775, January 1999.
\newblock ISSN 2469-9969.
\newblock \doi{10.1103/PhysRevB.59.1758}.

\bibitem[Dudarev et~al.(1998)Dudarev, Botton, Savrasov, Humphreys, and Sutton]{Dudarev1998Jan}
S.~L. Dudarev, G.~A. Botton, S.~Y. Savrasov, C.~J. Humphreys, and A.~P. Sutton.
\newblock {Electron-energy-loss spectra and the structural stability of nickel oxide: An LSDA+U study}.
\newblock \emph{Phys. Rev. B}, 57\penalty0 (3):\penalty0 1505--1509, January 1998.
\newblock ISSN 2469-9969.
\newblock \doi{10.1103/PhysRevB.57.1505}.

\bibitem[Jain et~al.(2011)Jain, Hautier, Moore, Ping~Ong, Fischer, Mueller, Persson, and Ceder]{Jain2011Jun}
Anubhav Jain, Geoffroy Hautier, Charles~J. Moore, Shyue Ping~Ong, Christopher~C. Fischer, Tim Mueller, Kristin~A. Persson, and Gerbrand Ceder.
\newblock {A high-throughput infrastructure for density functional theory calculations}.
\newblock \emph{Comput. Mater. Sci.}, 50\penalty0 (8):\penalty0 2295--2310, June 2011.
\newblock ISSN 0927-0256.
\newblock \doi{10.1016/j.commatsci.2011.02.023}.

\bibitem[Trinquet et~al.(2025{\natexlab{b}})Trinquet, Evans, Hargreaves, De~Breuck, and Rignanese]{Trinquet2025Jan}
Victor Trinquet, Matthew~L. Evans, Cameron~J. Hargreaves, Pierre-Paul De~Breuck, and Gian-Marco Rignanese.
\newblock {Optical materials discovery and design with federated databases and machine learning}.
\newblock \emph{Faraday Discuss.}, 256\penalty0 (0):\penalty0 459--482, January 2025{\natexlab{b}}.
\newblock ISSN 1359-6640.
\newblock \doi{10.1039/D4FD00092G}.

\bibitem[De~Breuck et~al.(2021{\natexlab{a}})De~Breuck, Hautier, and Rignanese]{DeBreuck2021}
Pierre-Paul De~Breuck, Geoffroy Hautier, and Gian-Marco Rignanese.
\newblock Materials property prediction for limited datasets enabled by feature selection and joint learning with {{MODNet}}.
\newblock \emph{npj Computational Materials}, 7\penalty0 (1):\penalty0 83, December 2021{\natexlab{a}}.
\newblock ISSN 2057-3960.
\newblock \doi{10.1038/s41524-021-00552-2}.

\bibitem[De~Breuck et~al.(2021{\natexlab{b}})De~Breuck, Evans, and Rignanese]{DeBreuck2021a}
Pierre-Paul De~Breuck, Matthew~L Evans, and Gian-Marco Rignanese.
\newblock Robust model benchmarking and bias-imbalance in data-driven materials science: A case study on {{MODNet}}.
\newblock \emph{Journal of Physics: Condensed Matter}, 33\penalty0 (40):\penalty0 404002, October 2021{\natexlab{b}}.
\newblock ISSN 0953-8984, 1361-648X.
\newblock \doi{10.1088/1361-648X/ac1280}.

\bibitem[De~Breuck et~al.(2022)De~Breuck, Heymans, and Rignanese]{DeBreuck2022}
Pierre-Paul De~Breuck, Grégoire Heymans, and Gian-Marco Rignanese.
\newblock Accurate experimental band gap predictions with multifidelity correction learning.
\newblock \emph{Journal of Materials Informatics}, 2:\penalty0 10, 2022.
\newblock ISSN 2770-372X.
\newblock \doi{10.20517/jmi.2022.13}.

\bibitem[Ward et~al.(2018)Ward, Dunn, Faghaninia, Zimmermann, Bajaj, Wang, Montoya, Chen, Bystrom, Dylla, Chard, Asta, Persson, Snyder, Foster, and Jain]{Ward2018}
Logan Ward, Alexander Dunn, Alireza Faghaninia, Nils E.~R. Zimmermann, Saurabh Bajaj, Qi~Wang, Joseph Montoya, Jiming Chen, Kyle Bystrom, Maxwell Dylla, Kyle Chard, Mark Asta, Kristin~A. Persson, G.~Jeffrey Snyder, Ian Foster, and Anubhav Jain.
\newblock Matminer: {{An}} open source toolkit for materials data mining.
\newblock \emph{Computational Materials Science}, 152:\penalty0 60--69, September 2018.
\newblock ISSN 0927-0256.
\newblock \doi{10.1016/j.commatsci.2018.05.018}.

\bibitem[Gouvêa et~al.()]{Gouvea2025}
Rogério Gouvêa et~al.
\newblock URL \url{https://github.com/rogeriog/pGNN}.
\newblock In preparation.

\bibitem[Miller(1964)]{Miller1964Jul}
Robert~C. Miller.
\newblock {Optical second harmonic generation in piezeoelectric crystals}.
\newblock \emph{Appl. Phys. Lett.}, 5\penalty0 (1):\penalty0 17--19, July 1964.
\newblock ISSN 0003-6951.
\newblock \doi{10.1063/1.1754022}.

\bibitem[Schmidt et~al.(2023)Schmidt, Hoffmann, Wang, Borlido, Carriço, Cerqueira, Botti, and Marques]{Schmidt2023}
Jonathan Schmidt, Noah Hoffmann, Hai-Chen Wang, Pedro Borlido, Pedro J. M.~A. Carriço, Tiago F.~T. Cerqueira, Silvana Botti, and Miguel A.~L. Marques.
\newblock Machine-{{Learning-Assisted Determination}} of the {{Global Zero-Temperature Phase Diagram}} of {{Materials}}.
\newblock \emph{Advanced Materials}, 35\penalty0 (22):\penalty0 2210788, 2023.
\newblock ISSN 1521-4095.
\newblock \doi{10.1002/adma.202210788}.

\bibitem[Ong et~al.(2013)Ong, Richards, Jain, Hautier, Kocher, Cholia, Gunter, Chevrier, Persson, and Ceder]{ong2013python}
Shyue~Ping Ong, William~Davidson Richards, Anubhav Jain, Geoffroy Hautier, Michael Kocher, Shreyas Cholia, Dan Gunter, Vincent~L. Chevrier, Kristin~A. Persson, and Gerbrand Ceder.
\newblock {Python Materials Genomics (pymatgen): A robust, open-source python library for materials analysis}.
\newblock \emph{Comput. Mater. Sci.}, 68:\penalty0 314--319, February 2013.
\newblock ISSN 0927-0256.
\newblock \doi{10.1016/j.commatsci.2012.10.028}.

\bibitem[Pedregosa et~al.(2011)Pedregosa, Varoquaux, Gramfort, Michel, Thirion, Grisel, Blondel, Prettenhofer, Weiss, Dubourg, Vanderplas, Passos, Cournapeau, Brucher, Perrot, and Duchesnay]{scikit-learn}
F.~Pedregosa, G.~Varoquaux, A.~Gramfort, V.~Michel, B.~Thirion, O.~Grisel, M.~Blondel, P.~Prettenhofer, R.~Weiss, V.~Dubourg, J.~Vanderplas, A.~Passos, D.~Cournapeau, M.~Brucher, M.~Perrot, and E.~Duchesnay.
\newblock Scikit-learn: Machine learning in {P}ython.
\newblock \emph{Journal of Machine Learning Research}, 12:\penalty0 2825--2830, 2011.

\bibitem[Geurts et~al.(2006)Geurts, Ernst, and Wehenkel]{Geurts2006Apr}
Pierre Geurts, Damien Ernst, and Louis Wehenkel.
\newblock {Extremely randomized trees}.
\newblock \emph{Mach. Learn.}, 63\penalty0 (1):\penalty0 3--42, April 2006.
\newblock ISSN 1573-0565.
\newblock \doi{10.1007/s10994-006-6226-1}.

\bibitem[Ke et~al.(2017)Ke, Meng, Finley, Wang, Chen, Ma, Ye, and Liu]{Ke2017}
Guolin Ke, Qi~Meng, Thomas Finley, Taifeng Wang, Wei Chen, Weidong Ma, Qiwei Ye, and Tie-Yan Liu.
\newblock Lightgbm: A highly efficient gradient boosting decision tree.
\newblock In I.~Guyon, U.~Von Luxburg, S.~Bengio, H.~Wallach, R.~Fergus, S.~Vishwanathan, and R.~Garnett, editors, \emph{Advances in Neural Information Processing Systems}, volume~30. Curran Associates, Inc., 2017.
\newblock URL \url{https://proceedings.neurips.cc/paper_files/paper/2017/file/6449f44a102fde848669bdd9eb6b76fa-Paper.pdf}.

\bibitem[Ruff et~al.(2024)Ruff, Reiser, St{\ifmmode\ddot{u}\else\"{u}\fi}hmer, and Friederich]{Ruff2024Mar}
Robin Ruff, Patrick Reiser, Jan St{\ifmmode\ddot{u}\else\"{u}\fi}hmer, and Pascal Friederich.
\newblock {Connectivity optimized nested line graph networks for crystal structures}.
\newblock \emph{Digital Discovery}, 3\penalty0 (3):\penalty0 594--601, March 2024.
\newblock ISSN 2635-098X.
\newblock \doi{10.1039/D4DD00018H}.

\bibitem[Chen et~al.(2019)Chen, Ye, Zuo, Zheng, and Ong]{Chen2019}
Chi Chen, Weike Ye, Yunxing Zuo, Chen Zheng, and Shyue~Ping Ong.
\newblock Graph {{Networks}} as a {{Universal Machine Learning Framework}} for {{Molecules}} and {{Crystals}}.
\newblock \emph{Chemistry of Materials}, 31\penalty0 (9):\penalty0 3564--3572, May 2019.
\newblock \doi{10.1021/acs.chemmater.9b01294}.

\bibitem[Simeon and de~Fabritiis(2023)]{Simeon2023}
Guillem Simeon and Gianni de~Fabritiis.
\newblock {TensorNet: Cartesian Tensor Representations for Efficient Learning of Molecular Potentials}.
\newblock \emph{arXiv}, June 2023.
\newblock \doi{10.48550/arXiv.2306.06482}.

\bibitem[Wen et~al.(2024)Wen, Horton, Munro, Huck, and Persson]{Wen2024May}
Mingjian Wen, Matthew~K. Horton, Jason~M. Munro, Patrick Huck, and Kristin~A. Persson.
\newblock {An equivariant graph neural network for the elasticity tensors of all seven crystal systems}.
\newblock \emph{Digital Discovery}, 3\penalty0 (5):\penalty0 869--882, May 2024.
\newblock ISSN 2635-098X.
\newblock \doi{10.1039/D3DD00233K}.

\bibitem[Xie et~al.(2024)Xie, Wan, Liu, Zeng, Wang, Zhang, Grazian, Kit, Ouyang, Zhou, and Hoex]{Xie2025}
Tong Xie, Yuwei Wan, Yixuan Liu, Yuchen Zeng, Shaozhou Wang, Wenjie Zhang, Clara Grazian, Chunyu Kit, Wanli Ouyang, Dongzhan Zhou, and Bram Hoex.
\newblock {DARWIN 1.5: Large Language Models as Materials Science Adapted Learners}.
\newblock \emph{arXiv}, December 2024.
\newblock \doi{10.48550/arXiv.2412.11970}.

\bibitem[Huck et~al.(2015)Huck, Jain, Gunter, Winston, and Persson]{huck2015community}
Patrick Huck, Anubhav Jain, Dan Gunter, Donald Winston, and Kristin Persson.
\newblock {A Community Contribution Framework for Sharing Materials Data with Materials Project}.
\newblock In \emph{2015 IEEE 11th International Conference on e-Science}, pages 535--541, 2015.
\newblock \doi{10.1109/eScience.2015.75}.

\bibitem[An et~al.(2025)An, Wang, Xie, Wu, Chu, Jin, Li, Pan, and Yang]{An2025Mar}
Ran An, Hongshan Wang, Congwei Xie, Mengfan Wu, Dongdong Chu, Wenqi Jin, Junjie Li, Shilie Pan, and Zhihua Yang.
\newblock {New Ways to Discover Novel Nonlinear Optical Materials: Scaling Machine Learning with Chemical Descriptors Information}.
\newblock \emph{Small}, 21\penalty0 (11):\penalty0 2500540, March 2025.
\newblock ISSN 1613-6810.
\newblock \doi{10.1002/smll.202500540}.

\bibitem[Kim et~al.(2025)Kim, Kim, Kim, Lee, Park, Kang, and Han]{Kim2025Jan}
Jaesun Kim, Jisu Kim, Jaehoon Kim, Jiho Lee, Yutack Park, Youngho Kang, and Seungwu Han.
\newblock {Data-Efficient Multifidelity Training for High-Fidelity Machine Learning Interatomic Potentials}.
\newblock \emph{J. Am. Chem. Soc.}, 147\penalty0 (1):\penalty0 1042--1054, January 2025.
\newblock ISSN 0002-7863.
\newblock \doi{10.1021/jacs.4c14455}.

\bibitem[Dmitriev et~al.(1999)Dmitriev, Gurzadyan, and Nikogosyan]{Dmitriev1999}
Valentin~G. Dmitriev, Gagik~G. Gurzadyan, and David~N. Nikogosyan.
\newblock \emph{{Handbook of Nonlinear Optical Crystals}}.
\newblock Springer, Berlin, Germany, 1999.
\newblock ISBN 978-3-540-46793-9.
\newblock URL \url{https://link.springer.com/book/10.1007/978-3-540-46793-9}.

\bibitem[Herfindahl(1950)]{Herfindahl1950}
Orris~Clemens Herfindahl.
\newblock \emph{{Concentration In The Steel Industry}}.
\newblock PhD thesis, Columbia University, 1950.

\bibitem[Gaultois et~al.(2013)Gaultois, Sparks, Borg, Seshadri, Bonificio, and Clarke]{Gaultois2013}
Michael~W. Gaultois, Taylor~D. Sparks, Christopher K.~H. Borg, Ram Seshadri, William~D. Bonificio, and David~R. Clarke.
\newblock Data-{{Driven Review}} of {{Thermoelectric Materials}}: {{Performance}} and {{Resource Considerations}}.
\newblock \emph{Chemistry of Materials}, 25\penalty0 (15):\penalty0 2911--2920, August 2013.
\newblock \doi{10.1021/cm400893e}.

\bibitem[Kim et~al.(2023)Kim, Mok, Kim, and Back]{Kim2023}
Jongseung Kim, Dong~Hyeon Mok, Heejin Kim, and Seoin Back.
\newblock Accelerating the {{Search}} for {{New Solid Electrolytes}}: {{Exploring Vast Chemical Space}} with {{Machine Learning-Enabled Computational Calculations}}.
\newblock \emph{ACS Applied Materials \& Interfaces}, 15\penalty0 (45):\penalty0 52427--52435, November 2023.
\newblock \doi{10.1021/acsami.3c10798}.

\bibitem[Glass et~al.(2006)Glass, Oganov, and Hansen]{Glass2006Dec}
Colin~W. Glass, Artem~R. Oganov, and Nikolaus Hansen.
\newblock {USPEX{\ifmmode---\else\textemdash\fi}Evolutionary crystal structure prediction}.
\newblock \emph{Comput. Phys. Commun.}, 175\penalty0 (11):\penalty0 713--720, December 2006.
\newblock ISSN 0010-4655.
\newblock \doi{10.1016/j.cpc.2006.07.020}.

\bibitem[Noh et~al.(2020)Noh, Gu, Kim, and Jung]{Noh2020May}
Juhwan Noh, Geun~Ho Gu, Sungwon Kim, and Yousung Jung.
\newblock {Machine-enabled inverse design of inorganic solid materials: promises and challenges}.
\newblock \emph{Chem. Sci.}, 11\penalty0 (19):\penalty0 4871--4881, May 2020.
\newblock ISSN 2041-6520.
\newblock \doi{10.1039/D0SC00594K}.

\bibitem[Ren et~al.(2022)Ren, Tian, Noh, Oviedo, Xing, Li, Liang, Zhu, Aberle, Sun, Wang, Liu, Li, Jayavelu, Hippalgaonkar, Jung, and Buonassisi]{Ren2022Jan}
Zekun Ren, Siyu Isaac~Parker Tian, Juhwan Noh, Felipe Oviedo, Guangzong Xing, Jiali Li, Qiaohao Liang, Ruiming Zhu, Armin~G. Aberle, Shijing Sun, Xiaonan Wang, Yi~Liu, Qianxiao Li, Senthilnath Jayavelu, Kedar Hippalgaonkar, Yousung Jung, and Tonio Buonassisi.
\newblock {An invertible crystallographic representation for general inverse design of inorganic crystals with targeted properties}.
\newblock \emph{Matter}, 5\penalty0 (1):\penalty0 314--335, January 2022.
\newblock ISSN 2590-2393.
\newblock \doi{10.1016/j.matt.2021.11.032}.

\bibitem[Wu et~al.(2025)Wu, Dong, Kang, and Lin]{Wu2025Jan}
Qingchen Wu, Linfeng Dong, Lei Kang, and Zheshuai Lin.
\newblock {Prediction and Evaluation of Li2ZnS2 Crystals as Mid-Infrared Nonlinear Optical Material with High Thermal Conductivity}.
\newblock \emph{Adv. Opt. Mater.}, n/a\penalty0 (n/a):\penalty0 2402922, January 2025.
\newblock ISSN 2195-1071.
\newblock \doi{10.1002/adom.202402922}.

\bibitem[Liu et~al.(2025)Liu, An, Li, Chu, Zhao, Pan, and Yang]{Liu2025Jan}
Qingyu Liu, Ran An, Chenxu Li, Dongdong Chu, Wang Zhao, Shilie Pan, and Zhihua Yang.
\newblock {Accelerating Discovery of Infrared Nonlinear Optical Materials with High Lattice Thermal Conductivity: Combining Machine Learning and First-Principles Calculations}.
\newblock \emph{Adv. Opt. Mater.}, n/a\penalty0 (n/a):\penalty0 2403292, January 2025.
\newblock ISSN 2195-1071.
\newblock \doi{10.1002/adom.202403292}.

\bibitem[Zhang et~al.(2017)Zhang, Yu, Wu, and Halasyamani]{Zhang2017Apr}
Weiguo Zhang, Hongwei Yu, Hongping Wu, and P.~Shiv Halasyamani.
\newblock {Phase-Matching in Nonlinear Optical Compounds: A Materials Perspective}.
\newblock \emph{Chem. Mater.}, 29\penalty0 (7):\penalty0 2655--2668, April 2017.
\newblock ISSN 0897-4756.
\newblock \doi{10.1021/acs.chemmater.7b00243}.

\bibitem[Lou and Ganose(2025)]{Lou2025Jan}
Yuchen Lou and Alex~M. Ganose.
\newblock {Discovery of highly anisotropic dielectric crystals with equivariant graph neural networks}.
\newblock \emph{Faraday Discuss.}, 256\penalty0 (0):\penalty0 255--274, January 2025.
\newblock ISSN 1359-6640.
\newblock \doi{10.1039/D4FD00096J}.

\bibitem[Dunn et~al.(2020)Dunn, Wang, Ganose, Dopp, and Jain]{Dunn2020}
Alexander Dunn, Qi~Wang, Alex Ganose, Daniel Dopp, and Anubhav Jain.
\newblock Benchmarking materials property prediction methods: The {{Matbench}} test set and {{Automatminer}} reference algorithm.
\newblock \emph{npj Computational Materials}, 6\penalty0 (1):\penalty0 1--10, September 2020.
\newblock ISSN 2057-3960.
\newblock \doi{10.1038/s41524-020-00406-3}.

\bibitem[Reiser et~al.(2022)Reiser, Neubert, Eberhard, Torresi, Zhou, Shao, Metni, van Hoesel, Schopmans, Sommer, and Friederich]{Reiser2022Nov}
Patrick Reiser, Marlen Neubert, Andr{\ifmmode\acute{e}\else\'{e}\fi} Eberhard, Luca Torresi, Chen Zhou, Chen Shao, Houssam Metni, Clint van Hoesel, Henrik Schopmans, Timo Sommer, and Pascal Friederich.
\newblock {Graph neural networks for materials science and chemistry}.
\newblock \emph{Commun. Mater.}, 3\penalty0 (93):\penalty0 1--18, November 2022.
\newblock ISSN 2662-4443.
\newblock \doi{10.1038/s43246-022-00315-6}.

\bibitem[Kaplan et~al.(2025)Kaplan, Liu, Qi, Ko, Deng, Riebesell, Ceder, Persson, and Ong]{Kaplan2025}
Aaron~D. Kaplan, Runze Liu, Ji~Qi, Tsz~Wai Ko, Bowen Deng, Janosh Riebesell, Gerbrand Ceder, Kristin~A. Persson, and Shyue~Ping Ong.
\newblock {A Foundational Potential Energy Surface Dataset for Materials}.
\newblock \emph{arXiv}, March 2025.
\newblock \doi{10.48550/arXiv.2503.04070}.

\bibitem[Grisafi et~al.(2018)Grisafi, Wilkins, Cs{\ifmmode\acute{a}\else\'{a}\fi}nyi, and Ceriotti]{Grisafi2018Jan}
Andrea Grisafi, David~M. Wilkins, G{\ifmmode\acute{a}\else\'{a}\fi}bor Cs{\ifmmode\acute{a}\else\'{a}\fi}nyi, and Michele Ceriotti.
\newblock {Symmetry-Adapted Machine Learning for Tensorial Properties of Atomistic Systems}.
\newblock \emph{Phys. Rev. Lett.}, 120\penalty0 (3):\penalty0 036002, January 2018.
\newblock \doi{10.1103/PhysRevLett.120.036002}.

\bibitem[Batzner et~al.(2022)Batzner, Musaelian, Sun, Geiger, Mailoa, Kornbluth, Molinari, Smidt, and Kozinsky]{Batzner2022May}
Simon Batzner, Albert Musaelian, Lixin Sun, Mario Geiger, Jonathan~P. Mailoa, Mordechai Kornbluth, Nicola Molinari, Tess~E. Smidt, and Boris Kozinsky.
\newblock {E(3)-equivariant graph neural networks for data-efficient and accurate interatomic potentials}.
\newblock \emph{Nat. Commun.}, 13\penalty0 (2453):\penalty0 1--11, May 2022.
\newblock ISSN 2041-1723.
\newblock \doi{10.1038/s41467-022-29939-5}.

\bibitem[Neumann(1885)]{neumann1885vorlesungen}
Franz Neumann.
\newblock \emph{Vorlesungen {\"u}ber die Theorie der Elasticit{\"a}t der festen K{\"o}rper und des Licht{\"a}thers}, volume~5.
\newblock BG Teubner, 1885.

\bibitem[Kleinman(1962)]{Kleinman1962Jun}
D.~A. Kleinman.
\newblock {Nonlinear Dielectric Polarization in Optical Media}.
\newblock \emph{Phys. Rev.}, 126\penalty0 (6):\penalty0 1977--1979, June 1962.
\newblock \doi{10.1103/PhysRev.126.1977}.

\bibitem[Brown et~al.(2020)Brown, Mann, Ryder, Subbiah, Kaplan, Dhariwal, Neelakantan, Shyam, Sastry, Askell, Agarwal, Herbert-Voss, Krueger, Henighan, Child, Ramesh, Ziegler, Wu, Winter, Hesse, Chen, Sigler, Litwin, Gray, Chess, Clark, Berner, McCandlish, Radford, Sutskever, and Amodei]{Brown2020}
Tom~B. Brown, Benjamin Mann, Nick Ryder, Melanie Subbiah, Jared Kaplan, Prafulla Dhariwal, Arvind Neelakantan, Pranav Shyam, Girish Sastry, Amanda Askell, Sandhini Agarwal, Ariel Herbert-Voss, Gretchen Krueger, Tom Henighan, Rewon Child, Aditya Ramesh, Daniel~M. Ziegler, Jeffrey Wu, Clemens Winter, Christopher Hesse, Mark Chen, Eric Sigler, Mateusz Litwin, Scott Gray, Benjamin Chess, Jack Clark, Christopher Berner, Sam McCandlish, Alec Radford, Ilya Sutskever, and Dario Amodei.
\newblock {Language Models are Few-Shot Learners}.
\newblock \emph{arXiv}, May 2020.
\newblock \doi{10.48550/arXiv.2005.14165}.

\bibitem[White(2023)]{White2023}
Andrew~D. White.
\newblock The future of chemistry is language.
\newblock \emph{Nature Reviews Chemistry}, 7\penalty0 (7):\penalty0 457--458, 2023.
\newblock \doi{10.1038/s41570-023-00502-0}.

\bibitem[Jablonka et~al.(2024)Jablonka, Schwaller, {Ortega-Guerrero}, and Smit]{Jablonka2024}
Kevin~Maik Jablonka, Philippe Schwaller, Andres {Ortega-Guerrero}, and Berend Smit.
\newblock Leveraging large language models for predictive chemistry.
\newblock \emph{Nature Machine Intelligence}, 6\penalty0 (2):\penalty0 161--169, 2024.
\newblock \doi{10.1038/s42256-023-00788-1}.

\bibitem[Zimmermann et~al.(2024)Zimmermann, Bazgir, Afzal, Agbere, Ai, Alampara, Al-Feghali, Ansari, Antypov, Aswad, Bai, Baibakova, Biswajeet, Bitzek, Bocarsly, Borisova, Bran, Brinson, Calderon, Canalicchio, Chen, Chiang, Circi, Charmes, Chaudhary, Chen, Chiu, Clymo, Dabhadkar, Daelman, Datar, de~Jong, Evans, Fard, Fisicaro, Gangan, George, Gonzalez, G{\ifmmode\ddot{o}\else\"{o}\fi}tte, Gupta, Harb, Hong, Ibrahim, Ilyas, Imran, Ishimwe, Issa, Jablonka, Jones, Josephson, Juhasz, Kapoor, Kang, Khalighinejad, Khan, Klawohn, Kuman, Ladines, Leang, Lederbauer, Sheng-Lun, Liao, Liu, Liu, Lo, Madireddy, Maharana, Maheshwari, Mahjoubi, M{\ifmmode\acute{a}\else\'{a}\fi}rquez, Mills, Mohanty, Mohr, Moosavi, Mo{\ss}hammer, Naghdi, Naik, Narykov, N{\ifmmode\ddot{a}\else\"{a}\fi}sstr{\ifmmode\ddot{o}\else\"{o}\fi}m, Nguyen, Ni, O'Connor, Olayiwola, Ottomano, Ozhan, Pagel, Parida, Park, Patel, Patyukova, Petersen, Pinto, Pizarro, Plessers, Pradhan, Pratiush, Puli, Qin, Rajabi, Ricci, Risch,
  R{\ifmmode\acute{\imath}\else\'{\i}\fi}os-Garc{\ifmmode\acute{\imath}\else\'{\i}\fi}a, Roy, Rug, Sayeed, Scheidgen, Schilling-Wilhelmi, Schloz, Sch{\ifmmode\ddot{o}\else\"{o}\fi}ppach, Schumann, Schwaller, Schwarting, Sharlin, Shen, Shi, Si, D'Souza, Sparks, Sudhakar, Talirz, Tang, Taran, Terboven, Tropin, Tsymbal, Ueltzen, Unzueta, Vasan, Vinchurkar, Vo, Vogel, V{\ifmmode\ddot{o}\else\"{o}\fi}lker, Weinreich, Yang, Zaki, Zhang, Zhang, Zhang, Zhu, Zhu, Janssen, Li, Foster, and Blaiszik]{Zimmermann2024}
Yoel Zimmermann, Adib Bazgir, Zartashia Afzal, Fariha Agbere, Qianxiang Ai, Nawaf Alampara, Alexander Al-Feghali, Mehrad Ansari, Dmytro Antypov, Amro Aswad, Jiaru Bai, Viktoriia Baibakova, Devi~Dutta Biswajeet, Erik Bitzek, Joshua~D. Bocarsly, Anna Borisova, Andres~M. Bran, L.~Catherine Brinson, Marcel~Moran Calderon, Alessandro Canalicchio, Victor Chen, Yuan Chiang, Defne Circi, Benjamin Charmes, Vikrant Chaudhary, Zizhang Chen, Min-Hsueh Chiu, Judith Clymo, Kedar Dabhadkar, Nathan Daelman, Archit Datar, Wibe~A. de~Jong, Matthew~L. Evans, Maryam~Ghazizade Fard, Giuseppe Fisicaro, Abhijeet~Sadashiv Gangan, Janine George, Jose D.~Cojal Gonzalez, Michael G{\ifmmode\ddot{o}\else\"{o}\fi}tte, Ankur~K. Gupta, Hassan Harb, Pengyu Hong, Abdelrahman Ibrahim, Ahmed Ilyas, Alishba Imran, Kevin Ishimwe, Ramsey Issa, Kevin~Maik Jablonka, Colin Jones, Tyler~R. Josephson, Greg Juhasz, Sarthak Kapoor, Rongda Kang, Ghazal Khalighinejad, Sartaaj Khan, Sascha Klawohn, Suneel Kuman, Alvin~Noe Ladines, Sarom Leang, Magdalena
  Lederbauer, Sheng-Lun, Liao, Hao Liu, Xuefeng Liu, Stanley Lo, Sandeep Madireddy, Piyush~Ranjan Maharana, Shagun Maheshwari, Soroush Mahjoubi, Jos{\ifmmode\acute{e}\else\'{e}\fi}~A. M{\ifmmode\acute{a}\else\'{a}\fi}rquez, Rob Mills, Trupti Mohanty, Bernadette Mohr, Seyed~Mohamad Moosavi, Alexander Mo{\ss}hammer, Amirhossein~D. Naghdi, Aakash Naik, Oleksandr Narykov, Hampus N{\ifmmode\ddot{a}\else\"{a}\fi}sstr{\ifmmode\ddot{o}\else\"{o}\fi}m, Xuan~Vu Nguyen, Xinyi Ni, Dana O'Connor, Teslim Olayiwola, Federico Ottomano, Aleyna~Beste Ozhan, Sebastian Pagel, Chiku Parida, Jaehee Park, Vraj Patel, Elena Patyukova, Martin~Hoffmann Petersen, Luis Pinto, Jos{\ifmmode\acute{e}\else\'{e}\fi}~M. Pizarro, Dieter Plessers, Tapashree Pradhan, Utkarsh Pratiush, Charishma Puli, Andrew Qin, Mahyar Rajabi, Francesco Ricci, Elliot Risch, Marti{\ifmmode\tilde{n}\else\~{n}\fi}o R{\ifmmode\acute{\imath}\else\'{\i}\fi}os-Garc{\ifmmode\acute{\imath}\else\'{\i}\fi}a, Aritra Roy, Tehseen Rug, Hasan~M. Sayeed, Markus Scheidgen, Mara
  Schilling-Wilhelmi, Marcel Schloz, Fabian Sch{\ifmmode\ddot{o}\else\"{o}\fi}ppach, Julia Schumann, Philippe Schwaller, Marcus Schwarting, Samiha Sharlin, Kevin Shen, Jiale Shi, Pradip Si, Jennifer D'Souza, Taylor Sparks, Suraj Sudhakar, Leopold Talirz, Dandan Tang, Olga Taran, Carla Terboven, Mark Tropin, Anastasiia Tsymbal, Katharina Ueltzen, Pablo~Andres Unzueta, Archit Vasan, Tirtha Vinchurkar, Trung Vo, Gabriel Vogel, Christoph V{\ifmmode\ddot{o}\else\"{o}\fi}lker, Jan Weinreich, Faradawn Yang, Mohd Zaki, Chi Zhang, Sylvester Zhang, Weijie Zhang, Ruijie Zhu, Shang Zhu, Jan Janssen, Calvin Li, Ian Foster, and Ben Blaiszik.
\newblock {Reflections from the 2024 Large Language Model (LLM) Hackathon for Applications in Materials Science and Chemistry}.
\newblock \emph{arXiv}, November 2024.
\newblock \doi{10.48550/arXiv.2411.15221}.

\bibitem[Ganose and Jain(2019)]{Ganose2019}
Alex~M. Ganose and Anubhav Jain.
\newblock Robocrystallographer: automated crystal structure text descriptions and analysis.
\newblock \emph{MRS Communications}, 9\penalty0 (3):\penalty0 874--881, 2019.
\newblock \doi{10.1557/mrc.2019.94}.

\bibitem[Mirza et~al.(2024)Mirza, Alampara, Kunchapu, R{\ifmmode\acute{\imath}\else\'{\i}\fi}os-Garc{\ifmmode\acute{\imath}\else\'{\i}\fi}a, Emoekabu, Krishnan, Gupta, Schilling-Wilhelmi, Okereke, Aneesh, Elahi, Asgari, Eberhardt, Elbeheiry, Gil, Greiner, Holick, Glaubitz, Hoffmann, Ibrahim, Klepsch, K{\ifmmode\ddot{o}\else\"{o}\fi}ster, Kreth, Meyer, Miret, Peschel, Ringleb, Roesner, Schreiber, Schubert, Stafast, Wonanke, Pieler, Schwaller, and Jablonka]{Mirza2024}
Adrian Mirza, Nawaf Alampara, Sreekanth Kunchapu, Marti{\ifmmode\tilde{n}\else\~{n}\fi}o R{\ifmmode\acute{\imath}\else\'{\i}\fi}os-Garc{\ifmmode\acute{\imath}\else\'{\i}\fi}a, Benedict Emoekabu, Aswanth Krishnan, Tanya Gupta, Mara Schilling-Wilhelmi, Macjonathan Okereke, Anagha Aneesh, Amir~Mohammad Elahi, Mehrdad Asgari, Juliane Eberhardt, Hani~M. Elbeheiry, Mar{\ifmmode\acute{\imath}\else\'{\i}\fi}a~Victoria Gil, Maximilian Greiner, Caroline~T. Holick, Christina Glaubitz, Tim Hoffmann, Abdelrahman Ibrahim, Lea~C. Klepsch, Yannik K{\ifmmode\ddot{o}\else\"{o}\fi}ster, Fabian~Alexander Kreth, Jakob Meyer, Santiago Miret, Jan~Matthias Peschel, Michael Ringleb, Nicole Roesner, Johanna Schreiber, Ulrich~S. Schubert, Leanne~M. Stafast, Dinga Wonanke, Michael Pieler, Philippe Schwaller, and Kevin~Maik Jablonka.
\newblock {Are large language models superhuman chemists?}
\newblock \emph{arXiv}, April 2024.
\newblock \doi{10.48550/arXiv.2404.01475}.

\bibitem[Alampara et~al.(2024)Alampara, Schilling-Wilhelmi, R{\ifmmode\acute{\imath}\else\'{\i}\fi}os-Garc{\ifmmode\acute{\imath}\else\'{\i}\fi}a, Mandal, Khetarpal, Grover, Krishnan, and Jablonka]{Alampara2025}
Nawaf Alampara, Mara Schilling-Wilhelmi, Marti{\ifmmode\tilde{n}\else\~{n}\fi}o R{\ifmmode\acute{\imath}\else\'{\i}\fi}os-Garc{\ifmmode\acute{\imath}\else\'{\i}\fi}a, Indrajeet Mandal, Pranav Khetarpal, Hargun~Singh Grover, N.~M.~Anoop Krishnan, and Kevin~Maik Jablonka.
\newblock {Probing the limitations of multimodal language models for chemistry and materials research}.
\newblock \emph{arXiv}, November 2024.
\newblock \doi{10.48550/arXiv.2411.16955}.

\end{thebibliography}
\bibliographystyle{unsrtnat}

\appendix
\renewcommand\thefigure{\thesection.\arabic{figure}}
\setcounter{figure}{0}
\section{Appendix}
\subsection{Details of ML benchmarking}\label{sec:ml-appendix}

\paragraph*{MODNet}
This neural-network is the model adopted throughout the AL loops. Although fairly simple, its performance on small datasets has been proven in the MatBench suite as it leads 5 out of 7 tasks with fewer than 10,000 data points~\cite{Dunn2020}. 
It comes with useful methods to optimize its hyperparameters and perform feature selection without much overhead. 
In this benchmark, the same parameters as for the AL loops were adopted.
The hyperparameters were optimized with the \texttt{FitGenetic} class of \emph{modnet} using 5-fold cross-validation. 
The \texttt{refit} parameter is set to 0 such that the top 10 models trained during the cross-validation are not refitted on the entire training set and instead constitute a final ensemble, composed of 50 individual MODNet models with 10 different sets of hyperparameters. 
\paragraph*{Tree-based methods: Extra-Trees and LGBM} 
Recently, Ref.~\cite{An2025Mar} modelled $d_\text{KP}$ with tree-based algorithms, in particlar Extra-Trees (ET) and Gradient-Boosted Machines (GBM). Implemented in the \emph{scikit-learn} Python package (\texttt{ExtraTreesRegressor}), ET is an ensemble of decision trees, whose nodes are split according to random cut-points~\cite{scikit-learn,Geurts2006Apr}. The whole dataset was used to grow the trees. GBM was used as implemented in the \emph{LightGBM} package (\texttt{LGBMRegressor})~\cite{Ke2017}. Its specialities are efficient feature reduction and robustness in low data regimes.
Both algorithms are well-known and popular, with fast training and inference times. Since they require all features for each entry, the missing features are imputed by their average in the training set. 
Three set of hyperparameters were investigated. The first one corresponds to the default values of the underlying libraries, the second is taken from ~\cite{An2025Mar}, which results from a Bayesian optimization process, and the third was found by a grid search on a validation set.
\paragraph*{Graph neural networks: co(N)GN, TensorNet and MEGNet}
Due to their inherent graphical structure, molecules and materials are increasingly modelled with graph neural networks (GNNs), where the atoms and bonds are naturally represented as nodes and edges. The literature is filled with different design choices~\cite{Reiser2022Nov}. Since an exhaustive test is not possible, we selected three popular open-source GNN models: co(N)GN, TensorNet and MEGNet. 
Introduced in early 2024, the connectivity-optimized crystal graph network (coGN) and its nested line graph network variant (coNGN) lead 5 out of the 6 MatBench tasks with more than 10,000 data points~\cite{Ruff2024Mar}. The default architecture and hyperparameters implemented in version 3.1 of the \href{https://github.com/aimat-lab/gcnn_keras/releases/tag/v3.1.0}{\emph{gcnn\_keras}} repository were adopted, which were obtained after optimization on the \emph{log\_gvrh} MatBench task and used throughout the MatBench submission.
The equivariant TensorNet~\cite{Simeon2023} and MEGNet~\cite{Chen2019} were benchmarked as implemented in the \href{https://github.com/materialsvirtuallab/matgl}{\emph{matgl}} Python package, with the default architectures and hyperparameters previously used to train universal models. Two separate TensorNet models were trained with $\text{SO}(3)$ and $\text{O}(3)$ equivariance, respectively; both models were trained to only predict the scalar $d_\text{KP}$ rather than the full SHG tensor. Although TensorNet was developed to predict molecular properties, it has recently been shown to perform well for materials~\cite{Kaplan2025}.

\paragraph*{Tensor predictions: Matten}
While the effective KP coefficient is useful for screening and visualization, the full SHG tensor is still required for, e.g., the determination of the phase-matching. To that end, and to circumvent the need for first-principles calculations, GNNs have emerged with the ability to predict a tensor from the input structure~\cite{Grisafi2018Jan}. 
A key component of these models is equivariance, which ensures that the output tensor corresponds to the input structure, independently of the frame of reference~\cite{Batzner2022May}.
Another feature is their ability to reflect the material symmetry in the output tensor, thus respecting Neumann's principle~\cite{neumann1885vorlesungen}. 
This is especially important if one wishes to present tensors following the IEEE conventions.
Among the available models, Matten ~\cite{Wen2024May} was chosen for this benchmark because its training can easily be generalized to tensors of any order and symmetry. 
Indeed, it only requires adapting the indicial notation reflecting the tensor symmetry in the configuration file. 
In the static limit, Kleinman symmetry is respected and the SHG tensor is reduced to 10 independent components~\cite{Kleinman1962Jun}. 
In indicial notation, this is expressed as \texttt{ijk=ikj=jik}.
In addition to the default hyperparameters, a restricted grid search using the validation set was performed for the \emph{distribution\_125} holdout test set and the resulting set of hyperparameters were then adopted for the other training runs.

\paragraph*{Large language models}

Large language models (LLMs) are attracting significant interest as zero-shot regressors for many scientific tasks~\cite{Brown2020, White2023, Jablonka2024, Zimmermann2024}.
Here, we benchmarked the leading closed models from Anthropic (Claude Sonnet 3.5) and OpenAI (GPT-4o) via their respective APIs, as well as the open weights model DARWIN 1.5 \cite{Xie2025}, a materials science-focused fine-tune of Meta's Llama 3.1-7b (i.e., two orders of magnitude fewer parameters than suspected of Sonnet and GPT-4o). This model was not further fine-tuned on our specific task.

A system prompt was crafted to explain the task to each model, and three different textual representations of the input structures were used: composition (using \emph{pymatgen's} reduced formula), composition and space group symbol, and the full output of RoboCrystallographer ~\cite{Ganose2019} for each structure. For the former two descriptions, benchmark runs were performed both with and without in-context learning, where the models were exposed to a subset of the training set in their prompt. An example prompt required to make one prediction is shown in \autoref{fig:example-llm-prompt}.
For a more exhaustive benchmark of more models across multiple chemical and materials tasks, we refer the reader to the recent MaCBench and ChemBench initiatives~\cite{Mirza2024, Alampara2025}.

\begin{figure}[H]\centering
\includegraphics[width=0.4\linewidth]{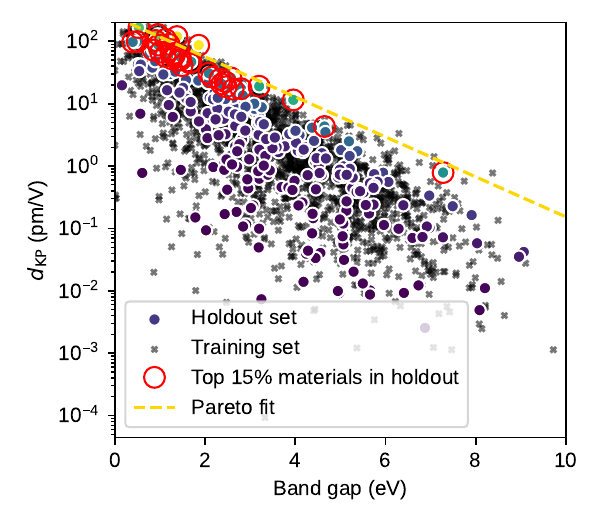}
\caption{The holdout set \emph{distribution\_250} plotted alongside the overall training set, with the top 15\% of materials according to the $T_0$-fitted Pareto front (gold) FOM highlighted in red.\label{fig:top-10pc-distribution_250}}
\end{figure}

\begin{figure}[H]
\begin{Verbatim}[fontsize=\small,frame=single,xleftmargin=10mm,xrightmargin=10mm]
Given a description of a crystal structure (composition), predict its 
second-harmonic generation (SHG) coefficient in the Kurtz-Perry form in pm/V.
    
All the structures you see will be non-centrosymmetric.
In our dataset, the SHG coefficients are computed with DFPT at the PBE level.

Most structures exhibit low SHG coefficients (below 10 pm/V), with exemplary 
materials ranging up to 170 pm/V.
Simply respond with the value which will be read as a raw float, do not provide 
any explanation.

Low SHG examples include:

    formula_reduced spg_symbol  dKP_full_neum
              YCuS2 P2_12_12_1       0.010081
               KCuS     Pna2_1       0.038218
              CsNO2     P3_121       0.038580
            Be4TeO7      F-43m       0.082648
        CdH4(BrO4)2 P2_12_12_1       0.035272
             SbIrSe      P2_13       0.141097
            RbGeIO6       P312       0.069342
        LiAs(XeF4)3       P2_1       0.059930
        K2Zn(SiO3)2     C222_1       0.046831
            Na2SiO3     Cmc2_1       0.056783

High SHG examples include:

    formula_reduced spg_symbol  dKP_full_neum
                GaP      F-43m      48.751564
              InPS4        I-4      33.199950
            Hg2P2S7         C2      45.587665
               K2S3     Cmc2_1      62.721255
           Al2ZnTe4        I-4      55.794952
             NbFeSb      F-43m     124.694574
              BiIrS      P2_13     134.980430
           Al2CdTe4        I-4      57.683547
            LiInTe2      I-42d      68.552174
            MgSiAs2      I-42d      72.369350

Input: TeO2 in P2_12_12_1

Response: 3.201152
\end{Verbatim}
\caption{An example prompt provided to LLMs for the prediction of the material \ce{TeO2} with space group $P2_{1}2_{1}2_{1}$ with the composition-space group structural description and the degree of in-context learning (ICL) set to "10" (i.e., take 10 low SHG examples and 10 high SHG examples); the real ICL benchmark used values up to 200.}
\label{fig:example-llm-prompt}
\end{figure}


\subsection{Supplementary figures for the active learning procedure}
\label{sec:al-appendix}
\begin{figure}[H]\centering
\includegraphics[width=0.45\linewidth]{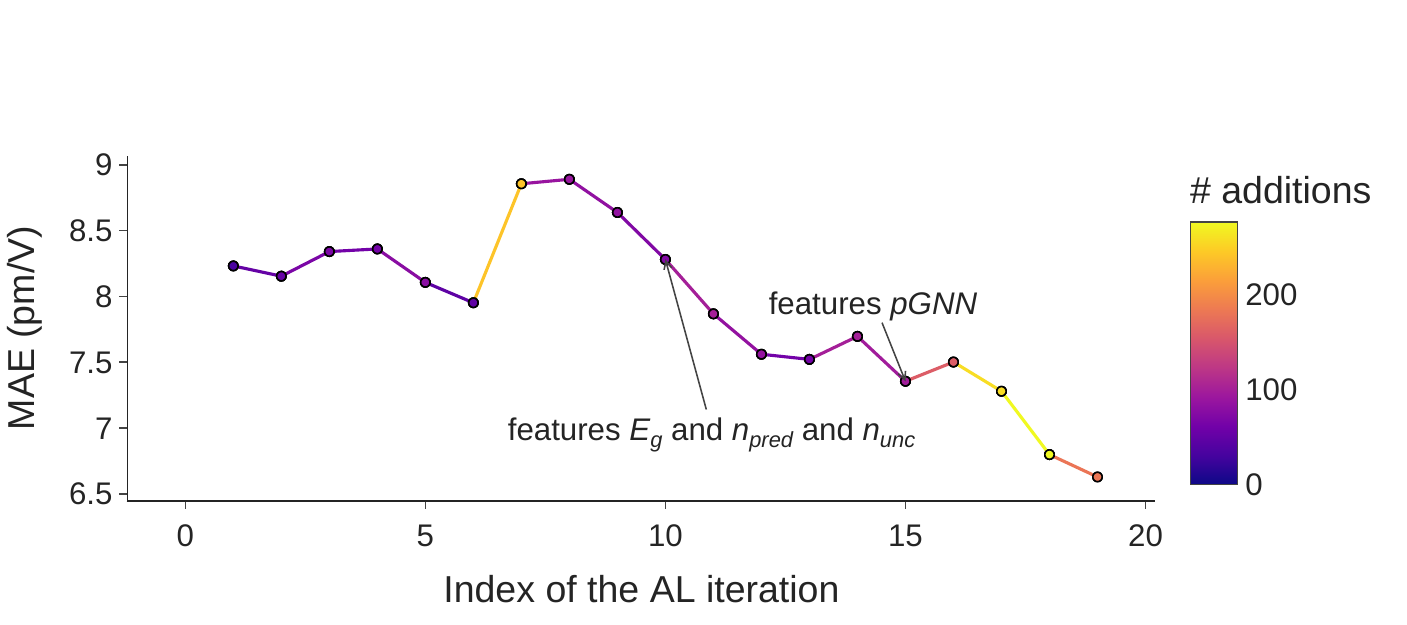}
\caption{Evolution of the average MAE (pm/V) from a nested 5-folds cross-validation during the active learning procedure. This plot was built at the time of the active learning using the raw data. The iteration 0 was not benchmarked with the same method so it is omitted here.}
\label{fig: raw_bmk_al_mae}
\end{figure}
\begin{figure}[H]\centering
\includegraphics[width=0.45\linewidth]{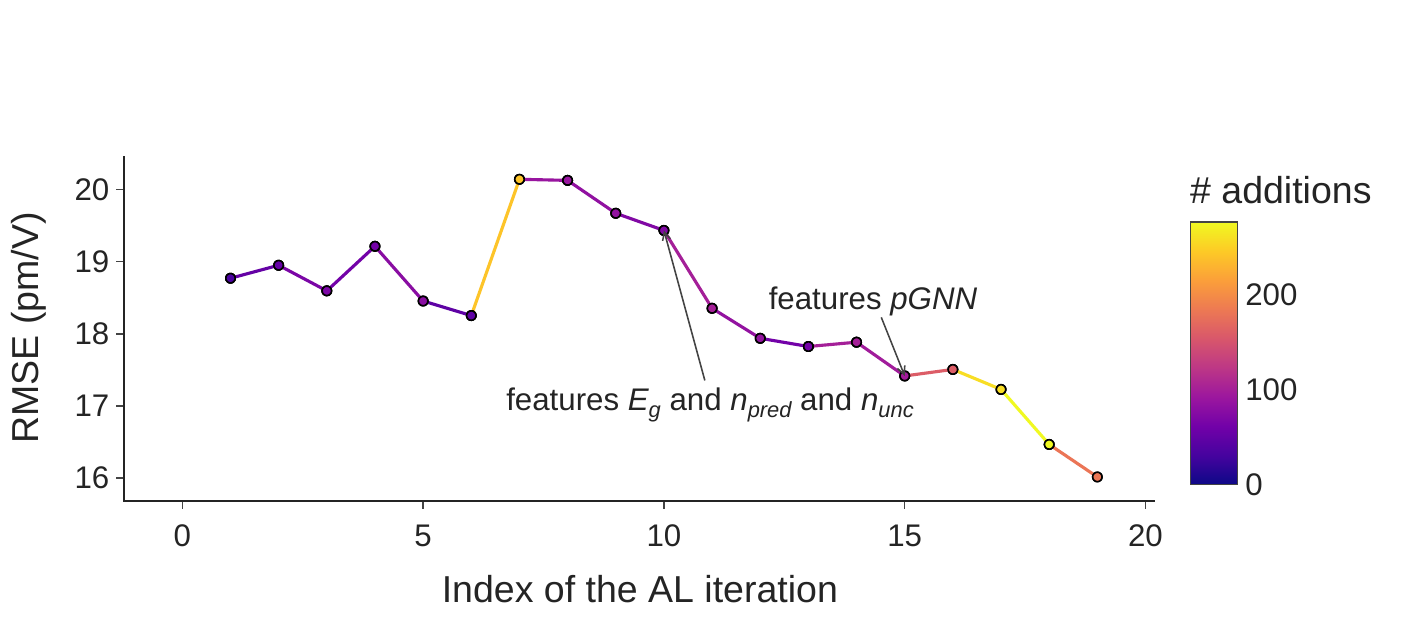}
\caption{Evolution of the average RMSE (pm/V) from a nested 5-folds cross-validation during the active learning procedure. This plot was built at the time of the active learning using the raw data. The iteration 0 was not benchmarked with the same method so it is omitted here.}
\label{fig: raw_bmk_al_rmse}
\end{figure}
\begin{figure}[H]\centering
\includegraphics[width=0.45\linewidth]{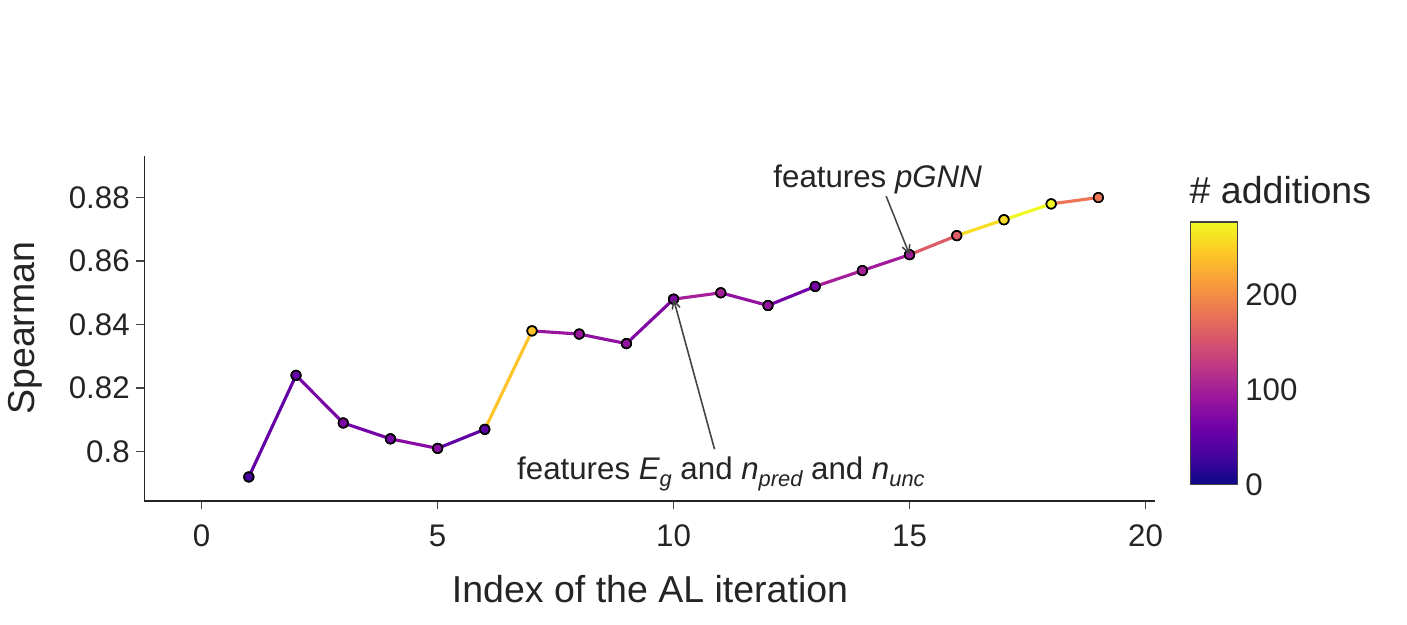}
\caption{Evolution of the average Spearman from a nested 5-folds cross-validation during the active learning procedure. This plot was built at the time of the active learning using the raw data. The iteration 0 was not benchmarked with the same method so it is omitted here.}
\label{fig: raw_bmk_al_spr}
\end{figure}
\begin{figure}[H]\centering
\includegraphics[width=0.45\linewidth]{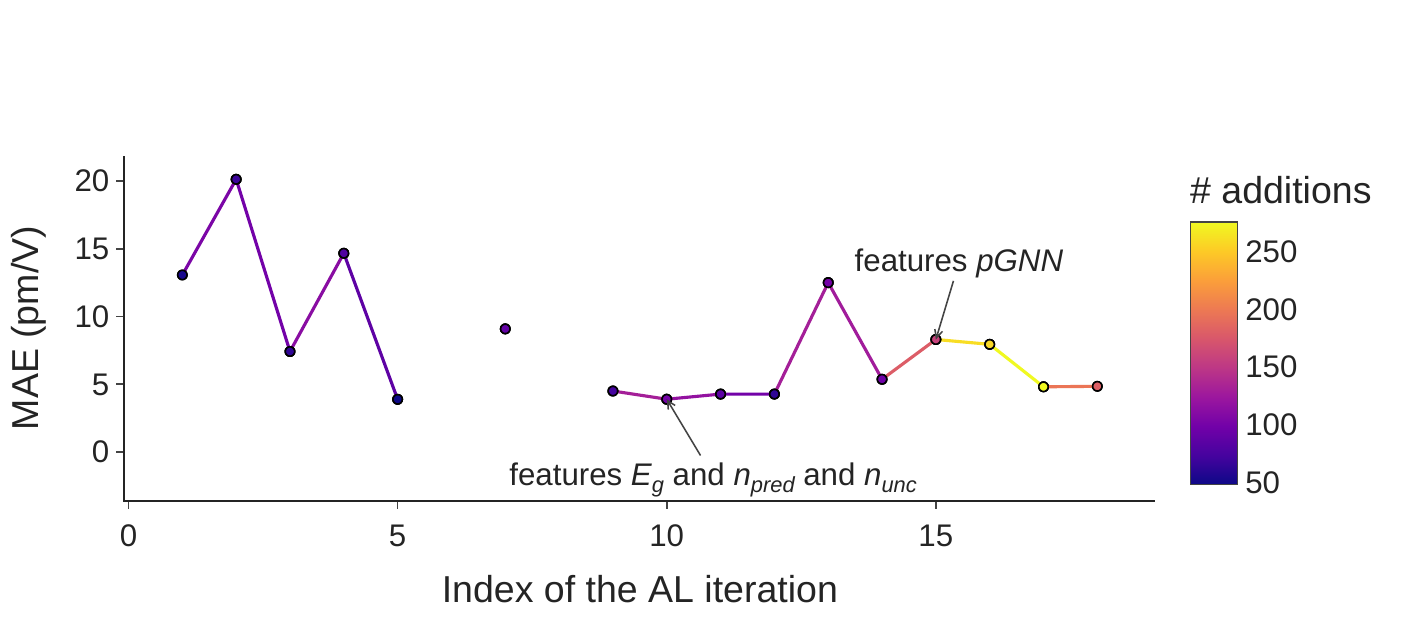}
\caption{MAE (pm/V) on the predictions of the selected compounds at the $i^\text{th}$ AL iteration after they had been computed. This plot was built at the time of the active learning using the raw data. The iterations 6 and 8 are missing because they respectively correspond to the addition of materials from \citet{Trinquet2025Jan} without any predictions and to a replacement of unconverged values in the dataset by correct ones.}
\label{fig: raw_bmk_al_mae_selection}
\end{figure}
\begin{figure}[H]\centering
\includegraphics[width=0.45\linewidth]{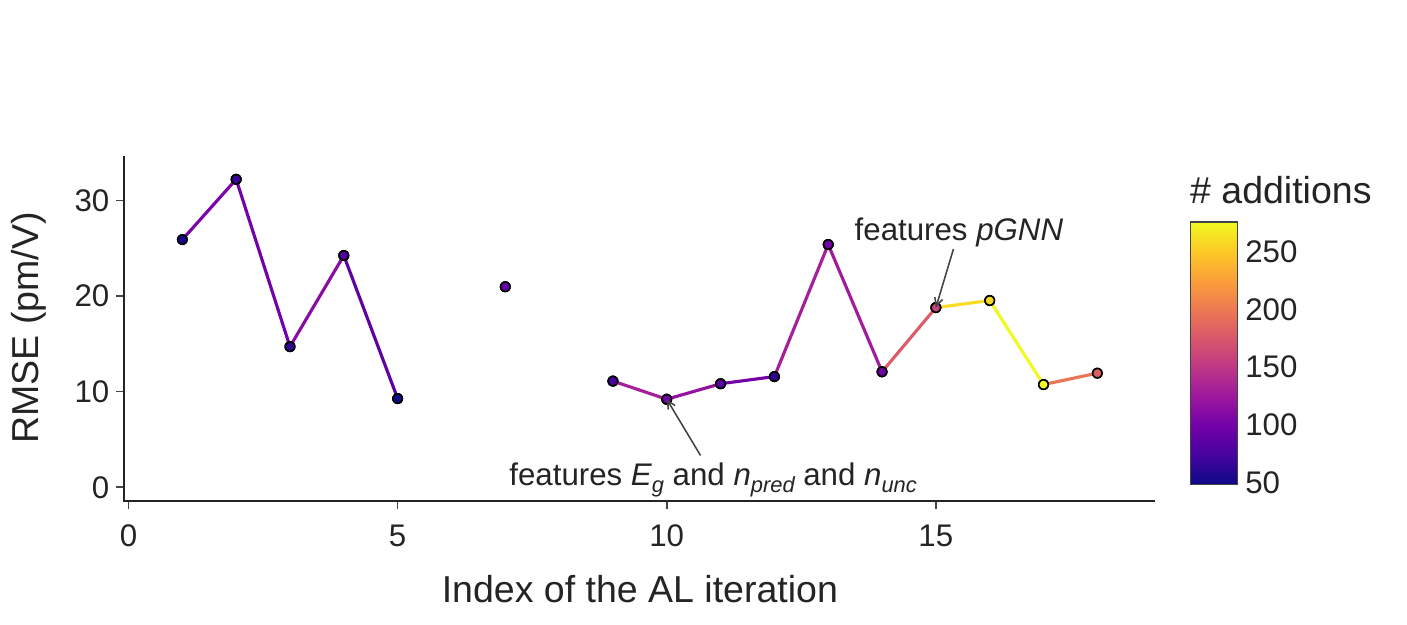}
\caption{RMSE (pm/V) on the predictions of the selected compounds at the $i^\text{th}$ AL iteration after they had been computed. This plot was built at the time of the active learning using the raw data. The iterations 6 and 8 are missing because they respectively correspond to the addition of materials from \citet{Trinquet2025Jan} without any predictions and to a replacement of unconverged values in the dataset by correct ones.}
\label{fig: raw_bmk_al_rmse_selection}
\end{figure}
\begin{figure}[H]\centering
\includegraphics[width=0.45\linewidth]{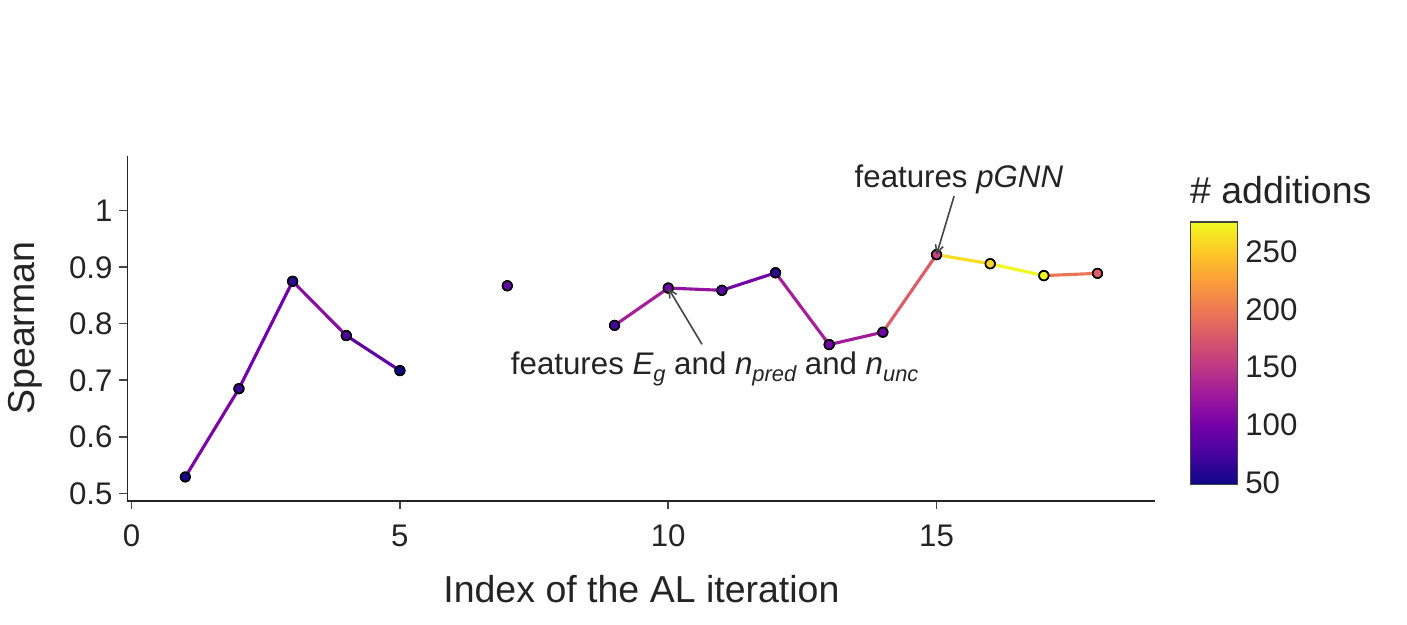}
\caption{Spearman on the predictions of the selected compounds at the $i^\text{th}$ AL iteration after they had been computed. This plot was built at the time of the active learning using the raw data. The iterations 6 and 8 are missing because they respectively correspond to the addition of materials from \citet{Trinquet2025Jan} without any predictions and to a replacement of unconverged values in the dataset by correct ones.}
\label{fig: raw_bmk_al_spr_selection}
\end{figure}

\begin{figure}[H]\centering
\includegraphics[width=0.5\linewidth]{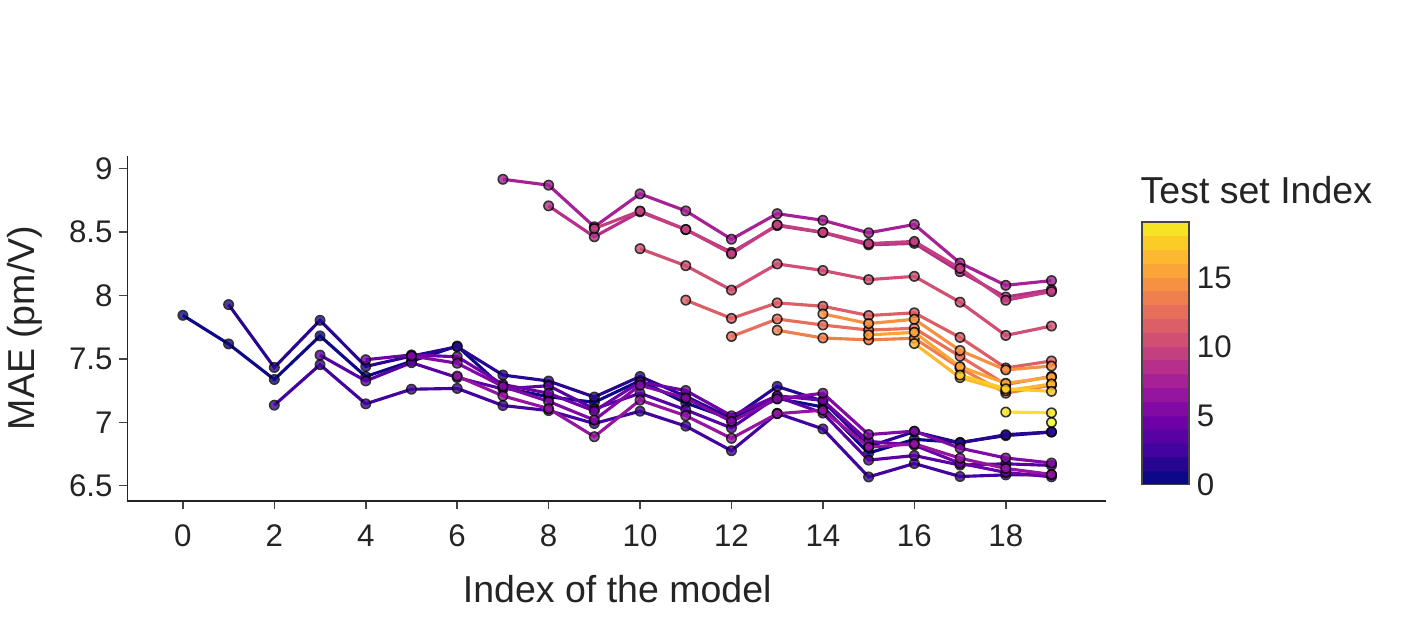}
\caption{Evolution of the average MAE (pm/V) over the AL process. The index of the model refers to its training set as the AL goes on. Each curve with index $i$ corresponds to the test sets of a 5-folds splitting of the dataset at the $i^\text{th}$ iteration of the AL procedure such that the same test sets are kept for the whole curve.}
\label{fig:bmk_al_mae}
\end{figure}
\begin{figure}[H]\centering
\includegraphics[width=0.5\linewidth]{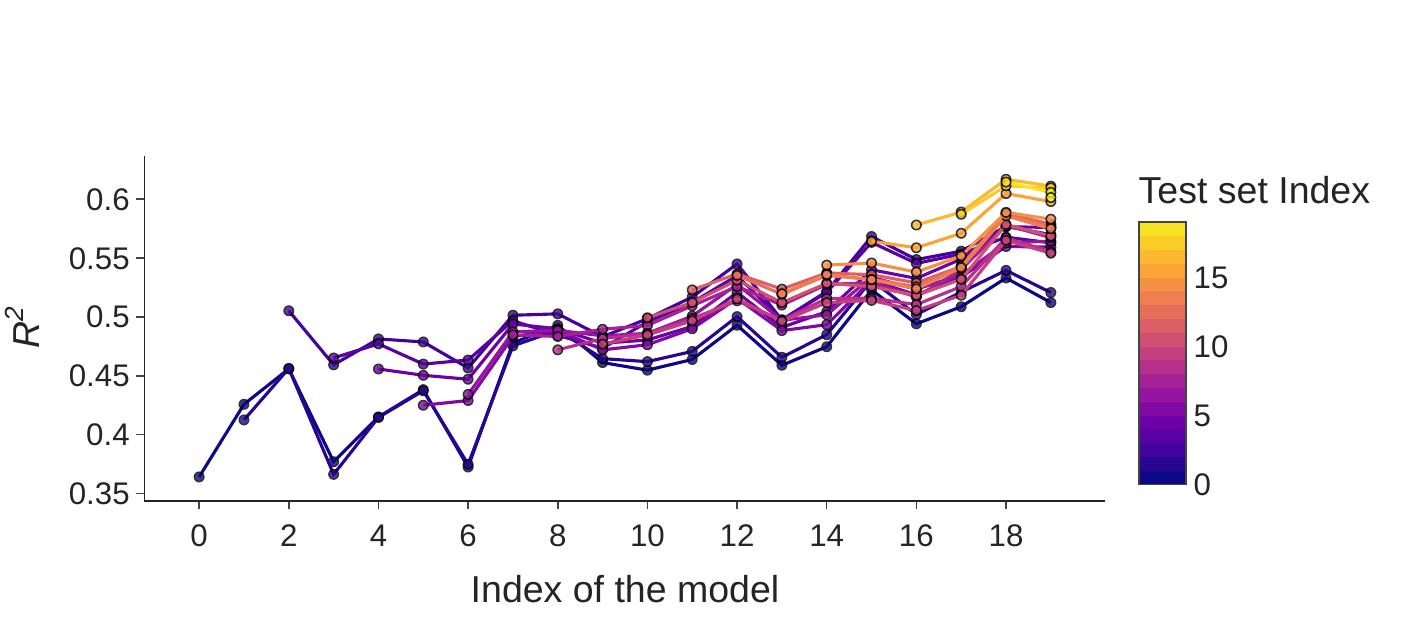}
\caption{Evolution of the average $R^2$ over the AL process. The index of the model refers to its training set as the AL goes on. Each curve with index $i$ corresponds to the test sets of a 5-folds splitting of the dataset at the $i^\text{th}$ iteration of the AL procedure such that the same test sets are kept for the whole curve.}
\label{fig:bmk_al_r2}
\end{figure}


\subsection{Supplementary figures for multi-fidelity correction learning}

\begin{table}[H]
\resizebox{\textwidth}{!}{
\centering
\begin{tabular}{lcccccc}
 & MAE (pm/V) & RMSE (pm/V) & Spearman & $R^2$ & $\eta$ (\%) & $\zeta$ (\%) \\
Linear regression & {\cellcolor[HTML]{54C568}} 1.6742 & {\cellcolor[HTML]{3FBC73}} \textcolor{white}{4.6544} & {\cellcolor[HTML]{FDE725}}0.9972 & {\cellcolor[HTML]{7FD34E}}0.9173 & {\cellcolor[HTML]{FDE725}} 0.0000 & {\cellcolor[HTML]{FDE725}} 0.0000 \\
\emph{mmf\_pgnn} & {\cellcolor[HTML]{440154}} \textcolor{white}{3.5342} & {\cellcolor[HTML]{460B5E}} \textcolor{white}{9.3651} & {\cellcolor[HTML]{440154}} \textcolor{white}{0.8895} & {\cellcolor[HTML]{48186A}} \textcolor{white}{0.6861} & {\cellcolor[HTML]{AADC32}} 0.1481 & {\cellcolor[HTML]{440154}} \textcolor{white}{10.3355} \\
\emph{mmf\_pgnn} $\cup$ $E_g^\text{HSE}$ & {\cellcolor[HTML]{481668}} \textcolor{white}{3.3949} & {\cellcolor[HTML]{440154}} \textcolor{white}{9.5840} & {\cellcolor[HTML]{3F4788}} \textcolor{white}{0.9127} & {\cellcolor[HTML]{440154}} \textcolor{white}{0.6659} & {\cellcolor[HTML]{FDE725}} 0.0000 & {\cellcolor[HTML]{443A83}} \textcolor{white}{8.5708} \\
\emph{mmf\_pgnn} $\cup$ $E_g^\text{LDA}$ & {\cellcolor[HTML]{3F4889}} \textcolor{white}{2.9861} & {\cellcolor[HTML]{3E4989}} \textcolor{white}{8.0167} & {\cellcolor[HTML]{2F6B8E}} \textcolor{white}{0.9266} & {\cellcolor[HTML]{355E8D}} \textcolor{white}{0.7574} & {\cellcolor[HTML]{2D718E}} \textcolor{white}{0.7407} & {\cellcolor[HTML]{3C508B}} \textcolor{white}{7.8235} \\
\emph{mmf\_pgnn} $\cup$ $\Delta E_g$ & {\cellcolor[HTML]{481C6E}} \textcolor{white}{3.3358} & {\cellcolor[HTML]{46085C}} \textcolor{white}{9.4404} & {\cellcolor[HTML]{3C4F8A}} \textcolor{white}{0.9154} & {\cellcolor[HTML]{46075A}} \textcolor{white}{0.6717} & {\cellcolor[HTML]{FDE725}} 0.0000 & {\cellcolor[HTML]{481F70}} \textcolor{white}{9.4521} \\
\emph{mmf\_pgnn} $\cup$ $d_\text{LDA}$ & {\cellcolor[HTML]{81D34D}} 1.4700 & {\cellcolor[HTML]{56C667}} 4.3213 & {\cellcolor[HTML]{AADC32}}0.9837 & {\cellcolor[HTML]{98D83E}}0.9285 & {\cellcolor[HTML]{2C718E}} \textcolor{white}{0.7397} & {\cellcolor[HTML]{26828E}} \textcolor{white}{5.7614} \\
\emph{mmf\_pgnn} $\cup$ $d_\text{LDA}$ $\cup$ $E_g^\text{LDA}$ & {\cellcolor[HTML]{9DD93B}} 1.3659 & {\cellcolor[HTML]{75D054}} 3.9336 & {\cellcolor[HTML]{C2DF23}}0.9872 & {\cellcolor[HTML]{B2DD2D}}0.9412 & {\cellcolor[HTML]{AADC32}} 0.1481 & {\cellcolor[HTML]{20938C}} \textcolor{white}{5.0218} \\
\emph{mmf\_pgnn} $\cup$ $d_\text{LDA}$ $\cup$ $E_g^\text{HSE}$ & {\cellcolor[HTML]{9DD93B}} 1.3611 & {\cellcolor[HTML]{5CC863}} 4.2510 & {\cellcolor[HTML]{DFE318}}0.9918 & {\cellcolor[HTML]{9DD93B}}0.9307 & {\cellcolor[HTML]{21908D}} \textcolor{white}{0.5893} & {\cellcolor[HTML]{44BF70}} \textcolor{white}{3.0969} \\
\emph{mmf\_pgnn} $\cup$ $d_\text{LDA}$ $\cup$ $\Delta E_g$ & {\cellcolor[HTML]{E5E419}} 1.0984 & {\cellcolor[HTML]{D5E21A}} 2.9003 & {\cellcolor[HTML]{C0DF25}}0.9870 & {\cellcolor[HTML]{EAE51A}}0.9659 & {\cellcolor[HTML]{21908D}} \textcolor{white}{0.5893} & {\cellcolor[HTML]{22A884}} \textcolor{white}{4.1307} \\
\emph{mmf\_pgnn} $\cup$ $d_\text{LDA}$ $\cup$ $E_g^\text{LDA}$ $\cup$ $\Delta E_g$ & {\cellcolor[HTML]{E7E419}} 1.0794 & {\cellcolor[HTML]{E5E419}} 2.7334 & {\cellcolor[HTML]{DDE318}}0.9916 & {\cellcolor[HTML]{EFE51C}}0.9687 & {\cellcolor[HTML]{3B518B}} \textcolor{white}{0.8845} & {\cellcolor[HTML]{37B878}} \textcolor{white}{3.3932} \\
\emph{mmf\_pgnn} $\cup$ $d_\text{LDA}$ $\cup$ $E_g^\text{HSE}$ $\cup$ $\Delta E_g$ & {\cellcolor[HTML]{FDE725}} 0.9895 & {\cellcolor[HTML]{FDE725}} 2.4398 & {\cellcolor[HTML]{C0DF25}}0.9869 & {\cellcolor[HTML]{FDE725}}0.9769 & {\cellcolor[HTML]{472C7A}} \textcolor{white}{1.0316} & {\cellcolor[HTML]{22A884}} \textcolor{white}{4.1307} \\
\emph{mmf\_pgnn} $\cup$ $d_\text{LDA}$ $\cup$ $E_g^\text{LDA}$ $\cup$ $E_g^\text{HSE}$ $\cup$ $\Delta E_g$ & {\cellcolor[HTML]{EAE51A}} 1.0766 & {\cellcolor[HTML]{E7E419}} 2.6977 & {\cellcolor[HTML]{C8E020}}0.9882 & {\cellcolor[HTML]{F4E61E}}0.9711 & {\cellcolor[HTML]{440154}} \textcolor{white}{1.1776} & {\cellcolor[HTML]{3DBC74}} \textcolor{white}{3.2484} \\
\end{tabular}
}
\caption{Performance of MODNet on the $d_\text{corr}$ task when using different set of features under a nested 5-folds cross-validation. The quantities $\eta$ (\%) and $\zeta$ (\%) correspond to the fraction of correction to $d_\text{KP}$ with a wrong sign and of negative ($d_\text{LDA}-d_\text{corr}$), respectively.}
\label{tab: correction learning full}
\end{table}

\begin{figure}[H]\centering
\includegraphics[width=0.45\linewidth]{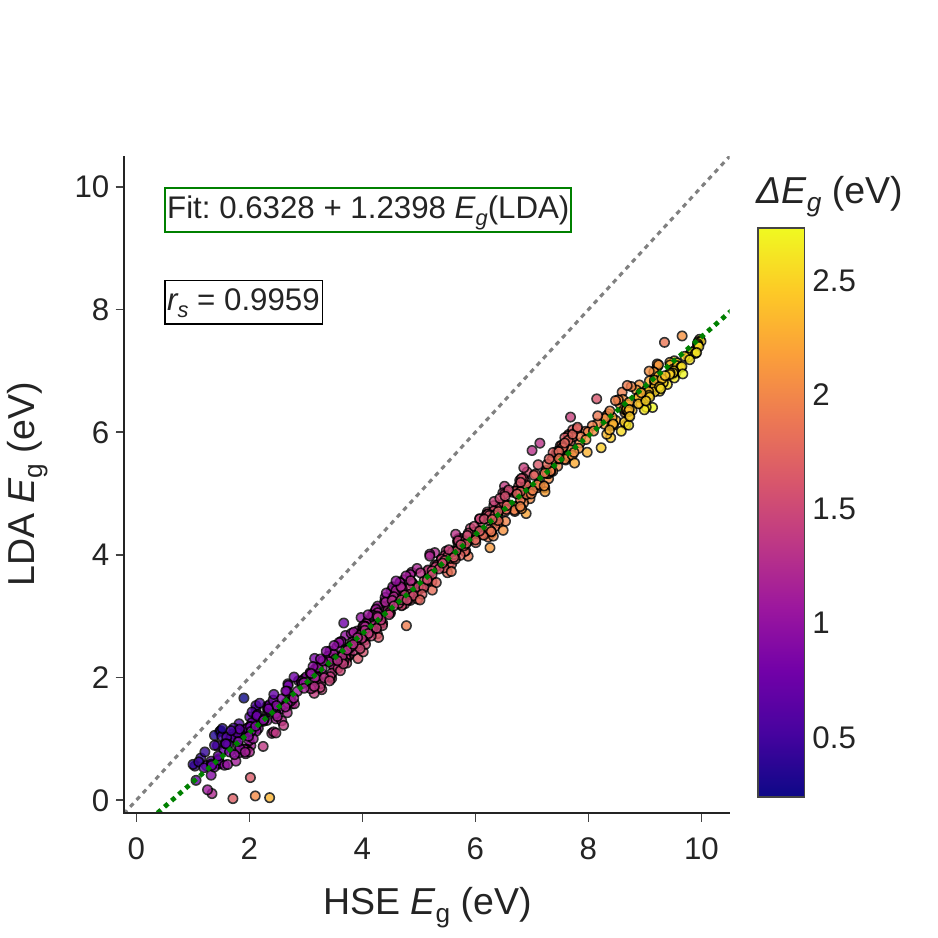}
\caption{Parity plot comparing the HSE and LDA band gaps. The colorbar indicates the scissor of each compound needed to match the HSE gap. A linear regression is fitted on those points as shown by the green dotted line and the formula in the box. The Spearman's rank correlation coefficient, $r_s$, is included as well.}
\label{fig: parity_plot_gaps}
\end{figure}
\begin{figure}[H]\centering
\includegraphics[width=0.45\linewidth]{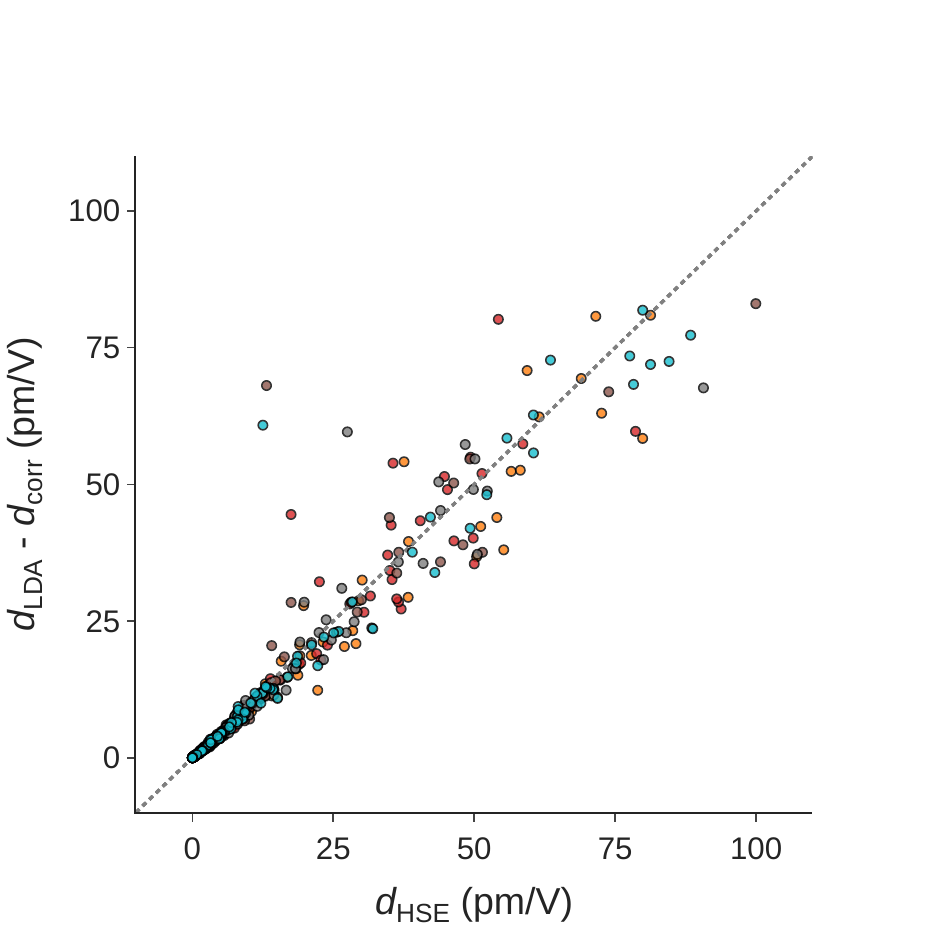}
\caption{Parity plot comparing the HSE KP coefficient and its predicted value via correction learning with the linear regression. The colours correspond to the different folds of the cross-validation scheme.}
\label{fig: parity_plot_dKP_corr_fit}
\end{figure}
\begin{figure}[H]\centering
\includegraphics[width=0.45\linewidth]{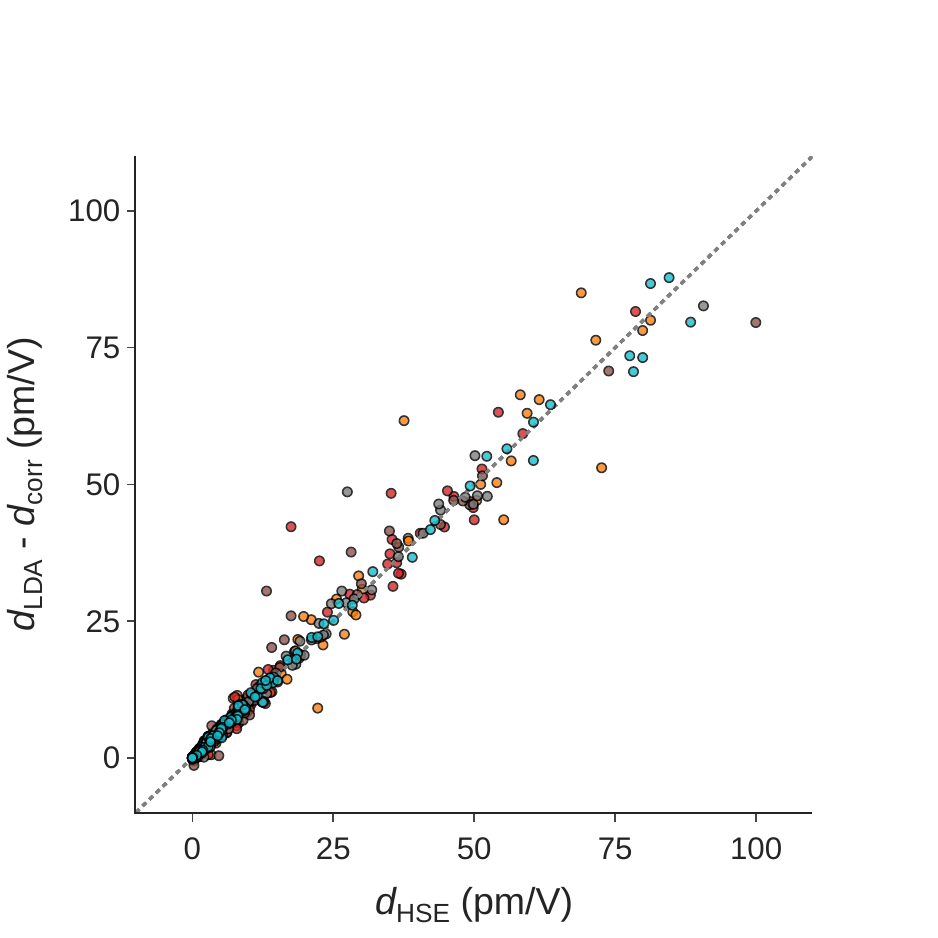}
\caption{Parity plot comparing the HSE KP coefficient and its predicted value via correction learning with MODNet and all custom features. The colours correspond to the different folds of the cross-validation scheme.}
\label{fig: parity_plot_dKP_corr_modnet}
\end{figure}
\begin{figure}[H]\centering
\includegraphics[width=0.45\linewidth]{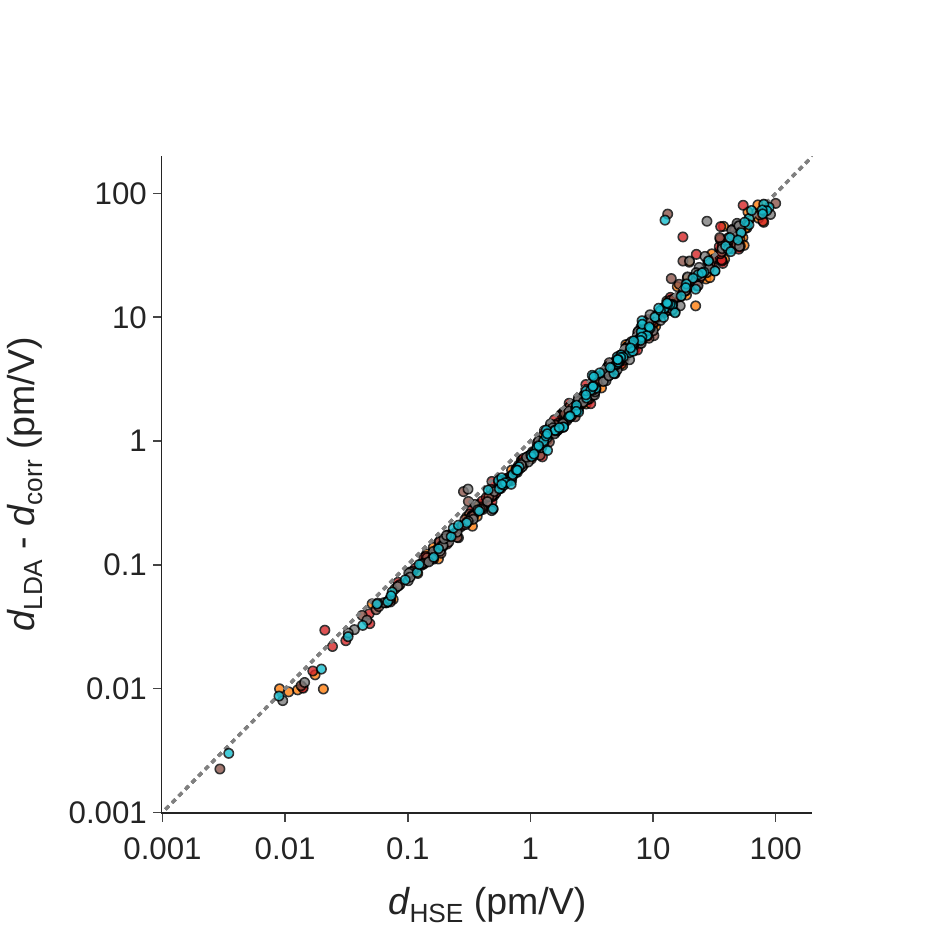}
\caption{Log-log parity plot comparing the HSE KP coefficient and its predicted value via correction learning with the linear regression. The colours correspond to the different folds of the cross-validation scheme.}
\label{fig: parity_plot_dKP_corr_fit_log}
\end{figure}
\begin{figure}[H]\centering
\includegraphics[width=0.45\linewidth]{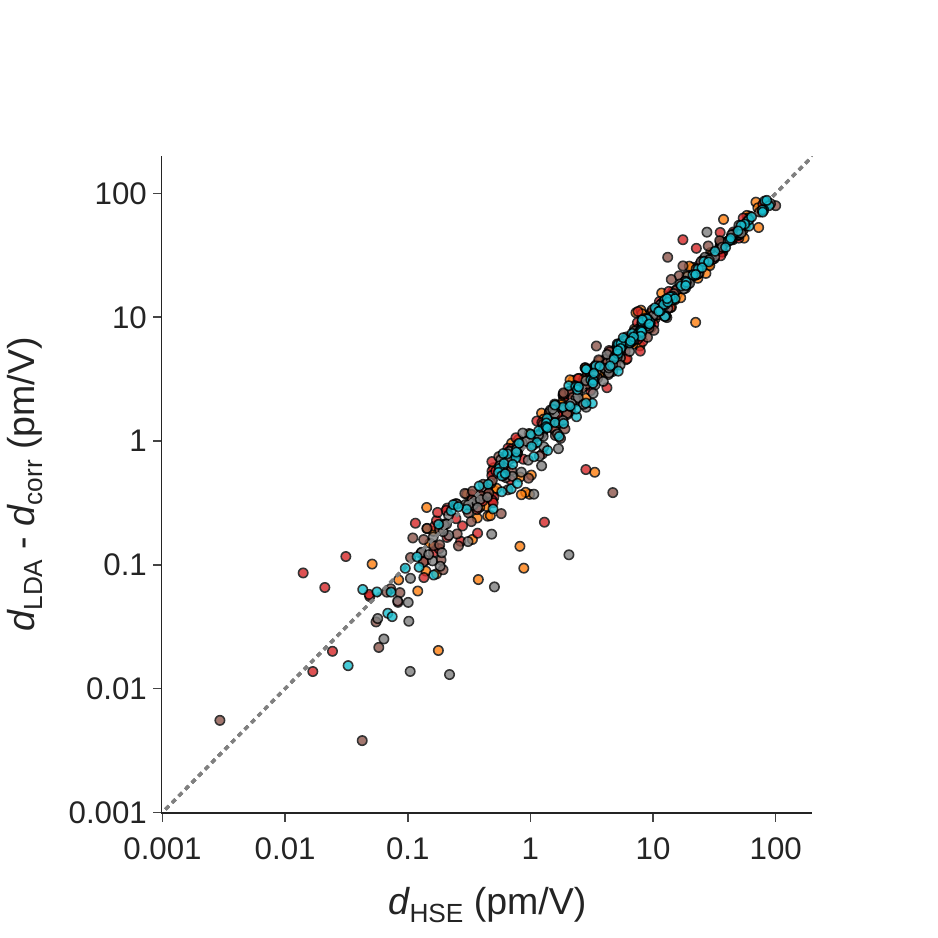}
\caption{Log-log parity plot comparing the HSE KP coefficient and its predicted value via correction learning with MODNet and all custom features. The colours correspond to the different folds of the cross-validation scheme.}
\label{fig: parity_plot_dKP_corr_modnet_log}
\end{figure}

\subsection{Promising materials}

\begin{table}[H]
\centering
\resizebox{0.8\textwidth}{!}{
\begin{tabular}{lccccccc}
\toprule
 Identifier & Formula & $E_g^\text{HSE}$ (eV) & $d_\text{HSE}$ (pm/V) & $\Delta n_\text{HSE}$ & Spacegroup & \citet{Wang2024Aug} & \citet{Chu2023Apr} \\
\midrule
mp-966800 & \ce{InP} & 1.261 & 27.499 & 0.108 & $P6_3mc$ & - & - \\
mp-1215429 & \ce{ZnSnP2} & 1.325 & 49.348 & 0.092 & $P\bar{4}m2$ & - & - \\
agm003450028 & \ce{Mg(InTe2)2} & 1.461 & 63.528 & 0.038 & $P\bar{4}2m$ & - & - \\
mp-35777 & \ce{Mg(InTe2)2} & 1.464 & 49.854 & 0.055 & $Cm$ & - & - \\
mp-571195 & \ce{ZnTe} & 1.938 & 46.356 & 0.035 & $P3_1$ & - & - \\
mp-1222182 & \ce{Mg(InTe2)2} & 1.960 & 51.373 & 0.041 & $I\bar{4}$ & - & - \\
agm002160623 & \ce{Mg(GaTe2)2} & 2.140 & 61.522 & 0.076 & $I\bar{4}$ & - & - \\
agm002088965 & \ce{InTeI} & 2.158 & 55.801 & 0.385 & $P2_1$ & - & - \\
agm002156796 & \ce{Li2SiSnSe4} & 2.306 & 30.451 & 0.257 & $Cm$ & - & - \\
agm005605697 & \ce{GaTeI} & 2.314 & 44.027 & 0.411 & $Pmn2_1$ & - & - \\
agm002790067 & \ce{AlInP2} & 2.399 & 27.310 & 0.039 & $I\bar{4}2d$ & - & - \\
agm005056337 & \ce{MgInGaSe4} & 2.497 & 21.144 & 0.052 & $I\bar{4}$ & - & - \\
agm002283412 & \ce{Li2SnGeS4} & 2.647 & 16.802 & 0.239 & $Cm$ & - & - \\
agm002160619 & \ce{Mg(GaSe2)2} & 2.748 & 21.159 & 0.064 & $I\bar{4}$ & - & - \\
agm002160138 & \ce{Mg(AlTe2)2} & 2.872 & 23.966 & 0.041 & $I\bar{4}$ & - & - \\
agm2000111340 & \ce{GaTeCl} & 3.128 & 8.074 & 0.536 & $Pmn2_1$ & - & - \\
mp-4586 & \ce{LiAlTe2} & 3.156 & 15.120 & 0.058 & $I\bar{4}2d$ & - & \checkmark \\
mp-27529 & \ce{PI3} & 3.199 & 8.224 & 0.265 & $P6_3$ & - & - \\
agm002793928 & \ce{NaBSe2} & 3.361 & 9.264 & 0.058 & $I\bar{4}2d$ & - & - \\
mp-690 & \ce{P4S5} & 3.413 & 6.937 & 0.164 & $P2_1$ & \checkmark & - \\
agm2000135800 & \ce{GaTeCl} & 3.499 & 7.012 & 0.507 & $Pca2_1$ & - & - \\
agm006047631 & \ce{Ga4SnS7} & 3.527 & 8.098 & 0.099 & $Pmn2_1$ & - & - \\
mp-20790 & \ce{InPS4} & 3.538 & 18.361 & 0.057 & $I\bar{4}$ & - & \checkmark \\
mp-30294 & \ce{Sr2SnS4} & 3.630 & 7.997 & 0.076 & $Ama2$ & \checkmark & \checkmark \\
agm005605595 & \ce{GaSeCl} & 3.653 & 9.915 & 0.384 & $Pmn2_1$ & - & - \\
agm002161193 & \ce{Mg(GaS2)2} & 3.773 & 8.978 & 0.051 & $I\bar{4}$ & - & - \\
mp-1227993 & \ce{BaGa2SiS6} & 3.909 & 10.020 & 0.063 & $P1$ & - & \checkmark \\
agm002157245 & \ce{LiAlSe2} & 3.962 & 5.415 & 0.041 & $I\bar{4}2d$ & - & - \\
mp-2646995 & \ce{Li3PS4} & 3.991 & 4.472 & 0.039 & $I\bar{4}2m$ & - & - \\
agm002158826 & \ce{LiGaS2} & 4.078 & 7.071 & 0.060 & $I\bar{4}2d$ & - & - \\
agm005605654 & \ce{GaSCl} & 4.264 & 5.416 & 0.260 & $Pmn2_1$ & - & - \\
mp-559065 & \ce{NaI3O8} & 4.474 & 4.325 & 0.149 & $P\bar{4}$ & - & - \\
agm002157243 & \ce{LiAlS2} & 4.620 & 2.641 & 0.031 & $I\bar{4}2d$ & - & - \\
mp-561104 & \ce{Ga(IO3)3} & 4.699 & 8.119 & 0.098 & $P6_3$ & - & - \\
mp-555903 & \ce{Al(IO3)3} & 4.747 & 7.942 & 0.082 & $P6_3$ & - & - \\
mp-559545 & \ce{SeO2} & 4.835 & 3.674 & 0.299 & $Pmc2_1$ & - & - \\
mp-27367 & \ce{SeOF2} & 5.537 & 2.830 & 0.051 & $Pca2_1$ & \checkmark & - \\
agm002163269 & \ce{CaSiN2} & 5.561 & 3.424 & 0.034 & $I\bar{4}2d$ & - & - \\
mp-22909 & \ce{ZnCl2} & 5.574 & 1.373 & 0.031 & $I\bar{4}2d$ & - & - \\
agm005604809 & \ce{AlSCl} & 5.743 & 1.164 & 0.179 & $Pca2_1$ & - & - \\
agm002137165 & \ce{CaAl2B2O7} & 6.405 & 0.713 & 0.066 & $R32$ & - & - \\
mp-5730 & \ce{Ba(BO2)2} & 6.418 & 1.365 & 0.115 & $R3c$ & - & - \\
mp-5853 & \ce{LiSi2N3} & 6.518 & 0.821 & 0.037 & $Cmc2_1$ & \checkmark & - \\
mp-557391 & \ce{Na2Ca2(CO3)3} & 6.533 & 0.631 & 0.034 & $Amm2$ & - & - \\
mp-753671 & \ce{PNO} & 6.705 & 1.844 & 0.121 & $I2_12_12_1$ & \checkmark & - \\
mp-36066 & \ce{PNO} & 6.736 & 1.612 & 0.151 & $Cc$ & \checkmark & - \\
agm005607967 & \ce{SiNF} & 6.737 & 0.471 & 0.188 & $Pmn2_1$ & - & - \\
mp-6524 & \ce{CaMg3(CO3)4} & 6.898 & 0.800 & 0.154 & $R32$ & - & - \\
mp-1195844 & \ce{Ba3B6O11F2} & 6.940 & 0.761 & 0.046 & $P2_1$ & - & - \\
mp-1202821 & \ce{Sr3B6O11F2} & 7.143 & 0.624 & 0.041 & $P2_1$ & - & - \\
mp-1020019 & \ce{Li2PNO2} & 7.271 & 0.449 & 0.103 & $Cmc2_1$ & - & - \\
mp-1200209 & \ce{Li2B6O9F2} & 7.993 & 0.485 & 0.061 & $Cc$ & - & - \\
mp-3660 & \ce{LiB3O5} & 8.166 & 0.561 & 0.042 & $Pna2_1$ & - & - \\
mp-1019509 & \ce{B2S2O9} & 9.158 & 0.656 & 0.034 & $C2$ & - & - \\
\bottomrule
\end{tabular}
}
\caption{List of the best materials boasting good theoretical stability ($E_\text{hull}\leq$ \SI{10}{\milli\electronvolt\per\atom}, versus the convex hull of known materials), an HSE gap greater than \SI{1.0}{\electronvolt}, a birefringence ($\Delta n_\text{HSE}$) larger than \SI{0.03}{}, and a scissor-corrected KP coefficient ($d_\text{HSE}$) greater than \SI{0.33}{\pico\meter\per\volt}. Only elements with an HHI lower than 6,000 were retained and toxic elements were removed (see text for more details). 
The MP entries appearing in the HSE datasets of~\citet{Wang2024Aug} and ~\citet{Chu2023Apr} are also flagged.}
\label{tab: best materials}
\end{table}
\end{document}